\documentclass[11pt,a4paper]{article}
\pdfoutput=1
\usepackage{jheppub}
\usepackage{amsmath,amscd}
\usepackage{amsfonts}
\usepackage{amssymb}
\usepackage{graphicx}
\usepackage{float}
\usepackage[caption = false]{subfig}
\usepackage{color}
\usepackage[normalem]{ulem}
\usepackage{hyperref}
\usepackage{linearb}

\newcommand{\ex}{{\rm e}}

\newcommand{\dif}{{\rm d}}

\newcommand{\RR}{\mathbb{R}} 
\newcommand{\ZZ}{\mathbb{Z}} 

\newcommand{\G}{\mathcal{G}}

\def\tr         {{\rm  tr}}

\def\calm         {{\cal M}}

\def\calo         {{\cal O}}

\def\calt         {{\cal T}}

\def\calv         {{\cal V}}

\def\be{\begin{equation}}
\def\ee{\end{equation}}
\def\bea{\begin{eqnarray}}
\def\eea{\end{eqnarray}}

\def\a{\alpha}

\def\h{\eta}

\def\G{\Gamma}
\def\d{\delta}
\def\e{\epsilon}
\def\D{\Delta}
\def\l{\lambda}
\def\L{\Lambda}

\def\f{\phi}

\def\m{\mu}
\def\n{\nu}
\def\o{\omega}

\def\p{\pi}
\def\r{\rho}

\def\x{\xi}
\def\s{\sigma}

\def\t{\tau}

\def\sF{{{ F}\!\!\!\!\hskip.8pt\hbox{\raise1pt\hbox{/}}\,}}
\def\som{{{ \omega}\!\!\!\!\hskip.8pt\hbox{\raise1pt\hbox{/}}\,}}
\def\sJ{{{\rm J}\!\!\!\!\hskip.8pt\hbox{\raise1pt\hbox{/}}\,}}

\def\F{\Phi}
\def\pa{\partial}

\def\to{\rightarrow}
\def\nonu{\nonumber \\{}}
\def\half{{1 \over 2}}

\title{Information recovery from pure state geometries in 3D}

\author[a,b]{Ond\v{r}ej Hul\'{i}k,}
\author[a]{Joris Raeymaekers}
\author[a]{and Orestis Vasilakis}
\affiliation[a]{Institute of Physics of the Czech Academy of Sciences, CEICO, \\ Na Slovance 2, 182 21 Prague 8, Czech Republic}

\affiliation[b]{Institute of Particle Physics and Nuclear Physics, Faculty of Mathematics and Physics, Charles University,  V Hole\v{s}ovi\v{c}k\'{a}ch 2, 180 00 Prague 8, Czech Republic}

\emailAdd{ondra.hulik@gmail.com}
\emailAdd{joris@fzu.cz}
\emailAdd{vasilakis@fzu.cz}

\abstract{It is a well-studied phenomenon in AdS$_3$/CFT$_2$ that pure states often appear `too thermal'  in the classical gravity limit, leading to a version of the information puzzle.  
One example is the case of a heavy scalar primary state, whose associated classical geometry is the BTZ black hole. 
Another example is provided by a heavy left-moving primary, which displays 
late time decay in chiral correlators.

In this paper we study a special class of pure state geometries which do not display such information loss. They describe heavy CFT states created by a collection of chiral operators at various positions on the complex plane. In the bulk, these take the form of multi-centered solutions from the backreaction of a collection of  spinning particles,
which we construct for  circular distributions of particles. We compute the two-point function of probe operators in these backgrounds and show that information is retrieved. 

We observe that the states for which our geometric picture is reliable are highly extended star-like objects in the bulk description. This may point to  limitations of  semiclassical  
microstate geometries for understanding the information puzzle and to the need for including quantum effects.}
\arxivnumber{}
\keywords{}
\begin{document}
\maketitle

\section{Introduction and summary}
The question of how information is returned from evaporating black holes \cite{Hawking:1974sw} 
in a unitary theory of quantum gravity continues to  challenge standard notions of the horizon and the validity of effective
field theory \cite{Mathur:2009hf,Almheiri:2012rt}.
For eternal black holes in Anti-de-Sitter (AdS) spacetimes,  a simple manifestation of the information puzzle was pointed out in \cite{Maldacena:2001kr} in the context of the AdS/CFT correspondence \cite{Maldacena:1997re}. In this case, the black hole geometry gives a description of the thermofield double state, yet it is
`too thermal' to correctly capture the late-time behaviour of correlators in the thermofield double state.
A related manifestation of information loss in the context of AdS$_3$/CFT$_2$ was investigated in \cite{Fitzpatrick:2016ive}: the classical geometry associated to a heavy primary state is the BTZ black hole \cite{Banados:1992wn}. This geometry is unreasonably thermal and leads to information-losing correlation functions which are not compatible with those of a pure state in a unitary CFT. 

In this work we will  simplify the setting even further by using two peculiarities of AdS$_3$: the fact that the symmetry algebra splits  in left- and right-moving parts and that in 2D CFTs  there is a gap between particle-like and black-hole-like states set by the Virasoro central charge. This allows one to consider states in the CFT Hilbert space which are above the gap in the left-moving sector but below it in the right-moving sector. A primary state with these properties is naively described by an  overrotating BTZ geometry with $M+J>0$ and $M-J<0$: it formally describes a temperature for the left-movers $T_L \sim \sqrt{M+J}$. The boundary-to-boundary propagator for a particle with mass $\tilde m$ and spin $\tilde s$ in this geometry is, in the saddle point approximation, obtained by evaluating the particle worldline action  $S_{wl}$ proposed in \cite{Castro:2014tta}. As we will show, the result is
\be 
e^{ - S_{wl} }= \left({2\sinh {\sqrt{M+J}\over 2}\Delta x_+  \over \sqrt{M+J}}\right)^{-(\tilde m+\tilde s)}
 \left({2\sinh {{\sqrt{M-J}\over 2}}\Delta x_-  \over\sqrt{M-J}} \right)^{-(\tilde m-\tilde s)}.\label{2ptgen}
 \ee
In the late-time regime where $\Delta x_+$ and $\Delta x_-$ are large,
this propagator decays exponentially for most values of the mass and spin,
and in the chiral limit $\tilde{m}=\tilde{s}$ it is even identical to the propagator in a black hole with the same value of $T_L$. 
This behaviour is again not compatible with a correlator probing a pure CFT state, leading to  a left-moving version  of the information puzzle. 
In \cite{Fitzpatrick:2016ive} it was argued that the corrections which restore unitarity are  non-perturbative effects in $1/c$, whose bulk interpretation has so far remained mysterious.

In this paper we will construct a different class of geometries with $M+J>0$ and $M-J<0$ which do not display information loss. They represent non-primary states in the dual CFT which are created by acting on the vacuum with a collection of light left-moving primaries inserted at various points within the unit disk. As we shall see, these correspond  in the bulk to multi-centered solutions from the backreaction of a collection of  spinning particles. The particles move on geodesics and rotate at constant distance from the center of $AdS_3$. Thus, in contrast to studies of the Vaidya metric \cite{Balasubramanian:2011ur,Balasubramanian:2012tu,Anous:2016kss} they do not collapse on each other. Using the methods of \cite{Hulik:2016ifr,Hulik:2018dpl} we construct the metric explicitly in the rotationally symmetric  limit where the centers are distributed  continuously on a circular shell. This geometry contains a throat region which is capped off inside the shell of matter.  We analyze two-point functions in this background and show that  they do not display the late-time decay characteristic of information loss.
This suggests that for these pure CFT states, the semiclassical geometry does give an accurate description of the physics. 

The idea of considering a shell of particles to mimic the black hole geometry has also been studied in a different context \cite{Lemos:2014eva, Lemos:2015xwa}, \cite{Danielsson:2017riq,Danielsson:2017pvl} and \cite{Crisostomo:2003xz}. In \cite{Lemos:2014eva, Lemos:2015xwa} the matter content of the shell is considered to be some fluid with pressure that stabilises the shell. In \cite{Crisostomo:2003xz} using the junction conditions for the case of a collapsing shell it was derived that the shell gives rise to the BTZ geometry.
The difference in our approach is that the particles are stabilised at fixed distance by angular momentum and that the consideration of the shell is only to make the problem analytically tractable. In principle one can have more complicated configurations with distinct particles placed at various distances from the center of $AdS_3$.

We should also point out that our classical geometries representing pure states are similar to  
the microstate geometries studied  within the fuzzball proposal for black holes
\cite{Mathur:2005zp,Mathur:2009hf,Bena:2007kg,Bena:2015bea} and that both approaches are qualitatively close in their description of   information retrieval in the capped throat geometry
(see \cite{Galliani:2016cai,Bombini:2017sge,Bena:2019azk} for work on information retrieval in the fuzzball program). One point of  difference with the microstate geometries picture, is that our multi-centered solutions are purely three-dimensional and as such somewhat more singular, having conical singularities corresponding to CFT vertex operator insertions. 
These singularities are expected to be resolved by quantum corrections in the UV completion of the theory. In the fuzzball paradigm on the other hand the microstate geometries are intrinsically higher dimensional, and feature the smoothening out of  singularities\footnote{Even so, conical singularities are still present in the fuzzball picture \cite{Lunin:2002bj}.} and the presence of nontrivial topology  from the higher dimensional point of view.
Inspiration for the current work also came  from the black hole deconstruction proposal of \cite{Denef:2007yt}, where 
microstates take the form of zero-entropy multi-centered D-brane configurations. 
 Indeed, the particular D0-brane centers in a D6-anti-D6 background considered in \cite{Denef:2007yt}  lift to spinning particles in AdS$_3$.  In that setting, the  backreaction on the non-gravitational fields in the 3D theory make the system hard to analyze \cite{Raeymaekers:2015sba,Raeymaekers:2017rdj}.

Our results may also give some insight into the limit  of validity of the  description of microstates
as semiclassical geometries. Indeed, we find that our information-preserving classical geometries
are  highly extended configurations in the bulk,  spread out over distances larger than the scale set by the minimum size of the throat. 
If we extrapolate  to the black hole regime, these would correspond to  star-like configurations spread out over distances larger than the horizon, while microstate geometries are more tightly bound solutions whose size is about a  Planck length  greater than horizon \cite{Guo:2017jmi}. This may point to  limitations on the usefulness of microstate geometries for understanding the information puzzle (somewhat similar to those pointed out in \cite{Raju:2018xue}), suggesting that typical microstates should rather be understood within the quantum regime. The importance of quantum effects has been stressed also within   the fuzzball paradigm, see e.g. \cite{MathurFAQ}.
 
The content of this paper is organised as follows. In section \ref{sec:eternal} we explain the general idea and introduce some necessary concepts.
These are the eternal BTZ black hole, its AdS/CFT interpretation and the role of correlation functions as indicators for the loss or restoration of information. In section \ref{sec:leftchiral} we examine geometries with left temperature. We present how these chiral geometries fit into an ansatz  
involving a  Liouville field, as well as their properties in terms of entropy and global geometry. We also examine the thermal behaviour of two-point functions, by evaluating the worldline action of spinning and spinless particles. In section \ref{sec:microstate} we construct a multi-centered solution with the same charges as the left-thermal geometry
by considering a shell of particles. We derive the Einstein's equations
with spinning particle sources within the Liouville ansatz and examine the properties of the shell solution. In section \ref{sec:geodesic} by examining geodesics we show how when the geodesic penetrates the shell information is restored.
We present our conclusions in section \ref{sec:Conclusions}.

 \section{Pure states and classical geometries}\label{sec:eternal}
We start this section by  recalling   some aspects of the 
correspondence between CFT states and classical bulk geometries in AdS$_3$/CFT$_2$. We review how  a heavy primary state appears unreasonably thermal in the classical  gravity regime, 
leading to a version of the information puzzle. In this case quantum corrections are expected to significantly alter the classical geometrical picture.  We then consider a different class of pure state geometries,  which describe pure states created by a collection of primary operators inserted at widely separated points on the complex plane. We give a qualitative argument that such states lead to a geometry where the black hole throat  is capped off and information should be returned. 
Though this argument is presently out of reach of computational verification, we will see in the following sections that a chiral version  of the  problem is in fact tractable, and demonstrate how information is preserved in a class of chiral  pure states.

\subsection{CFT vs geometry}
Consider a correlation function of primary operators in  a two-dimensional Euclidean CFT defined on the complex plane parametrized by $v$:
\be 
\langle \hat \calv_1 (v_1,\bar v_1 ) \ldots  \hat \calv_n (v_n,\bar v_n ) \rangle \, .
\ee
The conformal Ward identities imply that the expectation value of the stress tensor\footnote{To avoid carrying around factors of $c/24$ in what follows, our stress tensor is rescaled with respect to the standard normalization in the literature \cite{Belavin:1984vu}: $\hat T(v) = {24 \over c} \hat T_{standard} (v)$. Similarly, our   $h$ in (\ref{TWard})  is the rescaled conformal weight $h = {24 \over c} h_{standard}$. In particular,  $h$  of order one  corresponds at large $c$ to a heavy operator with weight of order $c$.} 
\bea 
T(v) &\equiv & {\langle \hat T (v)  \hat \calv (v_1,\bar v_1 ) \ldots  \hat \calv (v_n,\bar v_n )\rangle \over \langle \hat \calv (v_1,\bar v_1 ) \ldots  \hat \calv (v_n,\bar v_n )\rangle}\\
&=& \sum_{i = 1}^n \left(  { h_i \over (v - v_i)^2} + {c_i \over v-v_i} \right) \, ,\label{TWard}
\eea
and similarly for the antiholomorphic part $\bar T ( \bar v)$. Upon conformally mapping to the cylinder  and analytically continuing to Lorentzian signature, $v \to e^{i x_+}$, we obtain the Lorentzian cylinder  expectation value
\be 
T_{++} (x_+ ) = -1 +  e^{2 i x_+} T (e^{i x_+}) \, ,
\ee
and similarly for $T_{--}(x_-)$.

If the CFT under consideration has a holographic dual,
 meaning roughly that it has a large central charge $c$ and a sparse spectrum of low-lying states \cite{Heemskerk:2009pn}, 
the above stress tensor  expectation values translate into  a classical AdS$_3$ geometry 
due to Banados \cite{Banados:1998gg}. The metric takes the form of an all-order Fefferman-Graham 
(FG) expansion,
\be \label{FGmetric}
ds^2 =  \frac{dy^2}{y^2} - \frac{1}{4 y^2} dx_+ dx_- + \frac{T_{++}}{ 4} dx_+^2 + \frac{T_{--}}{ 4} dx_-^2 -  \frac{ y^2}{ 4} T_{++} T_{--} dx_+ dx_- \,. 
\ee
Here and in the rest of the paper, we have set the AdS radius to one. 
The metric (\ref{FGmetric}) solves the vacuum Einstein equations and satisfies asymptotic AdS$_3$ boundary conditions; the boundary is a cylinder and the boundary stress tensor \cite{Balasubramanian:1999re} computed from (\ref{FGmetric}) is precisely $T_{\pm \pm}$. The coordinate system is valid near the boundary at $y=0$. 
This geometry is expected to accurately capture the  physics only if the two following approximations are valid:
\begin{itemize}
 \item semiclassical approximation in the bulk, which corresponds to the limit of large central charge $c$ in the CFT. 
    \item  pure gravity approximation, where we neglect all 
  fields  in the bulk AdS$_3$ besides gravity. In the dual CFT, this is justified if the  contribution from the vacuum block in a specific channel dominates the amplitude, see \cite{Hartman:2013mia} for details.
   \end{itemize}
We will refer to this two-fold approximation as the `classical gravity approximation'.
When valid, we can compute observables in the bulk  by probing the metric (\ref{FGmetric}). For example, the correlator in the presence of two additional insertions of a  vertex operator $\calo$,  which is parametrically lighter so that it doesn't backreact on the metric, is given by
\be
\langle \hat \calv_1 \ldots  \hat \calv_n \hat \calo (x^+_1,x^-_1)
\hat \calo (x^+_2,x^-_2)\rangle = G_{| ds^2} ( \D x_+, \D x_-)
\sim e^{- S_{wl} ( \D x_+, \D x_-)} \, .\label{saddleprobe}
\ee
Here, $G_{| ds^2}$ means the boundary-to-boundary propagator for the field dual to $\calv$ evaluated in the geometry (\ref{FGmetric}). In the last equality we have indicated that in the saddle-point approximation, this propagator can be evaluated from the on-shell action $S_{wl}$ of a  worldline connecting the two boundary points. In what follows we will use bulk observables like (\ref{saddleprobe}) as a diagnostic to see whether the classical gravity approximation is justified, by comparing them with CFT expectations.

\subsection{Two heavy scalar operators}
Let us illustrate this classical gravity picture
in the simple case where we consider the two-point function of a scalar operator of rescaled dimension $\D \equiv h+ \bar h = 2 h$,
\be 
\langle \hat \calv_\D (\infty) \hat \calv_\D (0) \rangle.
\ee
In this situation, fluctuations around the geometry (\ref{FGmetric}) should  probe the state $|\calv \rangle$ created by $\calv$.
The stress tensor on the plane has second order poles in the origin and at infinity,
\be
T = {\D \over 2 v^2}, \qquad \bar T = {\D \over 2 \bar v^2}, \qquad
T_{++} =T_{--} =  \frac{\D}{2} -1.
\ee
Plugging the latter into the Fefferman-Graham metric (\ref{FGmetric}) yields a static, rotationally symmetric metric. Upon setting 
\be 
x_\pm = t \pm \f \, ,
\ee
and redefining the radial coordinate 
as 
\be 
r= \half (y ^{-1} + My )\, ,
\ee
it reduces to the non-rotating BTZ metric \cite{Banados:1992wn}
\be 
ds^2 = - \left( {r^2 } - M \right) dt^2 +  \left( {r^2} - M \right)^{-1} dr^2 + r^2 d \f^2 \, ,
\label{BTZ}\ee
with mass parameter \footnote{Our dimensionless parameter $M$ is essentially the ADM mass in Planck units, namely
$ M = 8 G M_{ADM}$.}
\be
M = {\D\over 2} -1.
\ee

In what follows it will be useful to use  different coordinates where we go to conformal gauge on  the constant time slices,
\be 
ds^2 = - N(z,\bar z)^2 dt^2 + 4  e^{-2 \F (z, \bar z )} dz d\bar z \, .\label{staticLanstaz}
\ee
Then the vacuum Einstein equation $R_{\m\n} + 2 g_{\m\n}=0$ implies that $\F$ satisfies the Liouville equation:
\be 
\pa_z \pa_{\bar z} \F + e^{-2 \F} =0.\label{Liouv}
\ee
One finds that the coordinates in (\ref{BTZ}) and (\ref{staticLanstaz}) are related as
\be 
r = - {\sqrt{M} \over \sin ( \sqrt{M} \log |z| )} , \qquad  \f = {\rm arg} z ,\label{ritoz}
\ee
while the Liouville solution describing the BTZ metric (\ref{BTZ}) is
\be 
e^{-2 \F} = {M \over 4 |z|^2 \sin^2 (\sqrt{M } \log |z| )}.\label{BTZLiouv}
\ee
We can also build  an  (anti-)holomorphic `stress tensor' from  the Liouville field,
\be 
\calt(z) = -4 (\pa_z \F )^2 - 4 \pa_z^2 \F
, \qquad \bar \calt(\bar z) = -4(\pa_{\bar z} \F )^2 -4 \pa_{\bar z}^2 \F \, .
\label{TLiouv}
\ee
One easily checks that for the solutions (\ref{BTZLiouv}) it takes the same form as the Euclidean boundary stress tensor
\be 
\calt (z) = T(z), \qquad \bar \calt (\bar z) = \bar T(\bar z)\, . \label{caltvst}
\ee
This formal equality of the Liouville stress tensor $\calt (z)$ 
and the Euclidean boundary stress tensor $T (v)$ holds generally in the class of metrics (\ref{staticLanstaz}) \cite{Hulik:2016ifr}. One should keep in mind  that $\calt (z)$ and $T (v)$ were a priori quite different objects: the former characterizes the gravitational field in the interior of the Lorentzian  bulk while the latter lives on the Euclideanized boundary plane. The relation (\ref{caltvst}) should therefore be seen as a boundary-bulk map\footnote{In a similar vein, our bulk Liouville field $\F$ is a priori different from the boundary Liouville field considered in \cite{Coussaert:1995zp,Martinec:1998wm} which is associated to $T(z), \bar T( \bar z)$; yet the relation (\ref{caltvst}) shows that they take the same form.}, which tells us for example how primary insertions on the boundary (second order poles in $T(v)$) map to singularities in the gravitational field along worldline sources in the bulk.
In section \ref{sec:Liouvthroat} below we will discuss  a chiral version of the bulk-boundary map where we will make these statements more precise.  This map will play an important role in the rest of the paper. 

We can  also compute observables in the state $|\calv_\D \rangle$ by probing the BTZ geometry (\ref{BTZ}). For example,  by evaluating the worldline action of a particle of mass $\tilde m$ 
we obtain from (\ref{saddleprobe}) the saddle point approximation to the holographic 
 two-point function of a light scalar primary operator of dimension $\tilde m$
  (see \cite{Balasubramanian:1999zv} or section \ref{sec:2ptBTZ}  below for  details of this calculation),
\be 
e^{- S_{wl}}
= \left({4 \over M }\sinh \left( {\sqrt{M}\over 2}\D x_+\right) \sinh \left({\sqrt{M}\over 2}\D x_-\right)   \right)^{-\tilde m }.\label{2ptBTZ}
 \ee
\subsection{Particles vs.  black hole microstates}
We  now describe in more detail  the above classical gravity picture  of the state $|\calv_\D \rangle$ in  various regimes  of $\D$ and discuss its limitations. For $\D$ above the black hole threshold, we will see that our classical geometry is not a very reliable guide to the physics and  corrections are expected to be significant. This can be viewed as a  manifestation of the breakdown of effective field theory near black holes \cite{Mathur:2005zp,Almheiri:2012rt} in a simplified setting. 

\begin{figure}
\begin{center}
\subfloat[]{\includegraphics[height=110pt]{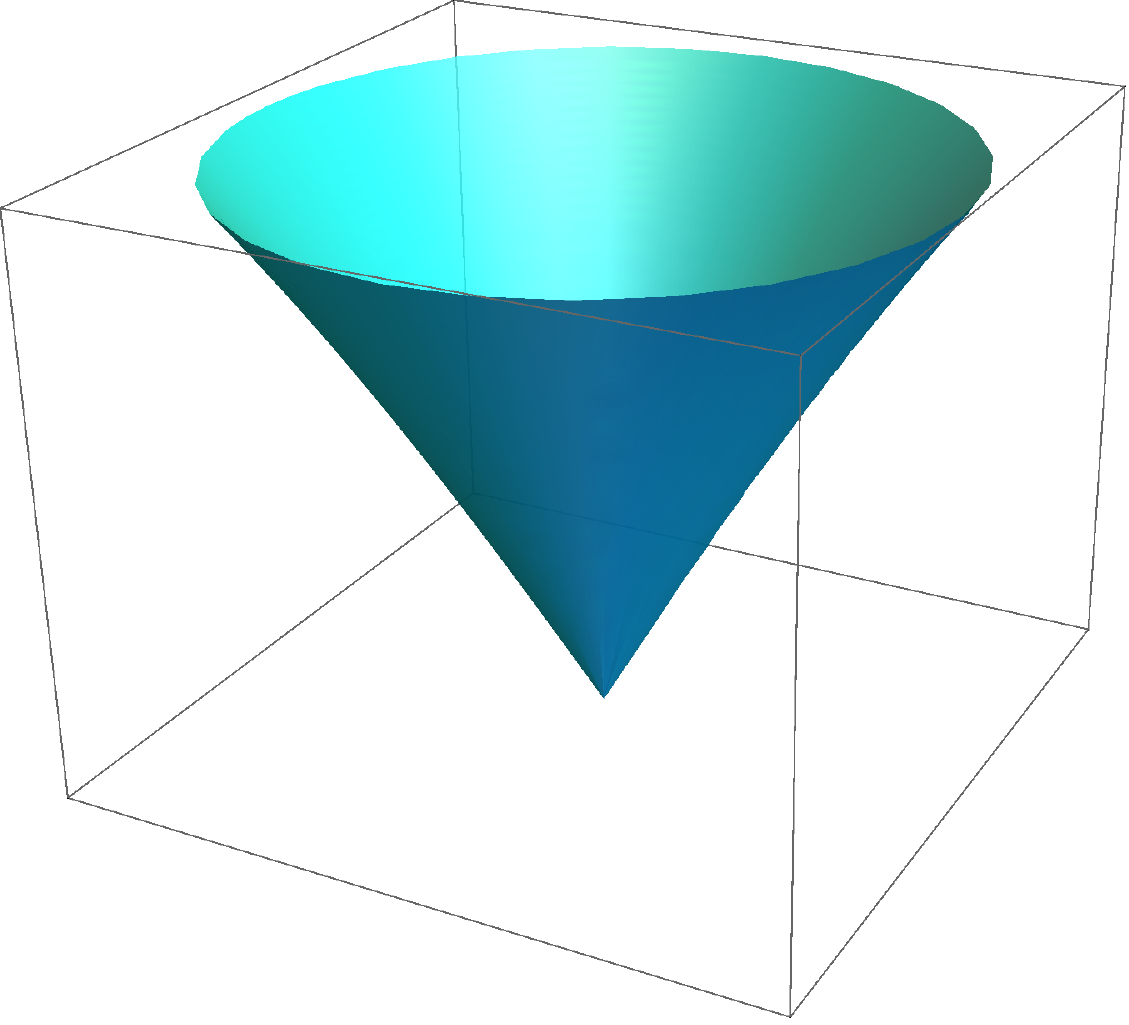}\label{FigERa}} 
\subfloat[]{\includegraphics[height=110pt]{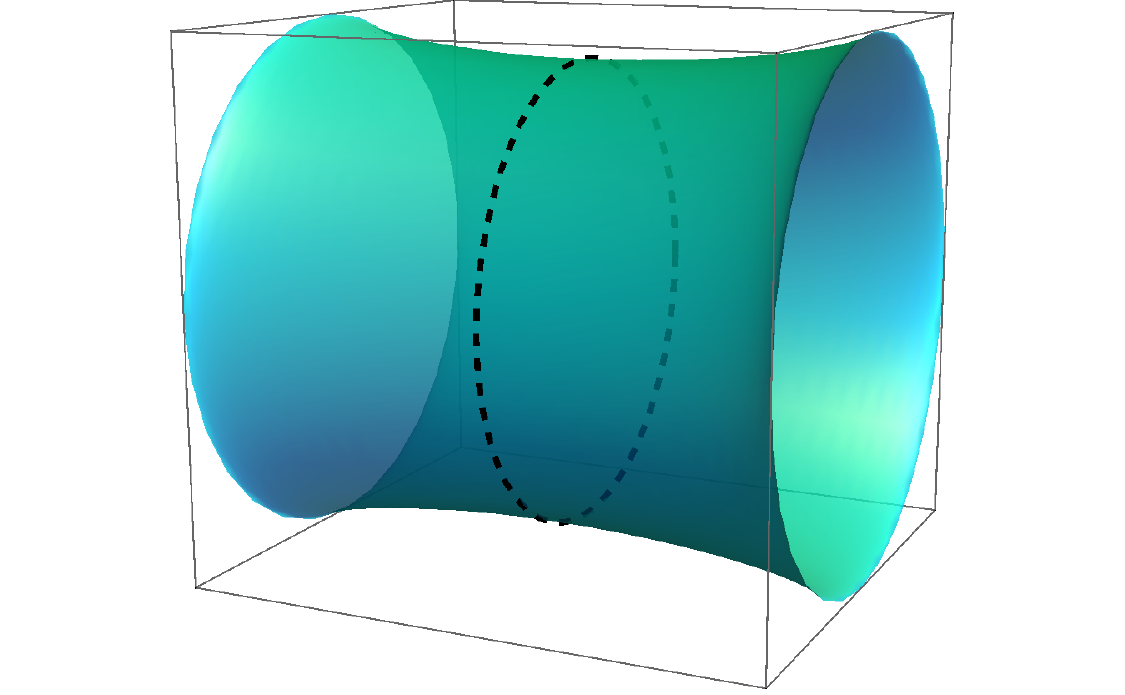}\label{FigERb}} 
\end{center}
\caption{Embedded spatial geometry of the metric representing (a) a light primary state and (b) a heavy primary above the black hole treshold.}
\label{FigER}
\end{figure}

For the vacuum with $\D=0$ (equivalently $M=-1$), the geometry is global AdS$_3$. For small enough $\D$ in the range $0 < \D < 2$ (equivalently $-1<M<0$), the geometry is a conical defect at the centre of AdS
and describes the backreaction of a spinless point particle \cite{Deser:1983nh}. A three dimensional embedding of the spatial geometry is shown in Figure \ref{FigERa}. 

While we do expect our classical gravity picture of this state to receive corrections near the conical singularity, observables computed in the naive geometry don't show any obvious pathologies. A good test is to see whether probe correlation functions display any forbidden singularities. In Euclidean signature, the only singularities that arise in CFT correlators come from OPE limits, when some of the operators collide.
Taking the result (\ref{2ptBTZ}) for $M<0$ and continuing to Euclidean signature $t\to - i t_E$ under which 
\be
\begin{split}
 & \D x_+ = \D t + \D\f \to -i ( \D t_E + i \D\f) \equiv  - i \D w\, ,\\
 & \D x_- = \D t - \D\f \to -i ( \D t_E - i \D\f) \equiv - i \bar \D w\, ,\label{analcont}
\end{split}
\ee
we see that  the correlator  has only an allowed OPE singularity for $\D w \to 0$, when the two probe vertex operators collide.

A quite different situation arises when we consider heavy states above  the black hole threshold,  in the regime $\D>2$ or $M>0$. The primary state $|\calv_\D \rangle$ is pure and is a black hole microstate, contributing to the microscopic entropy as counted by Cardy's formula.   The classical geometry (\ref{BTZ}) is  the BTZ black hole metric with horizon at  \be r_+ = \sqrt{M}.\ee 
In this case, one expects the  corrections to the classical gravity picture to be significant for the following reasons. 

First of all, the eternal black hole geometry does not satisfy the right boundary conditions  to  describe
 a   state in the dual CFT on the boundary $r \to \infty$, since the full geometry extended beyond the coordinate singularity at $r=r_+$ in fact contains a second conformal boundary. This can be seen in the standard way by going to Kruskal-like coordinates, but we will discuss it here from the perspective of our Liouville parametrization (\ref{staticLanstaz}).
When $M>0$, an important property of the conformal factor (\ref{BTZLiouv}) is the periodicity in $\log |z| \sim \log |z| + { 2\p \over \sqrt{M}}$.
The $r \to \infty$ conformal boundary can be taken to correspond to the unit circle $|z|=1$.  The  horizon  then  corresponds  to  $|z|=e^{− {\pi \over 2 \sqrt{M}}}$; this is the value for which the circle of   constant $|z|$  reaches its  smallest size.   The  geometry  extends  to  smaller  values of $|z|$  until one reaches the value $|z| = e^{- {\p \over \sqrt{M}}}$ where the conformal factor  $e^{-2 \F}$ blows up; this can be shown to be a second conformal boundary. The  Liouville solution (\ref{BTZLiouv}) on the annulus $e^{- {\p \over \sqrt{M}}} \leq |z|\leq 1$ describes the familiar
Einstein-Rosen throat geometry of the eternal Schwarzschild black hole. 
Part of the throat geometry can be isometrically embedded in three dimensional Euclidean space and is shown in figure \ref{FigERb}.

Another  reason  for  expecting  significant  corrections  to  the  classical  gravity  approximation comes from looking at probe correlators.  Considering again the  2-point function (\ref{2ptBTZ}), for $M>0$ it 
becomes  precisely the  thermal two-point function of a CFT on the infinite real line \cite{Cardy:1986ie} at temperature
 \be 
    T_H = {\sqrt{M} \over 2 \p } \, ,
    \ee
which agrees with the Hawking temperature  \cite{Hawking:1974sw} of the BTZ black hole.  
At large timelike separations, taking $\D x_+ = \D x_- = \D t \gg 1$, the correlator  (\ref{2ptBTZ}) tends to zero exponentially.
This is a manifestation of the information puzzle \cite{Fitzpatrick:2016ive} 
since, in a unitary CFT on the circle, the effects of the perturbation 
of the state $|\calv_\D \rangle$ at initial time $t=t_1$ should never get completely wiped out. 
This version of information loss is also closely related to the one discussed in \cite{Maldacena:2001kr} in the context of the classical gravity description of the thermofield double state.
A  further sign of pathologies is the fact that the two-point function, when analytically continued to Euclidean signature   using (\ref{analcont}), has forbidden singularities at $w= n/T_H$ for $n \in \ZZ_{ \neq  0  }$.

As was emphasized in \cite{Fitzpatrick:2016ive}, this overly thermal behaviour cannot arise in the exact four-point function of a unitary CFT on a compact space and with a finite amount of degrees of freedom. Therefore the corrections to our classical gravity approximation  are   guaranteed  to arrest the late-time decay of the correlator and return information    (or, in the Euclidean picture, to remove the forbidden singularities). From the CFT point of view, it was shown in \cite{Fitzpatrick:2014vua} that the approximate thermal behaviour is due to the large-$c$ approximation to the vacuum  conformal block. 
Furthermore it was argued in \cite{Fitzpatrick:2016ive} that including  corrections  to the vacuum block  which are nonperturbative  in $1/c$  modifies the late-time decay at time scales of order $S_{BH}$. Although we presently lack an understanding of these quantum effects in terms of geometry,   one would expect them to effectively cap off the Einstein-Rosen throat before the inner boundary at $|z| = e^{- {\p \over \sqrt{M}}}$ is reached. This constitutes a simple example where  the semiclassical approximation breaks down in a region where one might naively expect it to be accurate.

Before going on to consider more general pure states it is also worth recalling \cite{Martinec:1998wm} that the particle ($-1<M<0$) and black hole ($M>0$) regimes of the classical metric (\ref{BTZ}) are  distinguished by the type of Liouville solution describing the metric on spacelike slices. 
Any solution of the Liouville equation (\ref{Liouv}) gives rise to a 2D metric $e^{-2 \F} dz d \bar z$  which
can  locally be transformed  to the  Poincar\'e disk metric
\be 
ds^2 =  {d Z d \bar Z \over (1 - |Z|^2 )^2 }\, .\label{Poinc}
\ee
The map $Z(z)$ is however not single-valued in general and can have branch points. When encircling a branch point, $Z(z)$ undergoes a fractional linear transformation determined by a $2\times 2$ monodromy matrix $\calm$ in $PSU(1,1)$. The  branch points are classified by the conjugacy class of $\calm$ into hyperbolic, parabolic or elliptic types.

In the particle regime  $-1<M<0$ of the solution (\ref{BTZLiouv}) we have
\be Z(z) = z^{\sqrt{-M}}, \qquad \calm  = \left( \begin{matrix} e^{i \p \sqrt{-M}} & 0 \\ 0  &  e^{-i \p \sqrt{-M}} \end{matrix} \right) . \ee
Since  $\tr \calm >2$ the Liouville solution for this range of masses has elliptic monodromy.

On the other hand,  above the black hole threshold 
{\bf $ M >0$} we have: 
    \be Z(z) = {  z^{i \sqrt{M}}+ 1 \over   i z^{i \sqrt{M}} + 1} , \qquad \calm  = \left( \begin{matrix} \cosh \p \sqrt{M} & i \sinh \p \sqrt{M} \\ -i \sinh \p \sqrt{M}   &  \cosh \p \sqrt{M} \end{matrix} \right). \label{Zhyp} \ee
Since $\tr \calm >2$, the Liouville solution in this class  of the hyperbolic type.
For the metrics with $M=-1$ (global $AdS_3$) and $M=0$ the monodromy is trivial and of parabolic type respectively.

\subsection{Classical caps for extended microstates?}\label{sec:extendedcaps}
\begin{figure}
\begin{center}
	\begin{picture}(300,100)
	\put(-10,16){\includegraphics[height=73pt]{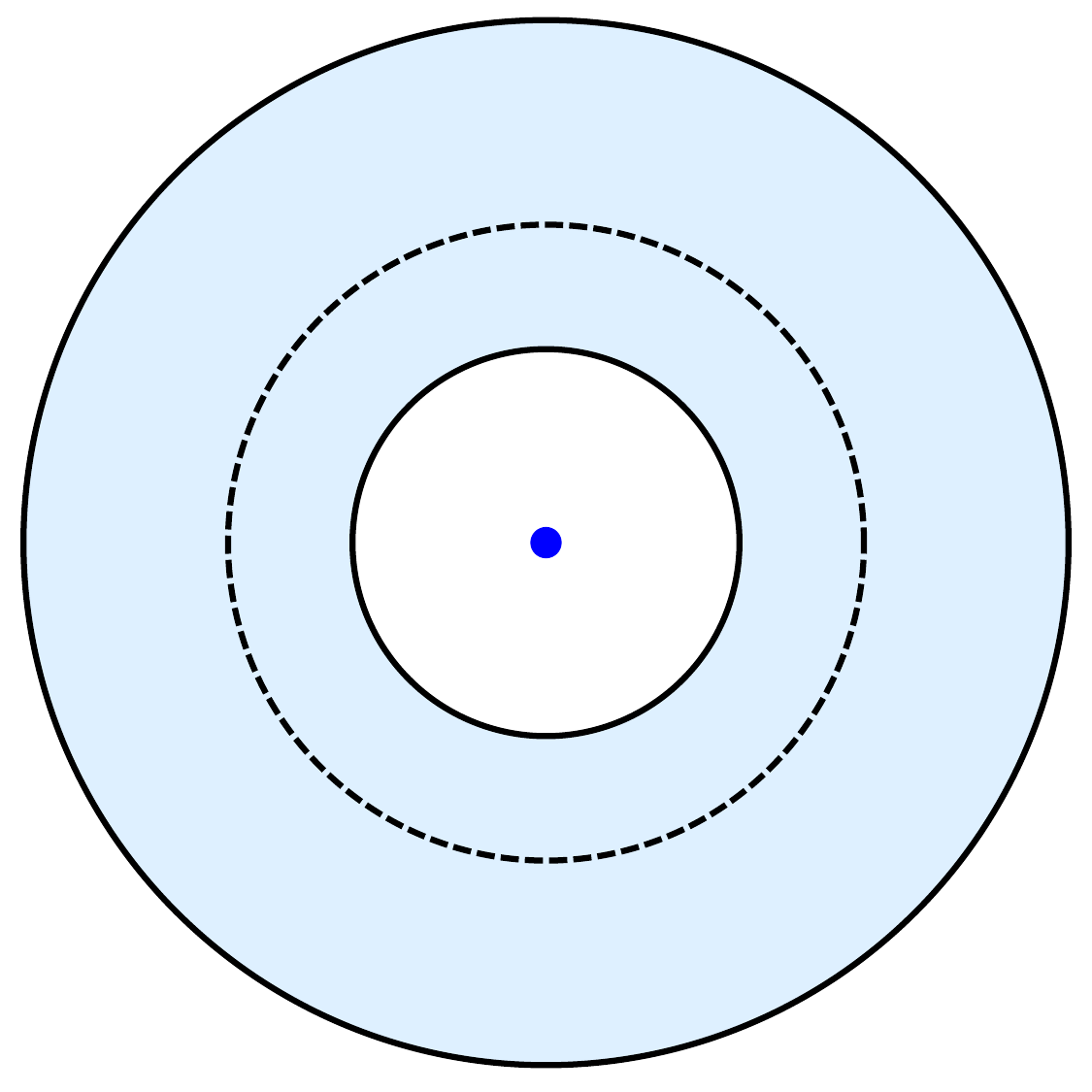}}
		\put(100,0){\includegraphics[height=100pt]{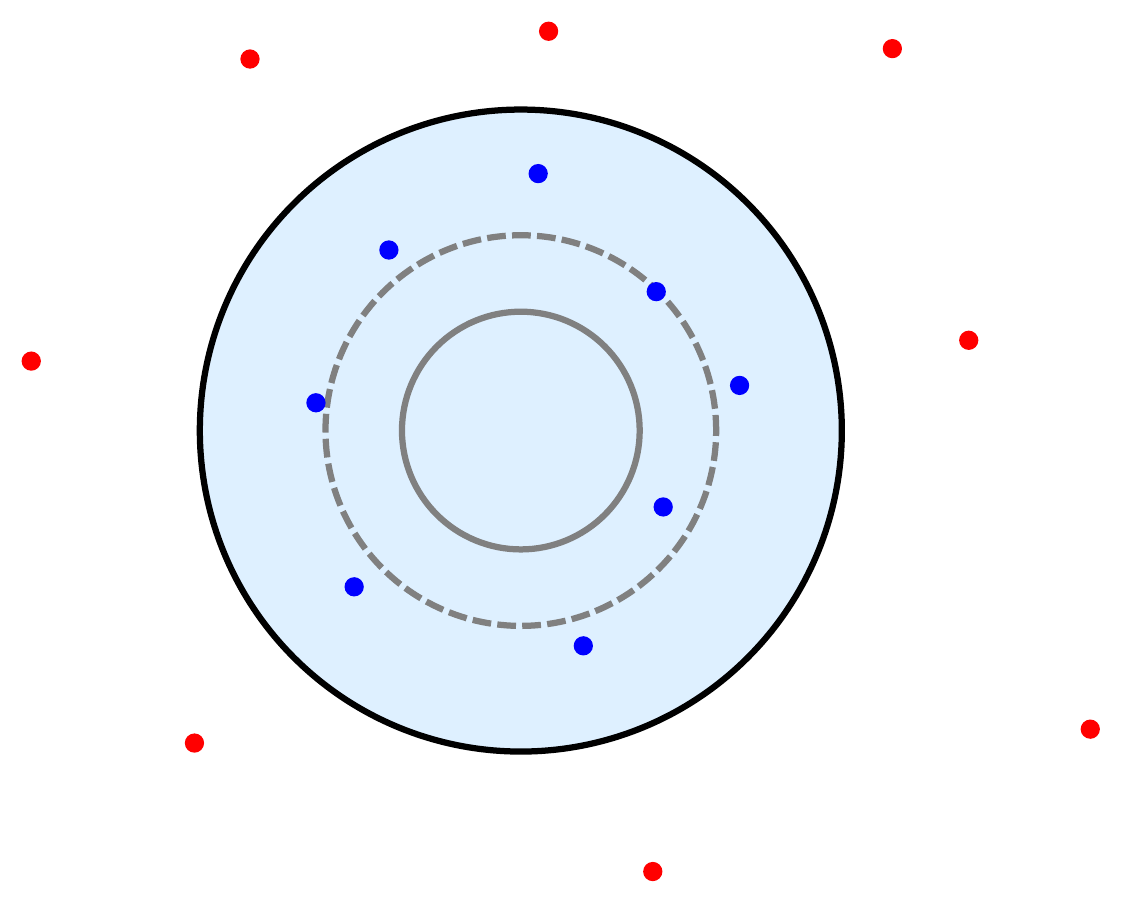}}
			\put(250,16){\includegraphics[height=73pt]{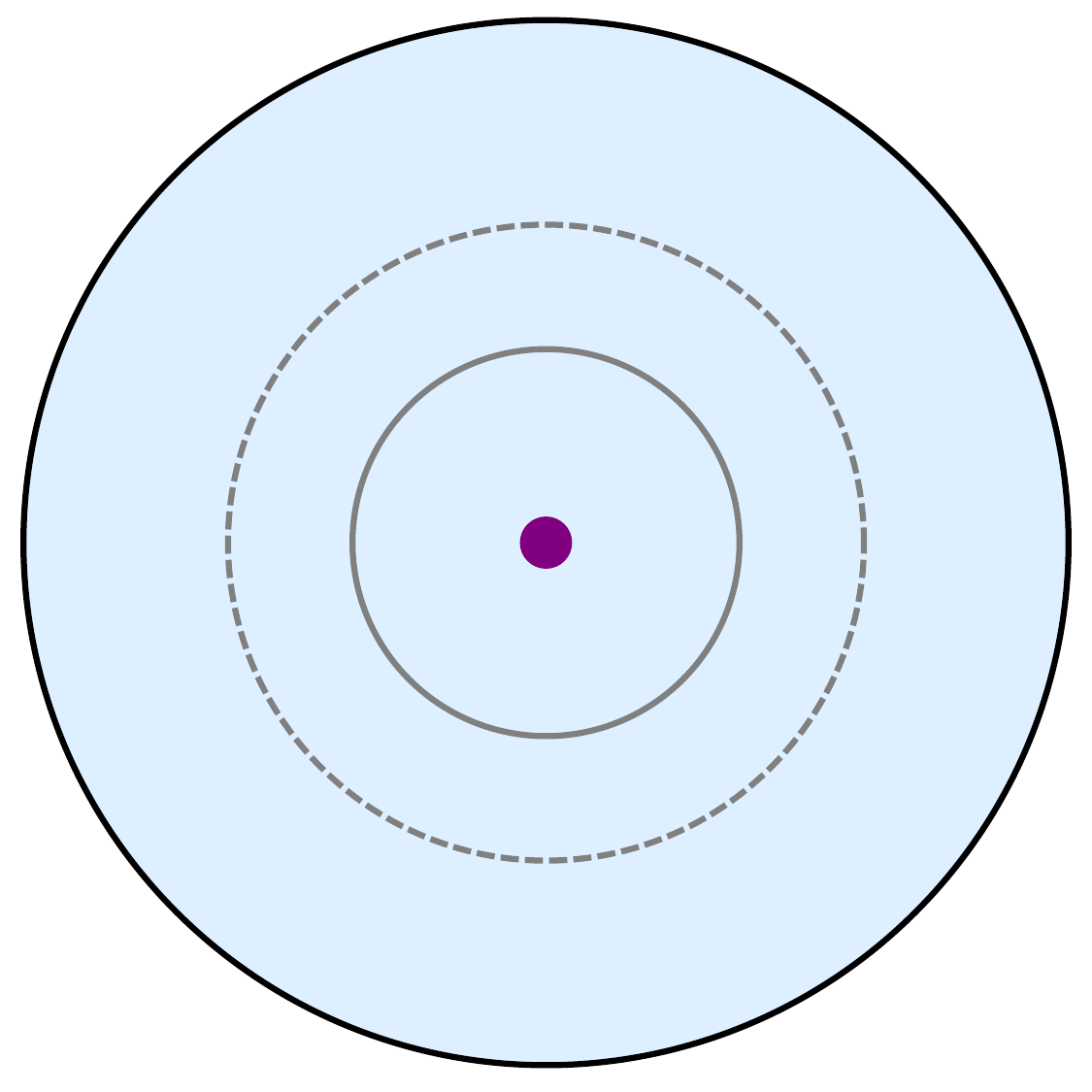}}
			\put(20,0){(a)}\put(150,0){(b)}\put(282,0){(c)}
\end{picture}
			\end{center}
			\caption{Positions of CFT vertex operators in the complex $v$-plane.  (a) Heavy primary, shown as a blue dot, in the origin. Only the shaded blue  region is mapped to the  region in spacetime between the conformal boundaries (the solid circles). The dotted circle maps to the horizon. (b) Collection of light primaries within the unit disk (blue dots), and in image points (red dots), which is expected to map to a capped geometry in the bulk. (c) From the CFT point of view the previous configuration is equivalent to inserting   a highly non-primary operator in the origin (purple dot).  }
\label{Figpure}\end{figure}
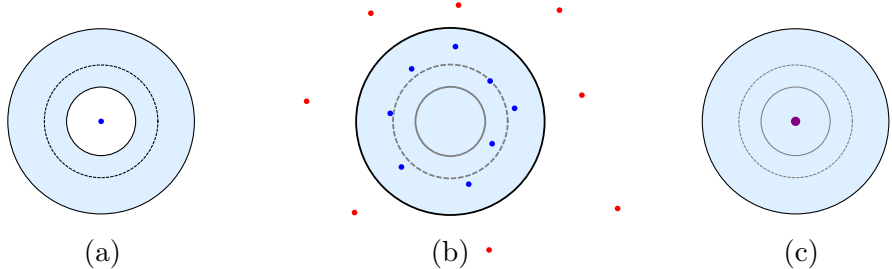

In the previous section we argued that the naive classical geometry of a heavy primary should receive  significant  corrections. Before moving on to discuss non-primary states, we want to point out  one further pathology of this classical description. The dual CFT  has  heavy primary operators inserted in $v=0$ and at $v=\infty$.
In the bulk, these map to sources in $z=0, \infty$ under the CFT-to-bulk map (\ref{caltvst}), yet these points `lie beyond the boundaries' and are not part of  the physical spacetime  where $e^{- {\p \over \sqrt{M}}} \leq |z|\leq 1$ (see Figure \ref{Figpure}(a)).
Therefore the points where energy is inserted in the CFT do not map to points in  spacetime, indeed   the black hole geometry  does not have any localized sources. Hence the   CFT region near the heavy vertex operator cannot be probed in the bulk. 
This is an additional reason to expect significant  corrections to the classical picture.

So far we focused on primary states in the CFT but we can also consider non-primary heavy states. In the rest of the paper we want to look at pure states which are created by acting on the vacuum with a collection of primary operators inserted  at various locations within the unit disk (see Figure \ref{Figpure}(b)):
\be 
\hat \calv_1 (v_1, \bar v_1) \ldots \hat \calv_n (v_n, \bar v_n) |0\rangle, \qquad |v_i| <1 .
\ee
Using the OPE one sees that  the resulting state is a  highly complicated mix of primaries and descendants\footnote{Recent work \cite{Datta:2019jeo} has emphasized the importance of descendants as black hole   microstates.  } (Figure \ref{Figpure}(c)).
We will take  each of the primary vertex operators to satisfy $\D_i< 2$ so that each of the individual centers has a mass below the black hole threshold, yet such that the total mass is above the black hole threshold, $\D_{tot} > 2$. 
In terms of the Liouville field, such configurations would correspond to solutions with several elliptic singularities, such that the monodromy encircling all the singularities is of hyperbolic type\footnote{It's straightforward to verify that the product of elliptic elements can indeed be hyperbolic, see also e.g. \cite{Anderson}.  }.  From the above comments on the bulk-boundary map,  one would expect that if these operators are inserted within the annulus roughly of size $e^{- {\p \over \sqrt{M}}} \leq |v|\leq 1$, they correspond to sources in points which are part of the bulk spacetime. These backreact classically  on the geometry, producing a multi-centered solution with conical defect singularity at each center. In this situation, the Einstein-Rosen bridge is capped off through classical backreaction effects, and we should be able to study the return of information by doing classical computations in the bulk.

The above argument is of course rather sketchy and hard  to make precise. 
For one, the static ansatz (\ref{staticLanstaz}) will fail to describe such multi-centered solutions, since each of the centers will tend to fall toward the center of AdS$_3$ due to the gravitational potential.\footnote{It is amusing to see what happens when trying to describe multicentered solutions within the static ansatz \cite{Hulik:2016ifr}: the equation for the lapse function $N(z,\bar z)$ is solvable only if the solution includes additional negative mass  `spurious' singularities (i.e. excess angles  of multiples of $2 \p$), which balance the attractive forces.} Instead  such bound states will be described by highly complicated time-dependent solutions. It might be possible to construct the  solution in the Fefferman-Graham patch (\ref{FGmetric}) provided one can determine the residues $c_i$ of the single poles in the boundary stress tensor (\ref{TWard}). This  amounts to knowledge of  the CFT correlator; we hope to return to this approach in the future. We would also like to mention that similar ideas have been put forth in \cite{Jackson:2014nla}, see also \cite{VerlindeCFTAdS}.

In the rest of this paper  we will instead consider the chiral version of the above idea, where we replace  all the scalar  CFT operators with purely chiral ones.

\section{Left-chiral primaries and overspinning BTZ}\label{sec:leftchiral}

In this section we will study the classical gravity geometry associated with a purely left-moving CFT primary, which 
is a BTZ geometry with angular momentum $J$, in the `overspinning' regime where $|J|>M$. 
Though this geometry doesn't have a  horizon and is not a black hole, we will show that it nevertheless  shares some features with a black hole.
For one,  the metric contains two conformal boundaries connected by  a negative curvature throat. A second property is that a subset of correlation functions in this background behave `too thermally' and display exponential late-time decay. For example the two-point function of chiral currents is the same as in a BTZ black hole with same temperature for the left-movers. These considerations lead to a chiral version of the information puzzle   and the conclusion that quantum effects must significantly alter the classical geometry.

\subsection{Overspinning BTZ metrics}

We start by considering the classical geometry associated to 
the two-point function of a  holomorphic primary  current with left-moving  dimension $h$,
\be 
\langle \hat \calv_{(h,0)} (\infty) \hat \calv_{(h,0)} (0) \rangle,
\ee
where we take $h$ to be above the `chiral  threshold',
\be  
h> 1.
\ee
 The stress tensor VEVs  on the plane and the cylinder are
 \be
T = {h \over  z^2}, \qquad \bar T = 0, \qquad
T_{++} =  h-1, \qquad T_{--} =  -1.\label{Tchir}
\ee
Plugging into the Fefferman-Graham metric (\ref{FGmetric}) and redefining the radial coordinate
we obtain an instance of 
the spinning BTZ metric with angular momentum $J \equiv {8 G\over l} J_{ADM}$:
\be 
ds^2 = - \left( {r^2 } - M   \right) dt^2 +  \left( {r^2 } - M  + {J^2 \over 4 r^2}  \right)^{-1} dr^2  +J dt d\phi +r^2 d\phi^2  .\label{BTZJ}
\ee
The boundary stress tensor for the metric (\ref{BTZJ}) is 
\be 
T_{++} = M + J , \qquad T_{--} = M - J  \, ,\label{TBTZ}
\ee
and therefore (\ref{Tchir}) corresponds to
\be 
M=  {h\over 2 } -1, \qquad J =  {h\over 2 }\, .  \label{chiralreg}\ee
Note that we are in the overspinning regime where  $|J|>M$ and where the geometry (\ref{BTZJ}) does not have a horizon.

Since this regime of the BTZ metric is not often discussed, we first discuss some   global aspects of the general overspinning BTZ metric.
Without loss of generality we   take
\be
M+J > 0, \qquad M-J <0 \, .\label{oversp}
\ee
We restrict to  $M-J \geq -1$ so that $T_{--}$ is not below the global AdS$_3$ vacuum value.
The $dr^2$ coefficient in (\ref{BTZ}), ${r^2 } - M  + {J^2 \over 4 r^2}$, is positive and finite in the range $0<r<\infty$. The limit $r \to  \infty$ is a conformal boundary, while  $r=0$ is a coordinate singularity where $g^{rr}$ diverges. The metric components become  regular there by passing  to the coordinate
\be 
u = r^2,
\ee
and one can show  that the metric is regular in $u=0$ and  can be extended to negative values of $u$. Indeed, in Appendix \ref{Appquotient} we describe the extended geometry, where $u$ is allowed to  take values on the real line, as a quotient of global $AdS_3$, and show that this quotient acts without fixed points, leading to a regular geometry. The overspinning regime in $AdS_3$ is therefore quite different from the one in higher dimensions where one encounters naked singularities.  However the negative $u$ region does contains pathologies in the form of closed timelike curves, since the Killing vector  generating the identification is timelike there.
 
A significant feature of this extended geometry is that when  $u \to - \infty$,  a second conformal boundary is reached. This boundary is again a cylinder, yet now the periodic coordinate $\f$ is a time coordinate while $t$ is a spatial coordinate.  Despite these causal pathologies on the inner boundary, we can still ask the question if signals can be sent between the two boundaries.   For this purpose we look for null geodesics which interpolate between the two conformal boundaries. The solution for $u (\t )$, where $\t$ is an affine parameter on the null geodesic, is
\be 
 u (\t) = (E^2 - L^2) (\t-\t_0)^2 \pm 2\sqrt{L (E M- L J)} (\t-\t_0).
 \ee
  where $E$ and $L$  are the conserved charges due to $t$ and $\f$ translation invariance respectively (to be identified with energy resp. angular momentum in the $u>0$ region).  
To interpolate between the to boundaries, the first term has to vanish so we need $|E|=|L|$.  In the overspinning regime (\ref{oversp}) of interest only the  
 \be L = -E
 \ee 
 geodesics are real. It  takes infinite affine parameter to go from one boundary to the other and consequently the boundaries are causally disconnected in this sense. It's also interesting to look at the behaviour of $x_\pm (\t)$, whose solution is
 \bea 
 x_+ (\t) &=& { \ln (J^2 - M^2 + (M + 4 E\sqrt{M+J} (\t-\t_0))^2 ) \over 2 \sqrt{M+J}} + x_{+0} \, ,\\
 x_- (\t) &=& - {\arctan {M+ 4 E \sqrt{M+J} (\t-\t_0) \over {J^2-M^2}} \over \sqrt{J-M} }+ x_{-0} \, .
 \eea
The interpolating geodesics   connect points with $x_+ \to \infty$ on each boundary  with $x_-$ values shifted by $\D x_- = - {\p \over \sqrt{J-M}}$.
 
Summarizing,  the overspinning metric resembles a black hole in that the extended geometry contains a second conformal boundary.
We will presently see that, as in the black hole regime, the two boundaries are connected by negative curvature throat geometry, though in this case it is   embedded in the 3D geometry in a more subtle manner.

\subsection{The Liouville throat }\label{sec:Liouvthroat}

In this subsection we will see that the left-thermal geometries contain a submanifold with constant negative curvature which is conveniently described by a solution of Liouville  theory \cite{Hulik:2016ifr,Hulik:2018dpl}. For simplicity we  will limit the discussion in this subsection  to the regime  of interest (\ref{chiralreg}) where the right-movers are in the ground state,  $J=M+1$  or $T_{--}=-1$, though it could easily be generalized to the regime where $ -1 \leq T_{--} \leq 0$.
It will be useful in what follows to keep left-moving stress tensor  general for the moment and consider solutions with arbitrary $T_{++} (x_+ )$. We will refer to this class of metrics as  the chiral sector,  since  the right moving sector of the dual theory is in the ground state. 

We first note that setting $T_{--} =-1$ in the Fefferman-Graham metric (\ref{FGmetric}) and changing to the coordinates  $(t, y, x_+)\sim (t, y, x_+ + 2 \p)$, the metric in the chiral sector can be written as 
\be 
ds^2 =- \left( dt  + {1 \over 4} ( y^{-2}-2 - T_{++} y^2)  dx_+\right)^2 + {dy^2\over y^2} 
+ {1 \over 16}\left( y^{-2} + T_{++} y^2\right)^2 dx_+^2 \,.\label{FGfibered}
\ee
In this form,  time is fibered over a 2D base manifold. 
One now observes that the  metric on the base, given by the last two terms in the above expression, has constant negative curvature. Therefore, if we make  a coordinate transformation which brings this base metric in conformal gauge, i.e. such that
\be 
 {dy^2\over y^2} 
+ {1 \over 16}\left( y^{-2} + T_{++} y^2\right)^2 dx_+^2 = e^{- 2 \F (z , \bar z)} dz d \bar z ,\label{confgauge}
\ee
the field $\F$ will satisfy Liouville's equation (\ref{Liouv}).
Summarizing, we have argued that the metrics in the chiral sector can be
brought into the form
\be 
ds^2 =  -  (dt -  A)^2 +  e^{-2\F} dz d\bar z \, ,\label{Liouvspin}
\ee
where $\F$ satisfies Liouville's equation (\ref{Liouv}). 
 Without loss of generality, we can choose the $z$ coordinate such that the $y\to 0$ conformal boundary in (\ref{FGfibered}) corresponds to $|z|=1$.
In \cite{Hulik:2016ifr}, to which we refer for more details, it was shown that the AdS$_3$ boundary conditions imply that the
the holomorphic `stress tensor' $\calt (z) = - (\pa_z \F)^2- \pa_z^2 \F$ constructed out of the bulk Liouville field can be extended to the full complex plane so as to satisfy the reflection condition
\be
\calt (z) = {1 \over z^4} \bar \calt  (1/z)\,.
\label{doubling}
\ee
The boundary stress tensor $T_{++}(x_+)$ can be found to be
\be 
T_{++} (x_+)=  - 1 +  e^{2 i x_+} \calt ( e^{ i x_+} )\,.\label{Tleft}
\ee
Analytically continuing to Euclidean signature and mapping the cylinder to the plane as in (\ref{analcont}) with complex coordinate $v$, we find that the holomorphic boundary stress tensor is formally identical to $\calt $:
\be 
T (v) = \calt (v), \qquad \bar T (\bar v) =0. \label{bulkbdychiral}
\ee
This should again be  seen as a bulk-boundary map in the chiral sector, showing e.g. how CFT vertex operator insertions (producing second order  poles in $T (v)$) translate into sources for the bulk gravitational field. We will study these sources in more detail in the next Section.

Returning to the special case of the BTZ metric in the chiral sector, the transformation to the form (\ref{Liouvspin}) is accomplished by rewriting the metric (\ref{BTZJ}) for $J = M+1$ in the fibered form
\be \label{BTZcompleted}
ds^2 = - \left( dt + \left(u - {J \over 2} \right) dx_+\right)^2 + {du^2 \over 4 \left(u^2 -  M u + {J^2 \over 4} \right)} +  \left(u^2 -  M u + {J^2 \over 4} \right) dx_+^2 \, ,
\ee
and transforming the base metric (i.e. the last two terms)  to conformal gauge. Explicitly one finds
\be
z = |z| e^{i x_+}, \qquad \ln |z| = {1 \over \sqrt{M+J} }\left( \arctan { 2 u - M \over \sqrt{M+J}} -{\p \over 2} \right) \, ,\label{LiouvtoBTZ}
\ee
and the Liouville  solution describing the chiral BTZ metric is
\be 
e^{-2 \F} = {M+J \over 4 |z|^2 \sin^2 (\sqrt{M+J } \log |z| )}\, .
\ee
Also the holomorphic Liouville stress tensor is
\be 
\calt = {M+J +1 \over  z^2} = {2(M+1) \over  z^2} .
\ee

As was the case for the static black hole, the overspinning geometry has a throat region described by a Liouville solution of hyperbolic type. The Liouville form (\ref{Liouvspin}) of the metric describes the fully extended solution including the region of negative $u$, and the throat  connects the outer boundary at $u \to \infty$ or $|z|=1$ with the inner boundary at $u \to -\infty$ or $|z |= e^{ - { \p \over \sqrt{ M+J}}}$. The narrowest point of the throat, where the horizon would be in the static black, is at $u={M\over 2}$ or $|z |= e^{ - { \p \over 2 \sqrt{ M+J}}}$.

Another class of solutions within the chiral  metrics (\ref{Liouvspin}), which will play an important role in what follows,  are those where the Liouville field has an elliptic singularity. Here the base metric is as illustrated in figure \ref{FigERa} and has only one boundary and a conical singularity in the interior. We will show below that these solutions arise from the backreaction of a spinning particle source.

\subsection{Late-time decay of  two-point  functions}\label{sec:2ptBTZ}

A second feature that the overspinning metrics share with black holes is the late-time decay of correlation functions probing the state, at least for a subset of correlators.  

Let us first recall some thermodynamics in the spinning BTZ black hole regime of the metric (\ref{BTZJ}), when $|J| \leq M, M\geq0$. One often formally defines the left- and right- moving temperatures
\be 
2 \p T_L = \sqrt{M + J}, \qquad 2 \p T_R = \sqrt{ M - J } \label{temps}
\ee
They are related to the Hawking temperature of the black hole as
\be 
T_H^{-1} = \half \left( T_L^{-1} +  T_R^{-1} \right),
\ee
and the entropies of  the outer and inner horizons  take a split form 
\begin{equation}
    S^{out,in} = S^+ \pm S^- = \frac{\pi }{3}\left(c_LT_L \pm c_RT_R\right)
\end{equation}

Though the overspinning  metric of interest (\ref{BTZJ})  doesn't have a horizon and is not in the black hole regime, we can from (\ref{temps})   formally associate a left-moving temperature $T_L = \sqrt{M+J}/2\pi$ to it.
The aim in the rest of this section is to  clarify what this statement  means concretely in terms of physical observables in the theory.

\subsubsection{Generic spacelike geodesics}

To study the thermal behaviour from the bulk side we want to consider particle probes in the overspinning BTZ geometry that start and end at timelike separated points at the boundary, while in between capture the characteristics of the geometry by diving into the bulk.
In Lorentzian signature, timelike geodesics don't reach the boundary, and the desired geodesics are actually spacelike geodesics of infinite proper length \cite{Balasubramanian:2011ur}.

Our starting point is the spinning  BTZ metric \eqref{BTZJ} with the  radial coordinate redefined to $u=r^2$,
\begin{equation}
    ds^2 = \left(M-u\right)d t^2 + \frac{d u^2}{4u\left(u-M+\frac{J^2}{4u}\right)} + u d\phi^2 +Jd t d\phi     .
\end{equation}
We define
\begin{equation}
    u_{\pm}=r_{\pm}^2 = \frac{1}{2}\left(M\pm\sqrt{M^2-J^2}\right)
\end{equation}
from which we infer the relations
\begin{equation}
    M=u_+ + u_-\, , \quad J^2=4u_+u_-\, ,\quad \sqrt{M\pm J} = \sqrt{u_+}\pm\sqrt{u_-}.
\end{equation}
For black holes  $u_\pm$ are real, while in the overspinning regime they are complex and conjugate to each other.

We are interested in spacelike geodesics denoted by $X^{\mu}(s) = \left( T(s), U(s), \Phi(s)  \right)$ and parametrised by proper length such that
\begin{equation}\label{spacelikecondition}
   \left({d X \over ds}\right)^2\equiv \dot{X}^2=1
\end{equation}
The BTZ metric has two killing vectors $V_t=\partial_t$ and $V_{\phi}=\partial_{\phi}$ from which get two conserved quantities
\begin{equation}\label{Echarge}
    \frac{d\sqrt{\dot{X}^2}}{d\dot{t}} = V^{\mu}_t\dot{X}^{\nu}g_{\mu\nu} = E
\end{equation}
\begin{equation}\label{Lcharge}
    \frac{d\sqrt{\dot{X}^2}}{d\dot{\phi}} = V^{\mu}_{\phi}\dot{X}^{\nu}g_{\mu\nu} = L
\end{equation}
From \eqref{spacelikecondition}, \eqref{Echarge} and \eqref{Lcharge} we derive the corresponding differential equations
\begin{equation}
    \begin{split}
        & E\dot{T}+L\dot{\Phi}+ \frac{\dot{U}^2}{4U\left(U-M+\frac{J^2}{4U}\right)}=1\\
        & E=(M-U)\dot{T}+\frac{J}{2}\dot{\Phi} \\
        & L = \frac{J}{2}\dot{T} + U\dot{\Phi} .
    \end{split}
\end{equation}
Solving these we find the spacelike geodesics 
\begin{equation}\label{3dBTZgeods}
    \begin{split}
        & U(\tau) = \left(\Lambda_+-\Lambda_- \right)\cosh^2(\tau-\tau_0)+\Lambda_-, \\
        & T(\tau) = \frac{1}{(u_+-u_-)}\Bigg(-\sqrt{u_-}\mathrm{arccoth}\left(\sqrt{\frac{u_--\Lambda_-}{u_--\Lambda_+}}\tanh\left(\tau-\tau_0\right)  \right) + \\
        & \sqrt{u_+}\mathrm{arccoth}\left(\sqrt{\frac{u_+-\Lambda_-}{u_+-\Lambda_+}}\tanh\left(\tau-\tau_0\right)  \right)\Bigg) + T_0, \\
        & \Phi(\tau) = \frac{1}{(u_+-u_-)}\Bigg(\sqrt{u_+}\mathrm{arccoth}\left(\sqrt{\frac{u_--\Lambda_-}{u_--\Lambda_+}}\tanh\left(\tau-\tau_0\right)  \right) - \\
        & \sqrt{u_-}\mathrm{arccoth}\left(\sqrt{\frac{u_+-\Lambda_-}{u_+-\Lambda_+}}\tanh\left(\tau-\tau_0\right)  \right)\Bigg) + \Phi_0,
    \end{split}
\end{equation}
where we defined
\begin{equation}
    \Lambda_{\pm}=\frac{1}{2}\left(L^2-E^2+M\pm\sqrt{\left(\left(E-L\right)^2-(M-J)  \right)\left( \left(E+L\right)^2-(M+J)  \right)}\right).\label{Lpm}
\end{equation}
Another class of solutions is similar to \eqref{3dBTZgeods} but with $\mathrm{arccoth}$ replaced by $\mathrm{arctanh}$. As we will see, in the black hole regime these solutions connect only spacelike separated points at the boundary. However they will be needed in the case of overspinning geometries. As long as $\L_\pm$ are real, the above geodesics are  real solutions both in the black hole and overspinning cases.

In the underspinning case ($|J|<M$), $u_{\pm}$ are real and thus $T$ and $\Phi$ blow up when the argument of $\textrm{arccoth}$ becomes one. This can be regularized by considering complexified geodesics as in   \cite{Balasubramanian:2012tu}.
However, in the overspinning case ($|J|>M$), $u_{\pm}$ are complex and thus $T$ and $\Phi$ remain finite. Thus no regularisation is needed and one can work with the geodesic equations as they are.

\paragraph{For AdS space:} We have $u_+=0$, $u_-=-1$ and
\begin{equation}
    \Lambda_{\pm}^{AdS}\equiv \Sigma_{\pm}=\frac{1}{2}\left(L^2-E^2-1\pm\sqrt{\left((E-L)^2+1 \right)\left((E+L)^2+1   \right)}\right)
\end{equation}
The $T$ and $\Phi$ geodesics then simplify as
\begin{equation}
    \begin{split}
        & T(\tau) = -  \mathrm{arccot}\left(\sqrt{\frac{-1-\Sigma_-}{1+\Sigma_+}}\tanh\left(\tau-\tau_0\right)  \right) + T_0 \\
        & \Phi(\tau) = - \mathrm{arccot}\left(\sqrt{\frac{-\Sigma_-}{\Sigma_+}}\tanh\left(\tau-\tau_0\right)  \right)\Bigg) + \Phi_0
    \end{split}
\end{equation}
\paragraph{For the maximally overspinning case:} When $J=M+1$, we have $u_+=\bar{u}_-=\frac{1}{2}\left(M + i \sqrt{2M+1}\right)$ and
\begin{equation}
    \Lambda_{\pm} = \frac{1}{2}\left(L^2-E^2+M\pm\sqrt{\left( (E-L)^2+1   \right)\left( (E+L)^2 -2M -1  \right) }\right)
\end{equation}

\subsubsection{Evaluation of geodesic length}\label{sec:GedLength}
For spinless particles evaluating the worldline action amounts  to finding the geodesic length
\begin{equation}
    \Delta s = \int_{s_i}^{s_f}\dif s=s_f-s_i
\end{equation}
We calculate the proper length of geodesics starting and ending on the boundary at infinite $u$. We regularise our computations by having the probe particle start and end at radial coordinate $u=\frac{1}{\epsilon}$ and take $\epsilon\rightarrow 0$ in the end. Also without loss of generality we will take $s_0=T_0=\Phi_0=0$. Thus we want to study geodesics between the points
\begin{equation}
    s_{f,i} = \pm \mathrm{arccosh}\left( \sqrt{\frac{\frac{1}{\epsilon}-\Lambda_-}{\Lambda_+-\Lambda_-}} \right)
\end{equation}
After expanding in terms of $\epsilon$ and keeping terms up to order one we have
\begin{equation}
    \Delta s = -\log\epsilon - \log\left( \frac{\Lambda_+-\Lambda_-}{4}  \right) + \ldots
\end{equation}
As explained in \cite{Balasubramanian:2011ur}, this divergent quantity in the limit $\e \to$ is to be renormalized by subtracting  the divergent part in the global AdS geometry. After removing the regulator we  then obtain
\begin{equation}\label{GLengthBTZ3dGen}
    \Delta s^{ren} = - \log\left( \frac{\Lambda_+-\Lambda_-}{4}  \right).
\end{equation}
For this expression to make sense  we must  require 
that the argument of the logarithm is greater than zero.\\
Our goal is to find the geodesic length as a function of $X_{\pm} = T \pm \Phi$. We have
\begin{equation}
    \Delta T = T\left(s_f\right) - T\left(s_i\right) = 2T\left(\infty\right) \, , \quad \Delta \Phi = \Phi\left(s_f\right) - \Phi\left(s_i\right) = 2\Phi\left(\infty\right)
\end{equation}
which for the $\mathrm{arccoth}$ branch is
\begin{equation}\label{3dDTDF}
    \begin{split}
        & \Delta T = \frac{2}{(u_+-u_-)}\Bigg(-\sqrt{u_-}\mathrm{arccoth}\left(\sqrt{\frac{u_--\Lambda_-}{u_--\Lambda_+}}  \right) + 
         \sqrt{u_+}\mathrm{arccoth}\left(\sqrt{\frac{u_+-\Lambda_-}{u_+-\Lambda_+}} \right)\Bigg)  \\
        & \Delta \Phi = \frac{2}{(u_+-u_-)}\Bigg(\sqrt{u_+}\mathrm{arccoth}\left(\sqrt{\frac{u_--\Lambda_-}{u_--\Lambda_+}}  \right) - 
         \sqrt{u_-}\mathrm{arccoth}\left(\sqrt{\frac{u_+-\Lambda_-}{u_+-\Lambda_+}}  \right)\Bigg) 
    \end{split}
\end{equation}
After solving the latter in terms of $\Lambda_{\pm}$ and expressing it in terms of $X_{\pm}$ we have
\begin{equation}\label{LambdaArccothBTZ}
    \Lambda_{\pm} = \frac{M}{2}\pm\frac{\sqrt{M^2-J^2}}{2}\left( \frac{1 \mp \cosh\left(\frac{\sqrt{M+J}}{2}\Delta X_+  \right)\cosh\left(\frac{\sqrt{M-J}}{2}\Delta X_-  \right)}{\sinh\left(\frac{\sqrt{M+J}}{2}\Delta X_+  \right)\sinh\left(\frac{\sqrt{M-J}}{2}\Delta X_-  \right)}  \right)
\end{equation}
For the $\mathrm{arctanh}$ branch we find the above equation with the signs in $\Lambda_{\pm}$ reversed
\begin{equation}\label{LambdaArctanhBTZ}
    \Lambda_{\pm} = \frac{M}{2}\mp\frac{\sqrt{M^2-J^2}}{2}\left( \frac{1 \pm \cosh\left(\frac{\sqrt{M+J}}{2}\Delta X_+  \right)\cosh\left(\frac{\sqrt{M-J}}{2}\Delta X_-  \right)}{\sinh\left(\frac{\sqrt{M+J}}{2}\Delta X_+  \right)\sinh\left(\frac{\sqrt{M-J}}{2}\Delta X_-  \right)}  \right)
\end{equation}
Let us now use the above to express the argument of the logarithm in \eqref{GLengthBTZ3dGen} as a function of $\Delta X_{\pm}$. From \eqref{LambdaArccothBTZ} and \eqref{LambdaArctanhBTZ} we have 
\begin{equation}\label{3dBTZCorr}
    \frac{\Lambda_+-\Lambda_-}{4} = \pm\frac{\sqrt{M^2-J^2}}{4\sinh\left( \frac{\sqrt{M+J}}{2}\Delta X_+ \right)\sinh\left( \frac{\sqrt{M-J}}{2}\Delta X_- \right)}>0
\end{equation}
where the plus sign is for the $\mathrm{arccoth}$ branch and the minus sign for the $\mathrm{arctanh}$ branch. Thus it is clear that in the black hole regime the above inequality is satisfied by the $\mathrm{arccoth}$ branch for timelike separated points, $(\Delta X_+\Delta X_->0)$, and by the $\mathrm{arctanh}$ for spacelike separated points, $(\Delta X_+\Delta X_-<0)$.\\
In the overspinning case with $J>M$ the above equation becomes
\begin{equation}\label{3dHyperbolicCorr}
    \frac{\Lambda_+-\Lambda_-}{4} = \pm\frac{\sqrt{M+J}\sqrt{|M-J|}}{4\sinh\left( \frac{\sqrt{M+J}}{2}\Delta X_+ \right)\sin\left( \frac{\sqrt{|M-J|}}{2}\Delta X_- \right)}>0
\end{equation}
In this case it is clear that once the  sine becomes negative then we need to switch to the $\mathrm{arctanh}$ for the inequality to be satisfied.
We also write the result for $AdS_3$ which will be useful later
\begin{equation}\label{3dAdS3Corr}
\frac{\Lambda_+-\Lambda_-}{4} = \pm\frac{1}{4\sin\left( \frac{1}{2}\Delta X_+ \right)\sin\left( \frac{1}{2}\Delta X_- \right)}>0
\end{equation}
%

\subsubsection{Spin contribution to the two point function}

Next we want to extend the above result for scalar particles  to the computation of the probe two-point function of operators with spin. From the bulk point of view, in the saddle-point approximation this involves evaluating the worldline action for a spinning particle in the background. The motion of a point particle with  intrinsic spin in general relativity is described by the Mathisson-Papapetrou-Dixon equations \cite{Mathisson:1937zz,Papapetrou:1951pa,Dixon:1970zza}. In three dimensions, these equations can be derived from an action principle.
This action was worked out in \cite{Castro:2014tta}, and we will review it in more detail in section \ref{sec:spinbackr} below. For now it suffices to mention that in three dimensions, standard geodesics are still solutions of the equations of motion of the spinning particle. Therefore, to compute the spin dependence of the two-point function  it suffices to evaluate the spin-dependent part of the action on the geodesics we constructed above. This spin-dependent part is given by
\begin{equation}
S_{s} = \tilde s \int d s  \tilde N^\m \nabla N^\n
\ee
where $\tilde s$ is the spin of the particle and $N$ and $\tilde N$ are orthonormal vectors to the tangent vector $\dot X$,
\be 
N^\m N_\m = 1,  \qquad \tilde N^\m \tilde N_\m= -1 , \qquad N^\m \tilde N_\m=N^\m \dot X_\m =\tilde N^\m \dot X_\m=0.
\ee
As shown in \cite{Castro:2014tta} this  spin term depends only on the boundary values $N_{i,f} = N(s_{i,f})$ of the normal vector and can be written as
\begin{equation} \label{spineval}
S_{s}  = \tilde{s} \log \left( \frac{(q_f -\tilde q_f) \cdot N_f}{(q_f -\tilde q_f) \cdot N_i}  \right) \, .
\end{equation}
where the vectors $q,\tilde{q}$ constitute a parallel transported  frame  normal to the geodesic and $q_{i,f}=q(s_{i,f})$, $\tilde{q}_{i,f}=\tilde{q}(s_{i,f})$.
The spin term then measures how strongly the normal frame
$(q, \tilde q)$ gets boosted  with respect to the frame  $(N, \tilde N)$ while being parallel transported from  the starting point to the endpoint of the geodesic. For simplicity, we will explicitly compute the spin term (\ref{spineval}) only in the black hole regime where $T_L$ and $T_R$ are real, and then analytically continue the result to the overspinning regime.

As argued in \cite{Castro:2014tta}, the appropriate boundary condition to impose is to take $N_i$ and $N_f$ to be equal
\be 
N_i=N_f= n.
\ee
In what follows we will take $n$ to take the form
\begin{equation}
n= \begin{pmatrix}
\ex^{T_R (\phi - t )} + \ex^{T_L (t +\phi )},
&
\ex^{T_R (\phi - t )} - \ex^{T_L (t +\phi )},
&
0
\end{pmatrix} \, .
\end{equation}
This vector is not of unit length, however the formula \eqref{spineval} is scale invariant and so the normalization is not necessary.
It can be shown  that the choosing a different $n$ results in a constant factor in the two-point function which can be absorbed in the normalization of the operator.

The reference frame can be chosen arbitrarily in the initial point and then parallel transported to the final point. However, instead of finding a set of normal vectors in BTZ spacetime directly we use the fact that AdS and BTZ spaces are locally isomorphic via \eqref{trans}. Thus we can translate normal vectors from AdS to BTZ space. 

In AdS space in lightcone Poincar\'e coordinates $(w_+,w_-,z)$ the parallel normal vectors are
\begin{gather}
q_{\rm AdS} = \begin{pmatrix}
\frac{z (w_{+ 2} - w_{+ 1})}{\sqrt{(w_{+ 2} - w_{+ 1})(w_{- 2} - w_{- 1})}} ,
&
-\frac{z (w_{- 2} - w_{- 1})}{\sqrt{(w_{+ 2} - w_{+ 1})(w_{- 2} - w_{- 1})}} ,
&
0
\end{pmatrix} \,,
\\
\tilde{q}_{\rm AdS} = \begin{pmatrix} 
\frac{ z l (w_{+ 2} - w_{+ 1})}{\sqrt{(w_{+ 2} - w_{+ 1})(w_{- 2} - w_{- 1})}} ,
&
 \frac{z l (w_{- 2} - w_{- 1})}{\sqrt{(w_{+ 2} - w_{+ 1})(w_{- 2} - w_{- 1})}} ,
&
z^2
\end{pmatrix}
\end{gather}
where $w_{\pm 1,2}$ are the initial and final coordinates of the boundary interval and the parameter $l$ varies from $-1$ in the initial point to $+1$ in the final point. In particular the regularized initial and final values of $q, \tilde{q}$ are

\begin{gather}
q^i_{\rm AdS} = \begin{pmatrix}
\frac{ (w_{+ 2} - w_{+ 1})}{\sqrt{(w_{+ 2} - w_{+ 1})(w_{- 2} - w_{- 1})}} ,
&
-\frac{ (w_{- 2} - w_{- 1})}{\sqrt{(w_{+ 2} - w_{+ 1})(w_{- 2} - w_{- 1})}} ,
&
0
\end{pmatrix} \,,
\\
\tilde{q}^i_{\rm AdS} = \begin{pmatrix} 
\frac{   (w_{+ 2} - w_{+ 1})}{\sqrt{(w_{+ 2} - w_{+ 1})(w_{- 2} - w_{- 1})}} ,
&
 \frac{  (w_{- 2} - w_{- 1})}{\sqrt{(w_{+ 2} - w_{+ 1})(w_{- 2} - w_{- 1})}} ,
&
0
\end{pmatrix} \,,
\\
q^f_{\rm AdS} = \begin{pmatrix}
\frac{ (w_{+ 2} - w_{+ 1})}{\sqrt{(w_{+ 2} - w_{+ 1})(w_{- 2} - w_{- 1})}} , 
&
-\frac{ (w_{- 2} - w_{- 1})}{\sqrt{(w_{+ 2} - w_{+ 1})(w_{- 2} - w_{- 1})}} ,
&
0
\end{pmatrix} \,,
\\
\tilde{q}^f_{\rm AdS} = \begin{pmatrix} 
\frac{  - (w_{+ 2} - w_{+ 1})}{\sqrt{(w_{+ 2} - w_{+ 1})(w_{- 2} - w_{- 1})}} ,
&
 \frac{ - (w_{- 2} - w_{- 1})}{\sqrt{(w_{+ 2} - w_{+ 1})(w_{- 2} - w_{- 1})}} ,
&
0
\end{pmatrix} \,.
\end{gather}
To transform them into BTZ we need to substitute $w_{\pm 1} =  \sqrt{\frac{r^2 - r_+^2}{r^2 - r_-^2}} \ex^{\pi T_{L,R } (-\phi \mp t)}$ and $w_{\pm 2} =  \sqrt{\frac{r^2 - r_+^2}{r^2 - r_-^2}} \ex^{\pi T_{L,R } (\phi \pm t)}$ for the endpoints. This  maps the endpoints of the CFT interval $(w_{+ 2} - w_{+ 1},w_{- 2} - w_{- 1})$ to the interval at the boundary of BTZ spacetime with $(-\tfrac{\phi}{2},-\tfrac{t}{2})$ for the starting point and $(\tfrac{\phi}{2},\tfrac{t}{2})$ for the endpoint. Furthermore, $q$ and $\tilde{q}$ get transformed by the  Jacobian
\begin{equation}
{\rm Jac} = \begin{pmatrix}
 e^{(t +\phi ) T_L} \sqrt{\frac{r^2-r_+^2}{r^2-r_-^2}} T_L & e^{(t +\phi ) T_L} \sqrt{\frac{r^2-r_+^2}{r^2-r_-^2}} T_L &
   -\frac{e^{(t +\phi ) T_L} r \left(r_-^2-r_+^2\right)}{\left(r^2-r_-^2\right){}^2 \sqrt{\frac{r^2-r_+^2}{r^2-r_-^2}}} \\
 -e^{(\phi -t ) T_R} T_R \sqrt{\frac{r^2-r_+^2}{r^2-r_-^2}} & e^{(\phi -t ) T_R} T_R \sqrt{\frac{r^2-r_+^2}{r^2-r_-^2}} &
   -\frac{e^{(\phi -t ) T_R} r \left(r_-^2-r_+^2\right)}{\left(r^2-r_-^2\right){}^2 \sqrt{\frac{r^2-r_+^2}{r^2-r_-^2}}} \\
 e^{t  r_-+\phi  r_+} r_- \sqrt{\frac{r_+^2-r_-^2}{r^2-r_-^2}} & e^{t  r_-+\phi  r_+} r_+
   \sqrt{\frac{r_+^2-r_-^2}{r^2-r_-^2}} & \frac{e^{t  r_-+\phi  r_+} r \sqrt{\frac{r_+^2-r_-^2}{r^2-r_-^2}}}{r_-^2-r^2}
\end{pmatrix} \,.
\end{equation}
Now we can evaluate \eqref{spineval},
which after simplifying the result and taking the exponential yields
\begin{equation}
    \ex^{ - S_{\tilde s} } = \left( \frac{\tfrac{\sqrt{M+J}}{2}\sinh\left( \frac{\sqrt{M-J}}{2}\Delta X_- \right)}{\tfrac{\sqrt{M-J}}{2} \sinh\left( \frac{\sqrt{M+J}}{2}\Delta X_+ \right)} \right)^{\tilde{s}} \, .
\end{equation}
Combining  the results of the last two subsections we obtain the result for the probe two-point function of spinning operators in BTZ backgrounds  anticipated in (\ref{2ptgen}).

\subsection{Further comments}
We have argued in this section that the overspinning BTZ geometry shares some features with black holes, namely the presence of two boundaries and the thermal behaviour of certain correlators, leading to forbidden singularities in the Euclidean theory. As in the black hole case, this suggests a breakdown of the classical gravity approximation for the description of the pure state created by a heavy left-moving primary, and
corrections  are expected to be significant.  Indeed,   as in \cite{Fitzpatrick:2016ive}, it can be argued that nonperturbative $1/c$ effects resolve the forbidden singularities in the correlator. 
In fact, the arguments presented to this effect in \cite{Fitzpatrick:2016ive} also apply to the current chiral situation, since they rely on an analysis of Virasoro blocks which are holomorphically factorized. 

Before moving on to analyze a class of pure states which have a reliable classical geometry with  a capped throat which returns information, we end this section with some further comments and puzzles related to the overspinning BTZ geometry.
\begin{itemize}
\item As argued in \cite{Maldacena:2001kr}, the extended BTZ black hole geometry can also be interpreted holographically as an (approximate) description of the thermofield double state \cite{Israel:1976ur}. The   
properties of the overspinning BTZ geometry we discussed above suggest that it might allow a similar interpretation as a state in  a product of two decoupled CFTs. In this case however the inner boundary has  pathologies   in the form of closed timelike boundary curves, which should be reflected in pathologies in  the 
CFT living there.
It's an interesting open question whether a  consistent   double copy interpretation for the  overspinning BTZ geometry is possible.

\item Although, as we saw,  the overspinning BTZ geometries behave thermally in some respects,  the thermodynamics of these solutions is not so clear.
They do not have a horizon and therefore no  macroscopic entropy. To see this from the dual CFT side, we first note that we cannot apply Cardy's formula
\be 
S = 2\p \left( \sqrt{c L_0 \over 6} + \sqrt{\bar c \bar L_0 \over 6} \right)\, ,
\ee
since for the left-thermal solution $\bar L_0$ is negative. As far as we are aware, there is no known universal expression for the  degeneracy of chiral states  of a holographic CFT in the limit of large dimension. However we can get some idea of the degeneracy from looking at examples. Consider the $N$-th symmetric orbifold of a free seed CFT at large $N$. The central charge of the orbifold CFT is $c = c_{seed} N$.
Purely chiral states which are in the right-moving ground state can come only from the untwisted sector (since the twisted sectors have non-vanishing right-moving zero-point energy) and can be seen to be in one-to-one correspondence with chiral states in the free seed theory. Their entropy is therefore
\be 
S \sim \sqrt{c_{seed} L_0} = {\sqrt{c L_0} \over \sqrt{N}} \, ,
\ee
and is down by a factor $\sqrt{N} \sim \sqrt{c}$ compared to the  entropy carried by a BTZ black hole and counted by the Cardy formula. 
\end{itemize}

\section{Pure chiral states with a cap}\label{sec:microstate}

In the previous section we discussed a form of the information paradox for chiral states: the classical geometry associated to a chiral primary state is `unreasonably thermal' and is expected to receive significant corrections. In the rest of the paper, we will construct a class of chiral, non-primary, pure states  which do not exhibit information loss in the classical gravity approximation. For this we  will  implement the chiral version of the idea outlined in section \ref{sec:extendedcaps}. The advantage of focusing on the chiral sector  is that it is more tractable, since the bulk metric lies in the class of stationary metrics (\ref{Liouvspin}). 

We want to study chiral, non-primary, pure states which are created by acting on the vacuum with a collection of primary chiral  currents which are inserted  at various locations within the unit disk,
\be 
\hat \calv_{h_1} (v_1) \ldots \hat \calv_{h_n} (v_n) |0\rangle, \qquad |v_i| <1 ,
\ee
where the rescaled weights satisfy $h_i \leq 1$ so as to correspond to particle-like rather than black hole-like excitations.
The classical bulk metric will be of the form (\ref{Liouvspin}) where the Liouville solution has elliptic singularities.

In the first part of this section, we will clarify the physical interpretation of these elliptic singularities in the bulk: they correspond to worldlines sources  of spinning particles with equal mass and spin. In the second part, we will solve for the classical gravity metric for a circular array of operators in the limit where they form a continuous distribution. In the  resulting metric, the Liouville throat region is capped off inside the matter shell. In the next section we will investigate whether correlation functions computed in this background display late-time decay.

\subsection{Chiral current insertions  from spinning particles}\label{sec:spinbackr}
In this subsection we will show that insertions of chiral vertex operators with rescaled weights $h_i<1$ correspond to worldline sources  of spinning particles in the bulk.
The motion of a point particle with  intrinsic spin in general relativity is described by the Mathisson-Papapetrou-Dixon equations \cite{Mathisson:1937zz,Papapetrou:1951pa,Dixon:1970zza}. In three dimensions, these equations can be derived from an action principle, which can  be used to derive the gravitational backreaction of spinning particles. This was worked out in \cite{Castro:2014tta}, to  which we refer for details.

The combined action for gravity and the spinning particles is (recall that we set $l=1$)
\bea
S =&=& S_{EH} + S_{wl}\, ,\\
S_{EH} &=& {1 \over 16 \p G} \int d^3 x \sqrt{-g} (R+2)\, ,\\
S_{wl} &=&- \int d\t \left( \tilde m \sqrt{-g_{\m\n} \dot X^\m \dot X^\m } + \tilde  s\, g_{\m\n} \tilde N^\m \nabla N^\n \right)\, .
\eea
Here, $ \tilde  m$ and $\tilde  s$ are  the mass and spin of the particle, $X^\m (\t)$ describes the particle wordline and the covariant derivative on the  worldline  is defined as
\be 
\nabla V^\m =  \dot V^\m + \G^\m_{\n\r} \dot X^\n V^\r  .
\ee 
The vectors $N^\m, \tilde N^\m$ are orthonormal  to the velocity $\dot X^\m$,
\be 
N^\m N_\m =\tilde N^\m \tilde N_\m=1, \qquad N^\m \tilde N_\m=N^\m \dot X_\m =\tilde N^\m \dot X_\m=0.
\ee
Following \cite{Castro:2014tta}, it is convenient to add  Lagrange multiplier terms to the action which enforce these constraints.

The equations of motion from varying the total action with respect to the metric and the worldline variables  are
\bea 
G^{\m\n}- g^{\m\n} &=&\p \int d\t {\d^3( x- X(\t)) \over \sqrt{- g}} \left[  m \dot X^\m \dot X^\m -  s \dot X^{(\m} \nabla S^{\n )\r } \dot X_\r \right]\label{Eeq1}\\
&& +\p  s \nabla_\r \left[ \int d\t {\d^3( x- X(\t)) \over \sqrt{- g}} S^{\r (\m} \dot X^{\n)}\right], \label{Eeq2}\\
0 &=&\nabla \left(  m \dot X^\m -s (\nabla S^{ \m \n}) \dot X_\n\right) \, . \label{geodspin}
\eea
Here, we defined reduced mass and spin parameters
\be 
m \equiv 8  G \tilde  m, \qquad  s \equiv 8  G \tilde s,
\ee
and have chosen the worldline parameter $\t$ to measure proper time, i.e.  such that $\dot X^\m \dot X_\m = -1$. The quantity $S^{\m\n}$ is the spin tensor\footnote{We normalized the $\e$-symbol such that $\e_{\m\n\r} \dot X^\m N^\m \tilde N^\n =1$.}
\be 
S^{ \m\n} \equiv 2 N^{[\m} \tilde N^{\n]} = \e^{\m\n\r}\dot X_\r.
\ee

Let us briefly comment on some important properties of these equations. First of all, the equation (\ref{geodspin}) for the particle trajectory is still solved by an ordinary geodesic $\nabla \dot X^\m =0$: in three dimensions, spinning particles are still  allowed to move on geodesics. There can also be non-geodesic solutions to (\ref{geodspin}), but we will not consider these in this work. It is also useful to note that, if the particle does move on a geodesic, the second term in (\ref{Eeq1}) vanishes. A second remark concerns the fact that the spin-dependent source term (\ref{Eeq2}) contains first derivatives of the delta function and is more singular than spinless particle sources. The inclusion of spin therefore leads to a more singular behaviour of the backreacted metric, as we will illustrate shortly. 

Next we want to work out the equations (\ref{Eeq2}) for the particular case of  the backreaction of spinning particles in $AdS_3$. We assume the metric ansatz (\ref{Liouvspin})
\be 
ds^2 = -  (dt -  A)^2 +  e^{-2\F} dz d\bar z \, ,\label{Liouvspin2}
\ee
and we want to backreact a particle on the trajectory at constant $z$:
\be 
X^\m (\t) =(T (\t) ,Z (\t) ,\bar Z (\t) )= ( \t , z_0, \bar z_0 )\, .
\ee
One easily shows that this trajectory is a geodesic, so that eq.  (\ref{geodspin}) is satisfied, and that $\t$ measures proper time. In global AdS$_3$, the constant $z$ trajectories are curves which spiral around the center, see Figure \ref{Figspiral}.
\begin{figure}
\begin{center}
	\begin{picture}(200,120)
	\put(0,15){\includegraphics[height=80pt]{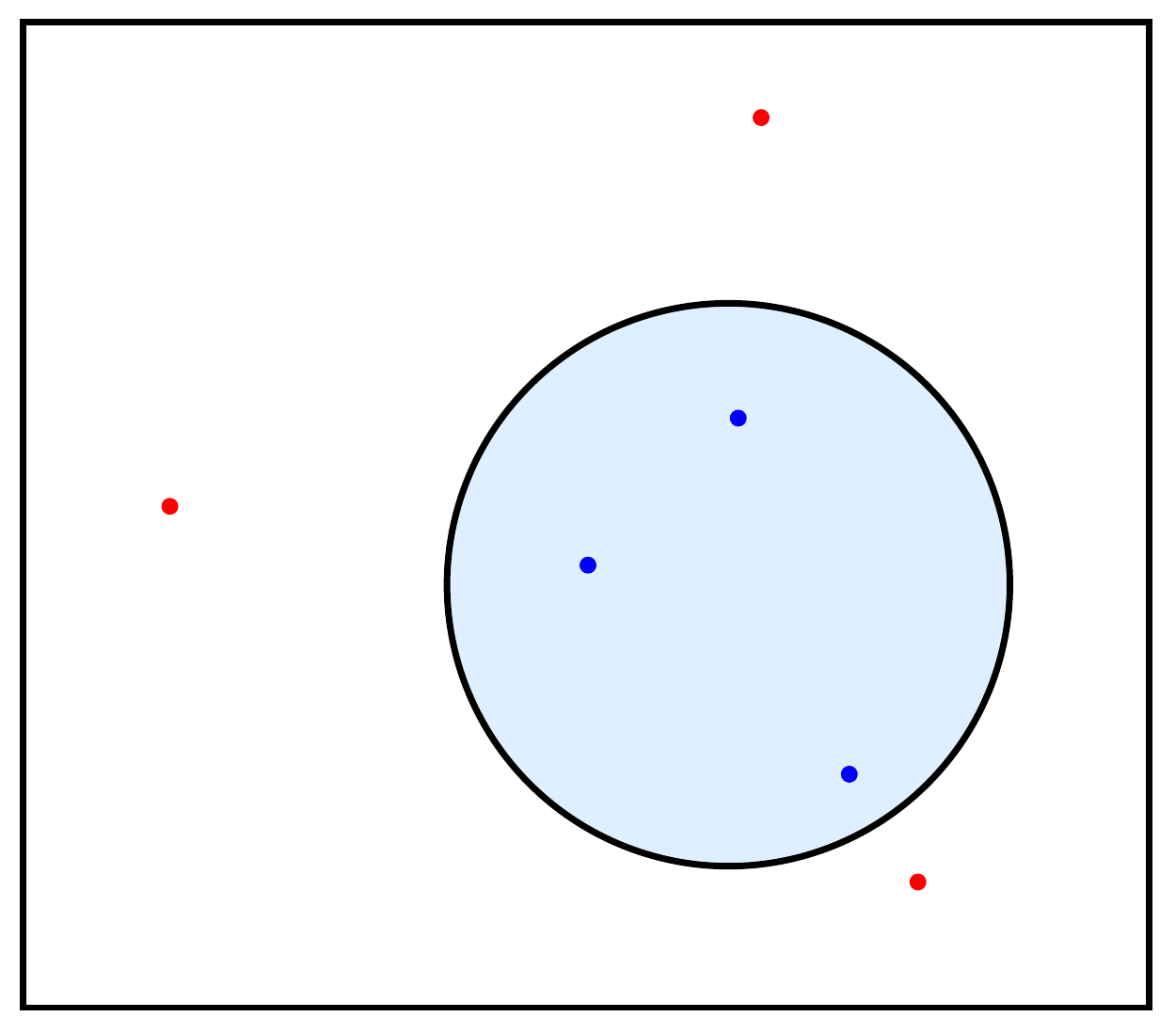}}
		\put(120,10){\includegraphics[height=120pt]{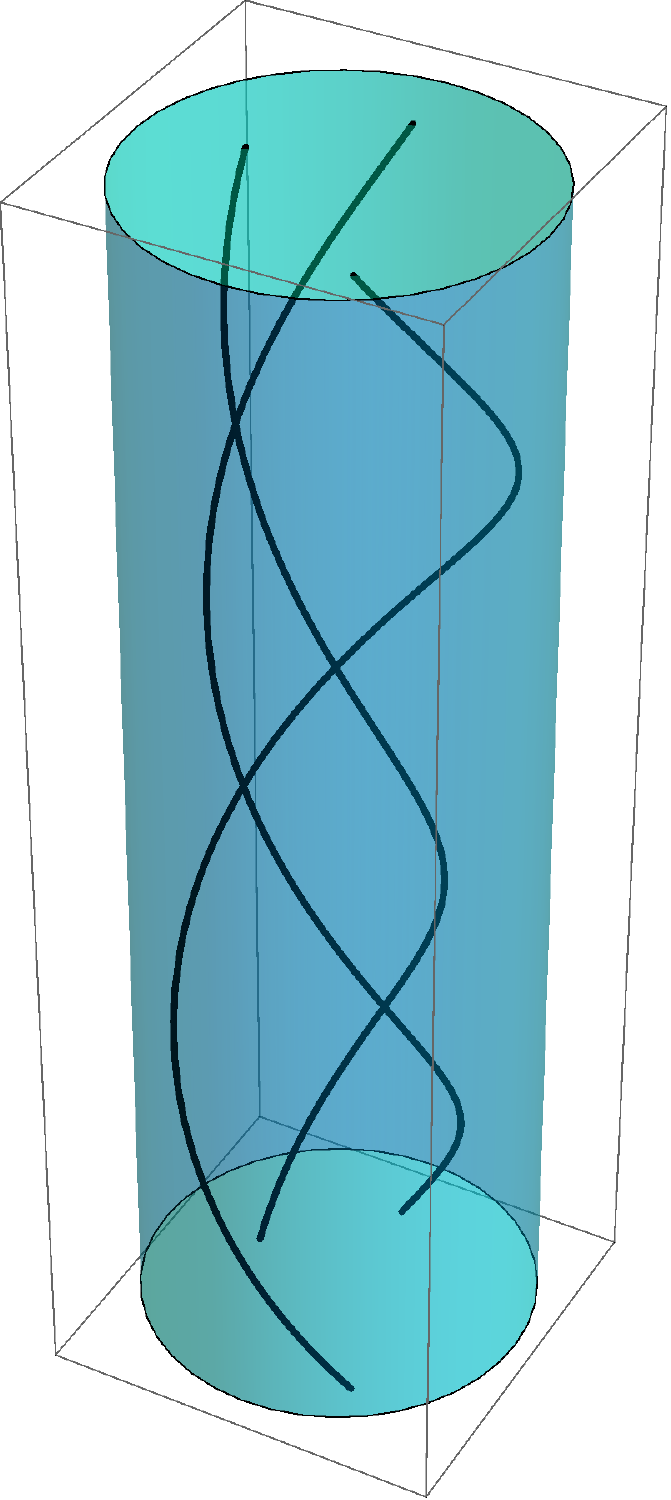}}
			\put(40,0){(a)}\put(140,0){(b)}
\end{picture}
			\end{center}
			\caption{(a) An insertion of three chiral CFT operators in the unit disk  (blue dots) and in their image points on the plane (red dots) corresponds (b) to a configuration of three spinning particles moving on helical geodesics in Lorentzian AdS$_3$. }
\label{Figspiral}\end{figure}

Next we turn to the equations (\ref{Eeq2}) for the backreacted geometry.
One finds that, in order for the terms containing derivatives of delta functions to cancel, 
the gauge field strength should have a delta-function singularity\footnote{Our delta-function is normalized such that $\int | dz d\bar z| \d^2(z,\bar z) =1$, in particular we have $ \pa_z \pa_{\bar z} \ln |z| = \p \d^2(z,\bar z)$ .}:
\be 
F_{z \bar z} = - i e^{-2  \F} + i \p s \d^2 (z- z_0, \bar z- z_0).\label{Feq}
\ee
The remaining Einstein equations reduce to\footnote{Upon substituting (\ref{Feq}) into Einstein's equation, one obtains indeterminate terms which are contain factors like  $e^{2 \F}$, which vanishes at the sources,  multiplied by a singular factor  from  the square of the  delta function. By writing the delta function as a limit of Gaussians one can show that these terms vanish. } the Liouville equation with a source term
\be
\pa_z \pa_{\bar z} \F + e^{-2 \F}= {m + s \over 2}\p \d^2 (z- z_0, \bar z- z_0).\label{Liouvsource}
\ee
Near the source $z=z_0$, the exponential term can be neglected for sufficiently small $m+s$, and the stress tensor constructed from the Liouville field has a double pole
\be 
\calt (z)  \sim 
(m +s) \left( 1- \frac{1}{4} (m+ s) \right) {1 \over (z-z_0)^2 }.\label{singsource}
\ee
The equation (\ref{Feq}) for the gauge potential can be solved in terms of $\F$ to give, up to regular gauge  transformations which can be reabsorbed in a redefinition of $t$,
\be 
A = \Im m \left( \pa_z \F d z - { m- s\over 4 \p} {dz \over z-z_0}\right)\, .\label{A1part}
\ee
The boundary stress tensor for such solutions, transformed to the Euclidean plane, is \cite{Hulik:2018dpl}
\be 
T (v) =\calt (v), \qquad  \bar T (\bar v) =  (m -s) \left( 1- \frac{1}{4} (m- s) \right) {1 \over \bar v^2 }.\label{Tschiral}
\ee 
From the second equality we see that `chiral' particles, for which the mass and spin are equal:
\be 
m = s\, ,
\ee
 do not excite the right-moving sector and produce a chiral metric in our earlier terminology. We will reσtrict our attention to such chiral particle sources in what follows. From the first equality in (\ref{Tschiral}) and from (\ref{singsource}) we see that inserting a  chiral vertex operator in the CFT corresponds to adding    a chiral particle source  in the bulk.
 
 The above comments on   the backreaction of spinning particles provide a direct derivation  in the metric formalism of properties  which were derived in \cite{Hulik:2018dpl} using Chern-Simons variables, where the coupling of spinning particles becomes particularly simple \cite{Castro:2014tta}. 
\subsection{Solution for a shell of spinning particles}\label{sec:shellsol}
The discussion in the previous subsection generalizes in a straightforward manner to the inclusion of several   spinning particle sources.
We will  presently derive the solution for a circular configuration of such particles in a continuum limit where they form a homogeneous    shell. 

We start from the Liouville equation (\ref{Liouvsource}) with sources from $N$ identical particles with mass $m$ and spin   $s=m$ placed symmetrically on a circle of radius $|z| = \r$:
\be
 \pa_z \pa_{\bar z} \F + e^{-2 \F}= m \p \sum_{j = 1}^N  \d^2 \left(z-\r e^{ 2 \p ij\over N}, \bar z- \r e^{-{2\p ij\over N}}\right).
 \ee
We take the continuum limit $N \to \infty$ keeping $m N$ fixed, in which the problem becomes rotationally symmetric and we can take $\F$ to be a function of $|z|$ alone. The equation reduces to 
\be 
{1 \over 4} \left( \F'' + {1 \over r}  \F' \right) + e^{-2 \F } = \a \d (|z| - \r) ,\label{sourceeq}
\ee 
where prime denotes a derivative with respect to $|z|$, and 
 $\a$ is  the   mass per unit radial coordinate of the shell :
 \be \a = \lim_{N \to \infty} { m N \over 4 \r  }.
 \ee 
 
 The solution to (\ref{sourceeq}) will be of the matched form
\be 
\F (r) = \F_{in}  (r) \Theta (\r - |z|) +  \F_{out} (r) \Theta (|z|-\r ),\label{matched}
\ee 
with $\F_{in},\F_{out}$ solutions of the Liouville equation without sources. The solution  $\F_{in} $ is  the vacuum AdS$_3$ solution. For the outside solution, we are interested in the case where it is above the black-hole-like threshold $M+J = 2M+1 >0$, so that  $\F_{out}$ is of the hyperbolic type and  describes a chiral BTZ geometry. 
In what follows we will solve for the density $\a$ of the shell  in terms of the mass $M$.
Concretely, the inner and outer solutions are of the form:
\bea 
\F_{out} &=& \log \left(-2 |z| {\sin (\sqrt{2 M+1} \log |z|) \over\sqrt{2 M+1}} \right) \, , \nonu
\F_{in}&=& \log \left( { 1 -\l^2 |z|^2 \over \l } \right)\, .\label{Liouvinout}
\eea
The as yet undetermined positive parameter $\l$ is introduced for the following reason: a priori the radial coordinates inside and outside of the shell are unrelated, and the shell could be at different values for the inside and outside radial coordinate. By adjusting the parameter $\l$ we can  assume that $r$ is continuous across the   shell, which is located at $r = \r$ both in the inside and outside coordinates.

Substituting (\ref{matched}) into (\ref{sourceeq}) we obtain  the equations 
\bea 
\F_{out} (\r)- \F_{in} (\r ) &=&0\, ,\\
 \F_{out}' (\r) - \F_{in}' (\r ) &=& 2\a \, .\label{matchcond}
\eea
These impose the continuity of $\F$ across the shell and relate the jump in the radial derivative to the source density $\a$. Using the first equation to solve for $\l$ we get
\be 
\l = {1 \over \r} \left( \sqrt{ 1+ \left({ \sin ( \sqrt{2 M+1} \ln \r ) \over \sqrt{2 M+1} }\right)^2 } + { \sin ( \sqrt{2 M+1} \ln \r ) \over \sqrt{2 M+1} } \right)\, .
\ee
Here, we have chosen the appropriate branch for which $\l$ is positive.
The matching condition (\ref{matchcond}) reduces to
\be 
\a = e^{- \F (\r )} \left( \sqrt{ 1+ \left({ \sin ( \sqrt{2 M+1} \ln \r ) \over \sqrt{2 M+1} }\right)^2}-\cos (\sqrt{2 M+1} \ln \r) \right)\, .\\
\ee
In these formulas, the shell radius should be taken to lie   between the inner and outer boundaries of the throat geometry, i.e. in the range $e^{- {\p \over \sqrt{2M+1}}} \leq \r \leq 1 $. One can show that, for fixed mass $M$, $\a (\r)$ is a positive, monotonically decreasing  function of $\r$ 
which interpolates between $+ \infty$ at the inner boundary   and 0 at the outer boundary. Therefore for every shell radius $\r$ in the range $e^{- {\p \over \sqrt{2M+1}}} \leq \r \leq 1 $    there is a unique value of the source density  $\a$ for which (\ref{sourceeq}) is solved.

When embedded in 3D Euclidean space, the solutions look like a throat region glued to a cap along the shell of matter, see Figure \ref{Figcap}. The deeper we make the throat, the greater is the required mass density $\a$ of the shell. For fixed  mass $M$ we can distinguish 3  cases, depending whether the
geometry is capped off before, precisely at, or beyond   the narrowest point of the throat. If we were describing, as in section \ref{sec:eternal}, an actual black hole throat where the narrowest point is the horizon,   the first case (Fig. \ref{Figcap}(a)) would be a star-like configuration spread out over an area greater than the horizon, while the second case (Fig. \ref{Figcap}(b)) would resemble\footnote{More precisely for a fuzzball  the structure is argued \cite{Guo:2017jmi} to appear just (i.e. about a Planck length) outside of the narrowest radius, though at the classical level it does not seem
meaningful to make this distinction.} a microstate geometry \cite{Mathur:2005zp}.
The   last case (Fig. \ref{Figcap}(c)), where the shell is placed beyond the narrowest point of the throat, does not have a stable counterpart in the black hole regime. We will argue below that in this case the
classical gravity picture is unreliable and subject to significant corrections.
\begin{figure}\begin{center}
	\begin{picture}(360,80)
	\put(0,0){\includegraphics[height=80pt]{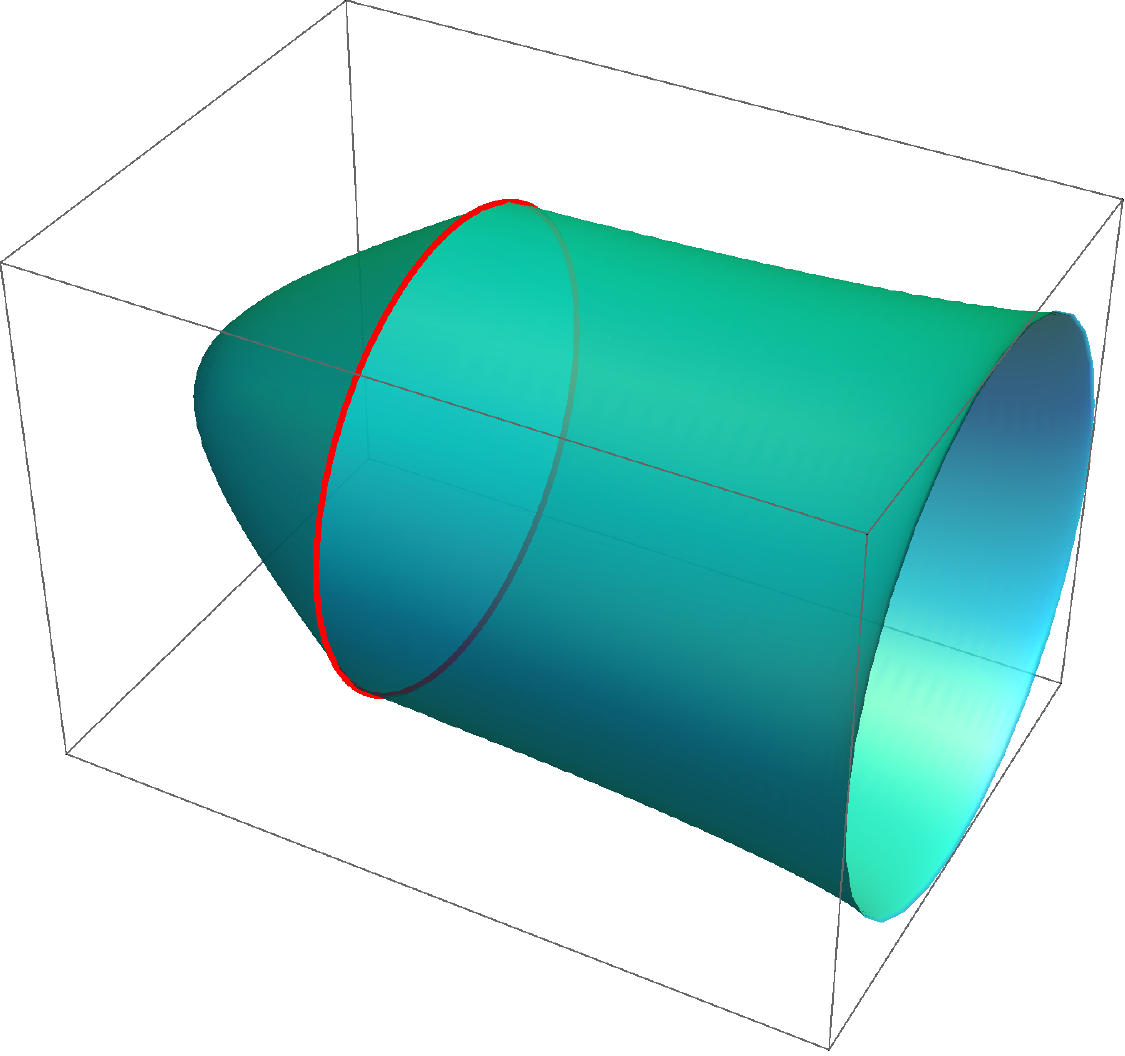}}
		\put(120,0){\includegraphics[height=80pt]{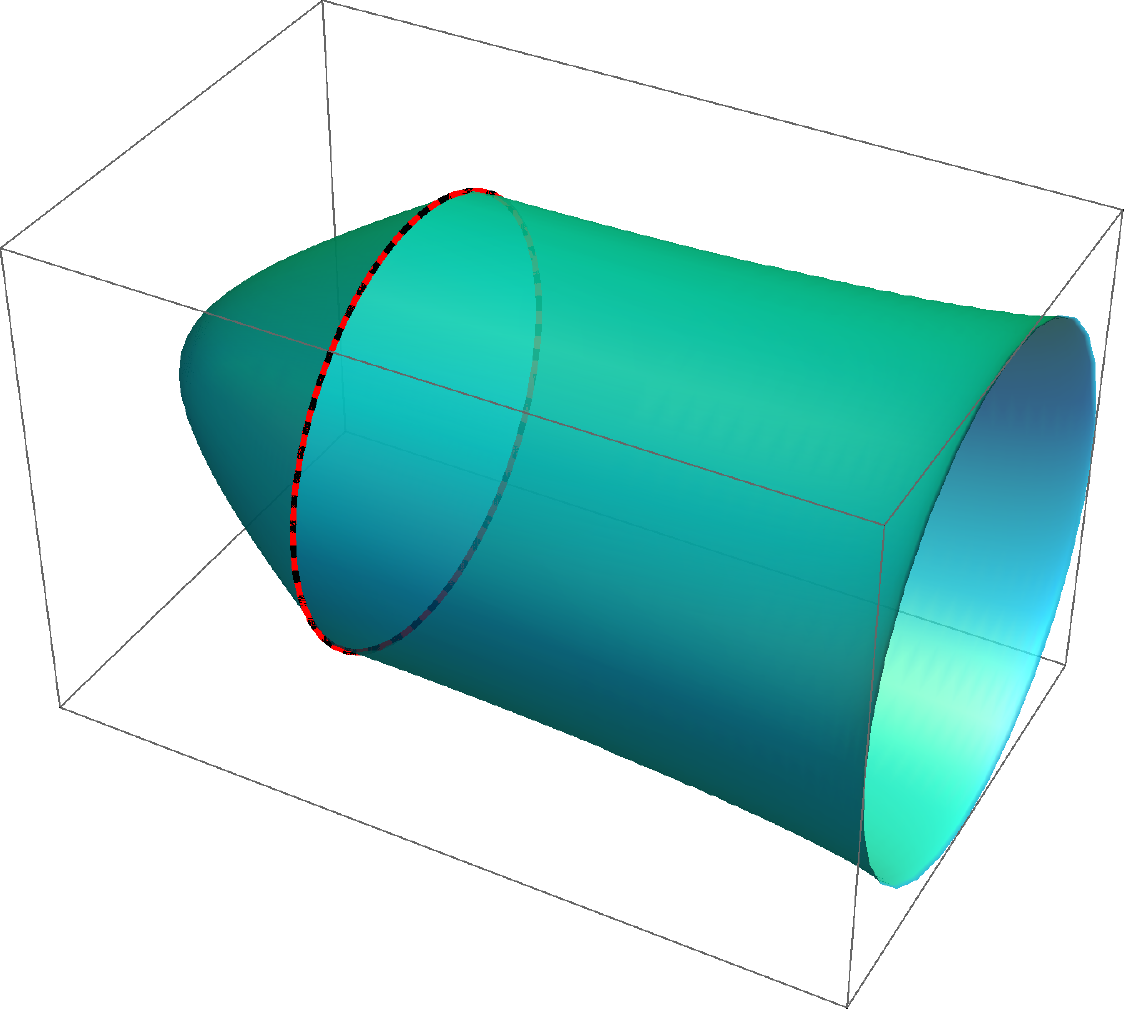}}
			\put(240,0){\includegraphics[height=80pt]{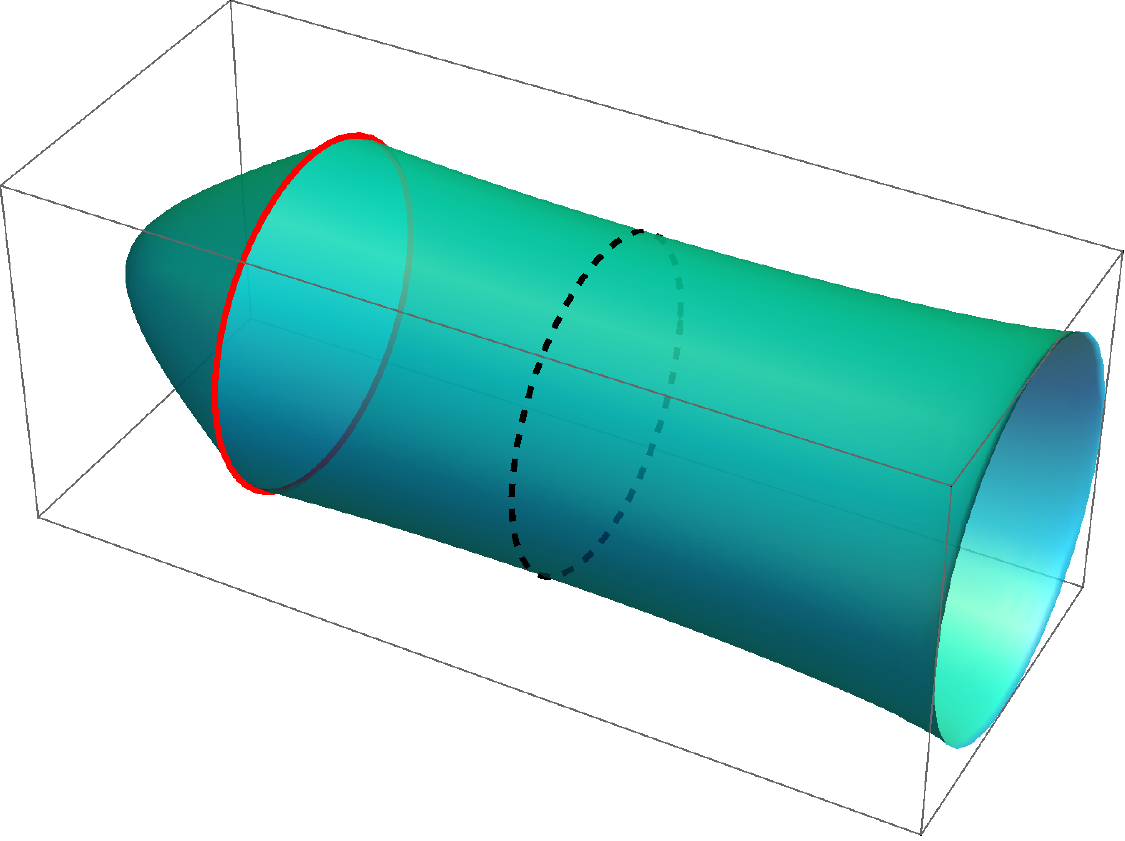}}
		\put(30,-5){(a)}\put(150,-5){(b)}\put(280,-5){(c)}
\end{picture}
			\end{center}
		\caption{Embedded geometry of the solution with matter shell (red circle) for three cases  with the same total mass. In case (a), we have a `star-like'  configuration where the throat is capped off before reaching the narrowest point. Case (b), were  the shell is placed at the narrowest point of the throat, resembles a microstate geometry. Case (c), where the throat is capped off beyond its narrowest point, does not have a counterpart in the black hole regime. }\label{Figcap}
		\end{figure}
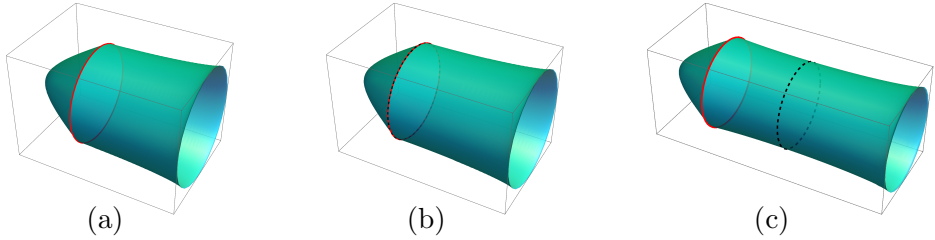

The Liouville stress tensor for the solution (\ref{Liouvinout}) is
\be 
\calt = { M+1 \over 2 z^2 } \Theta( |z|-\r  )\Theta(\r^{-1} - |z|) - {\a \r \over 2 z^2} \left( \r \d (|z|-\r) + \r^{-1} \d (|z|-\r^{-1})\right).\label{caltshell}
\ee
Here, we have extended $\calt$ to the complex plane so as to satisfy the reflection condition (\ref{doubling}). Due to the relation (\ref{bulkbdychiral}) this is also the expression for the  stress tensor in the Euclidean CFT. It's useful to rewrite it as
\be 
T(v)  = \int_0^{2 \p} {d \theta \over 2 \p} \left( {\e \over ( z - \r e^{i \theta} )^2} +  {\e \over ( z - \r^{-1}  e^{i \theta} )^2} + {c e^{-i \theta}\over  z - \r e^{i \theta}}  +  {\tilde c  e^{-i \theta}\over  z - \r^{-1} e^{i \theta}} \right) \, ,\label{Tshell}
\ee
with
\be
\e = {\a \r\over 2}, \qquad c = {M +1-  \a \r\over 2 \r} ,\qquad \tilde  c =- {(M +1+  \a \r)\r \over 2}\, .
\ee 
This form makes it clear that it is  the stress tensor of  a continuous distribution of vertex operators at $|z|= \r$ and their images at $|z| = \r^{-1}$.  The fact that this  results in an expression (\ref{caltshell}) which is  only piecewise holomorphic is familiar from   CFT studies of continuous vertex operator distributions  \cite{Anous:2016kss}.
Note that by considering a  rotationally symmetric configuration we were able to obtain an analytic solution, in particular it determined the accessory parameters $c$ and $\tilde c$ in (\ref{Tshell}). In less symmetric situations this would require solving a difficult monodromy problem \cite{Hulik:2016ifr}.

To write down the full 3D metric (\ref{Liouvspin2}) we also have to specify the one-form $A$, which we can read off from (\ref{A1part}) with $m=s$:
\bea 
A &=&  \Im m \left( \pa_z \F d z \right) = {|z| \over 2} \F' (|z|) d \arg z\nonu
 &=&  {|z| \over 2} \left(  \F'_{in}  (|z|) \Theta (\r - |z|) +  \F'_{out} (|z|) \Theta (|z|-\r ) \right)\, .
\eea
Using (\ref{matchcond}) we note that the gauge field (and hence the metric) is not continuous across the shell but jumps by a large gauge transformation:
\be \label{discont}
( A_{out})_{| |z|=\r} - ( A_{in})_{| |z|=\r} = \a \r\, d \arg z
\ee
Such a jump in the metric components does not occur in the more familiar case of thin shells of non-spinning matter \cite{Israel:1966rt}, and is a consequence of  the more-singular-than-usual  source term from the spin part of the action (\ref{Eeq2}).
 
In what follows we would also like to display the above shell solution in BTZ coordinates. In the outside part we can make the coordinate transformation (\ref{LiouvtoBTZ}).
On the inside we make the transformation
\be 
|z| = {1 \over \l} \sqrt { \s u \over \s u +1}, \qquad  |z|<\r \, ,
\ee
where $\s$ is a parameter similar to $\l$,  to be tuned such that the radial coordinates properly match up at the shell. From the above discussion we know that this requires that the base metric $e^{-2 \F} dz d \bar z$ is continuous across the shell (rather than the full 3D metric), which leads to
\be 
\s (\s u_* +1) =  u_* - M + {(M+1)^2 \over 4  u_*},
\ee
where $u=u_*$ is the location of the shell in BTZ coordinates. Solving for $\s$ gives
\be 
\s= {1 \over 2  u_*}\left( \sqrt{ (2  u_*-M)^2 + 2 (M+1)}-1 \right).\label{sigmadisk}
\ee
 In conclusion, in BTZ coordinates our shell geometry takes the form
\bea
ds^2_{out}&=& -\left(dt +\left(u-\frac{M+1}{2}\right)dx_+\right)^2 + \frac{du^2}{4u\left(u-M+\frac{\left(M+1\right)^2}{4u}\right)} + u\left(u-M+\frac{\left(M+1\right)^2}{4u}\right)dx_+^2\nonu
ds^2_{in}&=& -\left(dt +\s u d x_+\right)^2 + \frac{\s du^2}{4u\left(\s u+1\right)} + \s u\left(\s u+1\right)dx_+^2,\label{shellBTZ}
\eea
where the inner metric applies to $u<u_*$ and the outer one to $u>u_*$.

\section{Two-point function in shell background} \label{sec:geodesic}

In this section we  compute the boundary-to-boundary propagator in the shell background (\ref{shellBTZ}) and investigate its late-time behaviour. As before we work in the saddle-point approximation where we have to  evaluate the worldline action of boundary-to-boundary geodesics.  This leads to a straightforward qualitative picture: as long as the  geodesic lies completely outside of the shell the correlator behaves thermally, while  once  it starts penetrating the shell the behaviour will start resembling more and more the periodic answer in global AdS.

Despite being conceptually straightforward, the details of the calculation reveal interesting subtleties
which are due to the fact that several geodesics can connect the same endpoints in the presence of the shell. As we will see, these lead to `swallow-tail' phenomena and different  geodesics exchanging dominance in the correlator. These phenomena are more pronounced as the shell is placed deeper in the throat region. Similar  features were observed in other thin-shell computations such as \cite{Albash:2010mv,Balasubramanian:2012tu}.   

One important property that can be derived almost without computation is the following: when the shell  radius $u_*$ is below
the value $M/2$ where the throat geometry is at its narrowest (i.e. case (c) in Figure \ref{Figcap}), the geodesics never penetrate the shell and the correlator will still decay at late times. Indeed, it can be seen from (\ref{3dBTZgeods}, \ref{Lpm}) that none of the geodesics penetrates deeper into the bulk than
\be 
u_{min} = {M\over 2}
\ee
which is reached by geodesics with  $E = L = \sqrt{2M+1}/2$.
We will comment more on this result in the Discussion.

\subsection{Simplifying assumptions}
As we saw in the previous section, the presence of spinning matter on the shell results in jumps in the fiber (\ref{discont}), which makes the problem of matching geodesics across the shell nontrivial. Fortunately however, as we shall presently argue, symmetry arguments can be used to reduce the problem to a computation in the 2D base geometry in (\ref{shellBTZ}), where the metric is continuous.
A first simplification we will make is to restrict our attention here to the boundary-to-boundary propagator of a scalar particle with $s=0$; recall from (\ref{2ptgen}) that this correlator displays late-time decay in the overspinning BTZ background.
Secondly, since the shell is comprised of particles with $m=s$ (or, in the CFT language, the inserted vertex operators are chiral), the $x_-$-dependence of the two-point function should be the same as in the global AdS$_3$ background, namely the oscillating factor 
\be \left({2\sin {{\D X_-  \over 2}} } \right)^{-\tilde m} \label{xmdep}
\ee
It is therefore the dependence on the $\D X_+$ separation, for large  $\D X_+$, which decides whether the correlator decays at late times or not. To extract this dependence, we may just as well restrict  to the subset of geodesics which have a fixed value of $\D X_-$. In what follows, we will restrict attention to 
\be 
\D X_-=\p .
\ee
The correlator for general $\D X_-$ can the be obtained by multiplying our result with  
\be \left({\sin {{\D X_-  \over 2}} } \right)^{-m}
\ee
From our expressions for the geodesics in the  BTZ geometry, we find \eqref{3dDTDF} that, both in the global AdS$_3$ and in the $J=M+1$ overspinning geometries, the geodesics  with $\Delta X_-=\pi$ 
are the ones with equal energy and angular momentum,
\be 
E=L.
\ee
The   geodesics with $E=L$  are also the ones which have $\D X_-=\p$ in the shell geometry, for the following reason. Let's fix $E=L$ and first consider a shell with sufficiently small $u_*$ such that the geodesic does not penetrate it; it lies completely in the outer geometry and hence has $\Delta X_-=\pi$. Now let's move the shell outward until it crosses the geodesic; as we have argued this operation should not change the $x_-$-dependence of the two-point function which is still (\ref{xmdep}), and therefore we must still have $\D X_-=\p$.

Now we observe that the  geodesics with $E=L$ have the property that they live entirely in the two-dimensional base geometry, as   the pullback of the first term in the outside  metric (\ref{shellBTZ}) vanishes:
\be 
\dot T  +\left(U-\frac{M+1}{2}\right)\dot X =0
\ee
and similarly on the inside of the shell (note that the shell preserves $x_\pm$ translation invariance and therefore $E$ and $L$ are conserved also in the shell background).
It is therefore the two-dimensional throat geometry which controls the late-time behaviour of the correlator, and  we can reduce the problem to that of computing geodesic lengths in this 2D geometry.

\subsection{Geodesics in the throat}

In view of the above considerations, we first study
general geodesics on the Euclidean two-dimensional  base manifold
with coordinates $u$, $x_+ \sim x_+ + 2 \p$ and with metric
\begin{equation}\label{gendiskmetric}
    ds^2=\frac{du^2}{4u\left(u-M+\frac{\left(M+1\right)^2}{4u}\right)} + u\left(u-M+\frac{\left(M+1\right)^2}{4u}\right)dx_+^2 \, .
\end{equation}
This space has a Killing vector $V_+=\partial_{x_+}$ which results in the conserved quantity $Q_+=E+L$ given by
\begin{equation}\label{chargeQp}
    V^{\mu}_+\dot{X}^{\nu}g_{\mu\nu} = \dot{X}_+ U\left(U-M+\frac{\left(M+1\right)^2}{4U}\right) = \frac{Q_+}{2}\, .
\end{equation}
Parametrizing the geodesics by their proper length we have $\left(\dot{X}\right)^2=1$ which results in the equation
\begin{equation}\label{UeqnBase}
    \frac{1}{4}\left(\dot{U}^2 + Q_+^2\right) = U\left(U-M+\frac{\left(M+1\right)^2}{4U}\right) \, .
\end{equation}
From \eqref{chargeQp} and \eqref{UeqnBase} we find three branches of geodesics: one branch with $\cosh$ in the radial geodesic and two branches with $\sinh$. The latter will be needed to study geodesics that cross the shell, so with some foresight we include them here. For the $\cosh$ branch we have
\begin{equation}\label{geodeqnsBase}
    \begin{split}
        & U= \frac{M}{2} +  \frac{1}{2}\sqrt{Q_+^2-2M-1}\cosh\left(2\left(s-s_0\right)\right) \, ,\\
        & X_+ = \frac{\kappa}{\sqrt{2M+1}}\mathrm{arccoth}\left(\frac{Q_+}{\sqrt{2M+1}}\coth\left(2\left(s-s_0\right)\right)  \right)\, .
    \end{split}
\end{equation}
where $\kappa=\pm 1$.\\
For the two $\sinh$ branches we have
\begin{equation}
    \begin{split}
        & U= \frac{M}{2} \pm  \frac{1}{2}\sqrt{2M+1-Q_+^2}\sinh\left(2\left(s-s_0\right)\right) \, ,\\
        & X_+ = \frac{\kappa}{\sqrt{2M+1}}\mathrm{arctanh}\left(\frac{Q_+}{\sqrt{2M+1}}\tanh\left(2\left(s-s_0\right)\right)  \right)\, .
    \end{split}    
\end{equation}
We observe that the $\cosh$ branch is valid for $Q_+^2>2M+1$, while the $\sinh$ branches are valid for $Q_+^2<2M+1$. The $\cosh$ branch involves geodesics that start and end at the same boundary at $U\rightarrow \infty$ for $s \rightarrow \mp \infty$. However, the $\sinh$ branches extend from one boundary $U\rightarrow \infty$ for $s\rightarrow \pm \infty$ to the other boundary $U\rightarrow - \infty$ for $s\rightarrow \mp \infty$. In the case of the shell though this picture breaks down as the geometry caps off and there is only one boundary.
For the moment, we are not yet considering the  presence of the shell and we move forward by making use of the $\cosh$ branch. Also the solutions that differ by an overall sign through $\kappa$ are related by $Q_+\rightarrow -Q_+$. Again these solutions seem trivial at this point but they will play a significant role when we examine the shell.\\
For a geodesic starting from the boundary $U=1/\epsilon$ with $\epsilon\rightarrow 0^+$ we have
\begin{equation}
    \begin{split}
        & s_i = -\frac{1}{2}\mathrm{arccosh}\frac{\frac{2}{\epsilon}-M}{\sqrt{Q_+^2-2M-1}} + s_0 \approx +\frac{1}{2}\log\epsilon - \frac{1}{2}\log\left(\frac{4}{\sqrt{Q_+^2-2M-1}}\right) +s_0 \, ,\\
        & s_f = + \frac{1}{2}\mathrm{arccosh}\frac{\frac{2}{\epsilon}-M}{\sqrt{Q_+^2-2M-1}} + s_0 \approx - \frac{1}{2}\log\epsilon + \frac{1}{2}\log\left(\frac{4}{\sqrt{Q_+^2-2M-1}}\right) +s_0 \, .\\
    \end{split}
\end{equation}
Thus,
\begin{equation}\label{Dshyper}
    \Delta s = s_f-s_i = -\log\epsilon + \log\left(\frac{4}{\sqrt{Q_+^2-2M-1}}\right)\, .
\end{equation}
Similarly
\begin{equation}
    \Delta X_+ = X_+(s_f)-X_+(s_i) = \frac{2\kappa}{\sqrt{2M+1}}\mathrm{arccoth}\left(\frac{Q_+}{\sqrt{2M+1}}\right)\, .
\end{equation}
Solving the latter for $Q_+$ and substituting back in \eqref{Dshyper}, after renormalizing as in  Section \ref{sec:GedLength}  and letting $\e \to 0$, we get
\begin{equation}\label{2dHyperbolicCorr}
    \Delta s^{ren} = \log\left( \frac{4\kappa\sinh\left(\frac{\sqrt{2M+1}}{2}  \Delta X_+\right)}{\sqrt{2M+1}}  \right)\, .
\end{equation}
This expression decays exponentially for large $\D X_+$ as expected for the thermal solution.  As a check, we see that it indeed matches the three dimensional computation \eqref{3dBTZCorr} for $\Delta X_-=\pi>0$. In order to have timelike separated endpoints at the boundary we need $\Delta X_+ >0$ which requires $\kappa=+1$ for the argument of the logarithm to be positive.

For the AdS case, where $M=-1$, the hyperbolic sine turns into a sine and the behaviour becomes oscillatory, as expected for pure states. 
\begin{equation}
    \Delta s = \log\left( 4\kappa\sin\left(\frac{1}{2}  \Delta X_+\right)  \right) \, .
\end{equation}
Once again it matches the three dimensional result \eqref{3dAdS3Corr} for $\Delta X_-=\pi>0$. 
Again we need $\Delta X_+ >0$ for timelike separate points at the boundary. The choice of $\kappa$ depends on the sign of sine. For example we have $
\kappa=+1$ for $0<\Delta X_+<2\pi$ and $\kappa=-1$ for $2\pi<\Delta X_+<4\pi$. Thus even though the sine has period $4\pi$ in terms of $\Delta X_+$, the overall proper length has period $2\pi$.

\subsection{Geodesics in the presence of the shell}

We now look at geodesics with endpoints on the boundary in the presence of the shell placed at radius $u = u_*$.    For the  metric outside and inside  of the shell we respectively have
\begin{equation}\label{shellbasemetric}
    \begin{split}
        & ds^2_{out}=\frac{du^2}{4u\left(u-M+\frac{\left(M+1\right)^2}{4u}\right)} + u\left(u-M+\frac{\left(M+1\right)^2}{4u}\right)dx_+^2 \,, \qquad {\rm for \ } u >u_* ,\\
        & ds^2_{in} = \frac{\sigma du^2}{4u(\sigma u+1)} + \sigma u(\sigma u+1)dx_+^2\, , \qquad {\rm for \ } u <u_* ,
    \end{split}
\end{equation}
where $\s$ was given in (\ref{sigmadisk}).

Not all geodesics will cross the shell: for small enough endpoint separation $\D X_+$ the geodesic will lie completely outside of the shell. With increasing $\D X_+$ the geodesics reach deeper  into the bulk and eventually approach a minimum radius less than $u_*$. More precisely, the minimum radius a geodesic reaches can be read off from \eqref{geodeqnsBase}, 
\begin{equation}\label{Uminbase}
    U_{min} = \frac{M}{2}+\frac{1}{2}\sqrt{Q_+^2-2M-1}\, ,
\end{equation}
which because of \eqref{Dshyper} and \eqref{2dHyperbolicCorr} gives
\begin{equation}\label{UminDXpbase}
    U_{min} = \frac{M}{2} + \frac{\sqrt{2M+1}}{2|\sinh\left( \frac{\sqrt{2M+1}}{2}\Delta X_+  \right)|} \ge \frac{M}{2}\, .
\end{equation}
As already mentioned, for $u_* < \frac{M}{2}$ the geodesics never penetrate the shell and the geometry behaves as a left-thermal one. For $u_* > \frac{M}{2}$ the geodesic goes through the shell at $\Delta X_+$ given by
\begin{equation}
    \Delta X_+^* = \frac{2}{\sqrt{2M+1}}\mathrm{arccsch}\left( \frac{2u_* -M}{\sqrt{2M+1}}  \right) \, .
\end{equation}
Furthermore, the geodesic that touches the shell has charge
\begin{equation}
    \left(Q_+^*\right)^2 = \left(2u_* - M \right)^2 +2M+1 \, .
\end{equation}
We have that $Q_+^*>2M+1$ which means that it is the $\cosh$ branch of geodesics that initially crosses the shell.

\subsection{Shell 2-point function }
We now construct the geodesics crossing the shell, pictured schematically  in Figure \ref{Geodesic_Shell}.
We denote the parameter values where the geodesic enters and exits the shell by $s_1$ and $s_2$ respectively. Thus the presence of the shell splits the geodesic into three segments: 1) segment ``I" is for parameters between $s_i$ and $s_1$ before the geodesic crosses the shell, 2) segment ``II" is for parameters between $s_1$ and $s_2$ when the geodesic is inside the shell and 3) segment ``III" is for parameters between $s_2$ and $s_f$ when the geodesic comes out of the shell.
\begin{figure}
    \centering
    \includegraphics[scale=0.3]{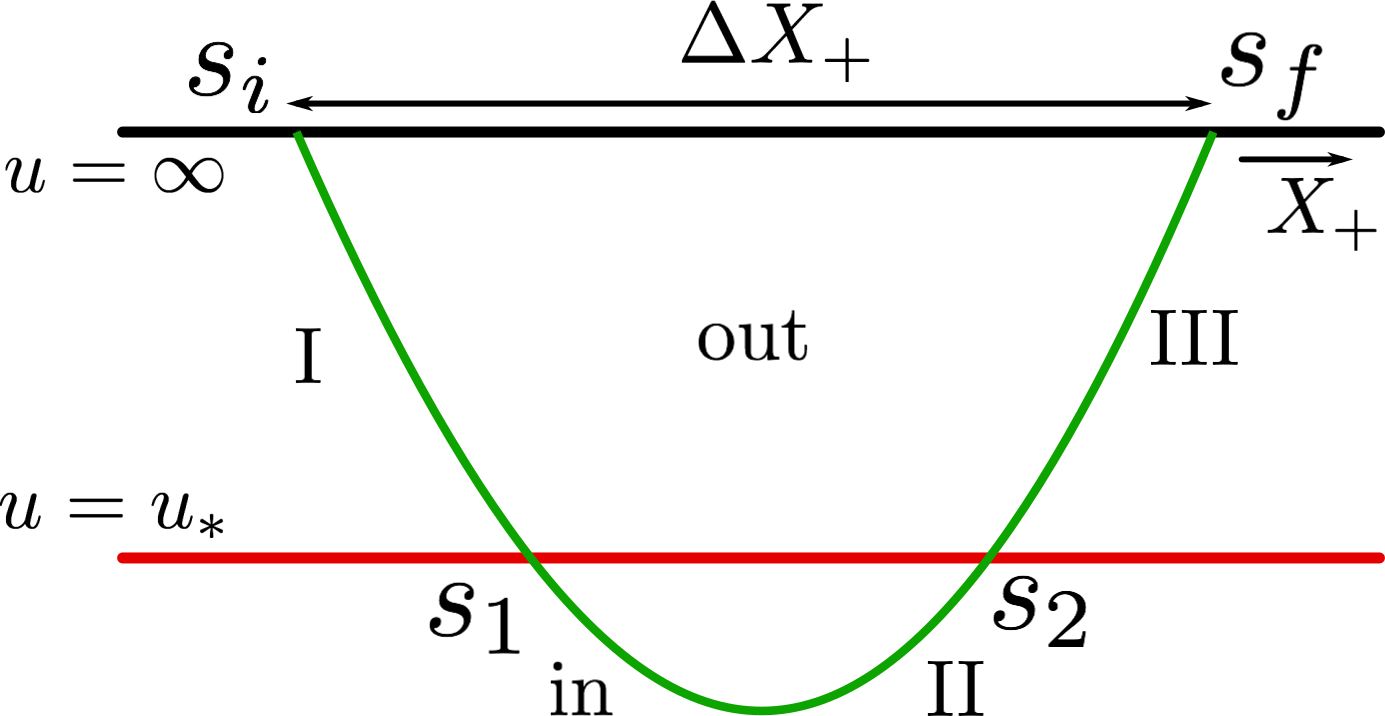}
    \caption{The geodesic (green line) starts from the boundary at time $s_i$, crosses the shell (red line) at times $s_1$ and $s_2$ and ends at the boundary at time $s_f$. The latin numbers denote different segments of the geodesic.}
    \label{Geodesic_Shell}
\end{figure}
Since the presence of the shell preserves $x_+$translation invariance, the associated conserved charge  $Q_+$ should be constant also in the presence of the shell, so that
\begin{equation}
    Q_+^I = Q_+^{II} = Q_+^{III}\equiv Q_+\, .
\end{equation}
As in \cite{Balasubramanian:2011ur} this could be rewritten as a form of Snell's law determining the refraction of geodesics when crossing the shell.\\
When $2M+1<Q_+^2<(Q_+^*)^2$ all the segments of the geodesic are given by the $\cosh$ branch and we have
\begin{equation}\label{ShellGeodBaseCosh}
    \begin{split}
        & U^{I,III} = \frac{M}{2} +  \frac{1}{2}\sqrt{Q_+^2-2M-1}\cosh\left(2\left(s-s_0^{I,III}\right)\right) \, ,\\
        & X_+^{I,III} = \frac{\kappa_c^{I,III}}{\sqrt{2M+1}}\mathrm{arccoth}\left(\frac{Q_+}{\sqrt{2M+1}}\coth\left(2\left(s-s_0^{I,III}\right)\right)  \right) + X_{+0}^{I,III} \, ,\\
        & U^{II} = \frac{1}{2\sigma}\left(-1 + \sqrt{Q_+^2+1}\cosh\left(2\left(s-s_0^{II}\right)\right)   \right) \, ,\\
        & X^{II} = \kappa^{II}\mathrm{arccot}\left(Q_+\coth\left(2\left(s-s_0^{II}\right)\right)   \right)\, .
    \end{split}
\end{equation}
For $Q_+^2<2M+1$ we have to use the $\sinh$ branch but only in the segments of the geodesic outside the shell. Within the shell we always have $Q_+^2>-1$ and thus it is the $\cosh$ branch that should be used. Furthermore, since we want the geodesic to start and end at the same boundary at $U\rightarrow +\infty$, for sections $I$ and $III$ we should choose $\sinh$ branches with different signs. Overall we have
\begin{equation}\label{ShellGeodBaseSinhU}
    \begin{split}
        & U^{I} = \frac{M}{2} -  \frac{1}{2}\sqrt{2M+1-Q_+^2}\sinh\left(2\left(s-s_0^{I}\right)\right) \, ,\\
        & U^{III} = \frac{M}{2} +  \frac{1}{2}\sqrt{2M+1-Q_+^2}\sinh\left(2\left(s-s_0^{III}\right)\right) \, ,\\
        & U^{II} = \frac{1}{2\sigma}\left(-1 + \sqrt{Q_+^2+1}\cosh\left(2\left(s-s_0^{II}\right)\right)  \right)  \, .
    \end{split}
\end{equation}
and
\begin{equation}\label{ShellGeodBaseSinhXp}
    \begin{split}
        & X_+^{I,III} = \frac{\kappa_s^{I,III}}{\sqrt{2M+1}}\mathrm{arctanh}\left(\frac{Q_+}{\sqrt{2M+1}}\tanh\left(2\left(s-s_0^{I,III}\right)\right)  \right) + X_{+0}^{I,III}\, , \\
        & X^{II} = \kappa^{II} \mathrm{arccot}\left(Q_+\coth\left(2\left(s-s_0^{II}\right)\right)   \right)\, .
    \end{split}
\end{equation}
For both cases without loss of generality we can choose the origin of $s$ such that $s_0^{II}=0$. Regarding the sign coefficients, since outside the shell we should match the geodesics of a left-thermal geometry we choose $\kappa_{c,s}^{I,III}=1$.   Inside the shell we keep the parameter $\kappa^{II}\equiv \kappa$, and we consider both signs similarly to pure AdS$_3$. 
Then, for each choice of $\kappa$, we overall we have ten parameters $\left(Q_+, \,s_0^{I,III}, \,X_{+0}^{I,II,III}, \, s_{i,f,1,2}    \right)$, which are fixed by the following ten boundary conditions and continuity equations. \\
The boundary conditions are 
\begin{equation}\label{BoundaryBase}
    U^I\left(s_i\right) = \frac{1}{\epsilon}\, , \quad U^{III}\left(s_f\right) = \frac{1}{\epsilon}\, , \quad X_+^I\left(s_i\right) = 0 \, , \quad X_+^{III}\left(s_f\right) = \Delta X_+\, ,
\end{equation}
and the continuity equations are
\begin{equation}\label{ContinuityBase}
\begin{split}
    & U^I\left(s_1\right) = U^{II}\left(s_1\right) = u_*\, , \quad U^{III}\left(s_2\right) = U^{II}\left(s_2\right) = u_* \, , \\
    & X_+^I\left(s_1\right) = X_+^{II}\left(s_1\right) \, , \quad X_+^{III}\left(s_2\right) = X_+^{II}\left(s_2\right)\, .
\end{split}
\end{equation}
Upon imposing these conditions we eventually find the geodesic length, renormalized as before in Section \ref{sec:GedLength}, for $2M+1<Q_+^2<(Q_+^*)^2$ 
\begin{equation}\label{DsShellBaseCosh}
    \Delta s_{ren}^{(1)} = \log\left( \frac{4}{\sqrt{Q_+^2-2M-1}}\right) - \mathrm{arccosh}\left(\frac{2u_*-M}{\sqrt{Q_+^2-2M-1}}  \right) + \mathrm{arccosh}\left( \frac{2\sigma u_*+1}{\sqrt{Q_+^2+1}} \right)\, ,
\end{equation}
and for $Q_+^2<2M+1$
\begin{equation}\label{DsShellBaseSinh}
    \Delta s_{ren}^{(2)} = \log\left( \frac{4}{\sqrt{2M+1-Q_+^2}}\right)  - \mathrm{arcsinh}\left(\frac{2u_*-M}{\sqrt{2M+1-Q_+^2}}  \right)+ \mathrm{arccosh}\left( \frac{2\sigma u_*+1}{\sqrt{Q_+^2+1}} \right)\, .
\end{equation}
In the above expressions the first term is the result (\ref{2dHyperbolicCorr}) for the left-thermal geometry, the second term subtracts out the length of the segment where $u<u_*$ in the left-thermal geometry, and the last term adds
the length of the segment with $u<u_*$ in the AdS$_3$ geometry.
One can check that in the infinite shell limit $u_*\rightarrow\infty$ we have 
\begin{equation}
    \Delta s_{ren}^{(1,2)} = \log\frac{4}{\sqrt{Q_+^2+1}} = \Delta s^{ren}_{AdS}\, .
\end{equation}
For the boundary separation of the endpoints $\D X_+$ we obtain
\begin{equation}\label{DXpShellBaseCosh}
    \begin{split}
        & \Delta X_+^{(1)} = -\frac{2}{\sqrt{2M+1}}\mathrm{arctanh}\left( \frac{\sqrt{2M+1}}{Q_+}\tanh{\left(\mathrm{arccosh}\left( \frac{2u_*-M}{\sqrt{Q_+^2-2M-1}} \right)\right)} \right) \\
        & +  \frac{2}{\sqrt{2M+1}}\mathrm{arccoth}\left( \frac{Q_+}{\sqrt{2M+1}} \right) \\
        & + 2\kappa\mathrm{arctan}\left( \frac{1}{Q_+} \tanh\left(\mathrm{arccosh}\left(\frac{2\sigma u_* +1}{\sqrt{Q_+^2+1}}\right)\right)\right) +2\pi n \, ,
    \end{split}
\end{equation}
where $n$ is a integer and
\begin{equation}\label{DXpShellBaseSinh}
    \begin{split}
        & \Delta X_+^{(2)} = -\frac{2}{\sqrt{2M+1}}\mathrm{arctanh}\left( \frac{\sqrt{Q_+}}{2M+1}\tanh{\left(\mathrm{arcsinh}\left( \frac{2u_*-M}{\sqrt{2M+1-Q_+^2}} \right)\right)} \right) \\
        & +  \frac{2}{\sqrt{2M+1}}\mathrm{arctanh}\left( \frac{Q_+}{\sqrt{2M+1}} \right) \\
        & + 2\kappa\mathrm{arctan}\left( \frac{1}{Q_+} \tanh\left(\mathrm{arccosh}\left(\frac{2\sigma u_* +1}{\sqrt{Q_+^2+1}}\right)\right)\right) +2\pi n \, .
    \end{split}
\end{equation}
Unlike in the geometry without shell, it is not possible to invert the dependence of $\D X_+ $  on $Q_+$  to  find an analytic expression\footnote{It is possible to  invert the dependence of $\D s $  on $Q_+$, by solving a cubic equation, and obtain $\D X_+ (\D s)$, though the result is not very enlightening.} for $\D s (\D X_+)$. 
In what follows we will rather parametrically plot $\Delta s$ in terms of $\Delta X_+$ using $Q_+$ as a parameter.
Before embarking into that, let us make here a few more analytical remarks.

Before crossing the shell the behaviour is that of a left thermal geometry and we have $Q_+>Q_+^*$. At $Q_+=Q_+^*$ the geodesic touches the shell. At this point we have
\begin{equation} \label{touchingDXp}
    \Delta X_+ = \Delta X_+^* = \Delta X_+^{(1)}\, .
\end{equation}
However, \eqref{DXpShellBaseCosh} is defined only for $2M+1<Q_+^2<(Q_+^*)^2$. Thus after crossing the shell, as it can be seen from \eqref{DXpShellBaseCosh} and \eqref{DXpShellBaseSinh} we are confined to $|Q_+|\le Q_+^*.$ Within this range we see that \eqref{DsShellBaseCosh} and \eqref{DsShellBaseSinh} are well behaved finite functions. A possible pathological point seems to be for $Q_+^2 = 2M+1$, but after carefully considering the limit we get
\begin{equation}\label{mDeltasmax}
    \Delta s = \frac{1}{2}\log{\frac{2}{1+M}}+\log{\left(1-\frac{\sqrt{2(1+M)+(M-2u_*)^2}}{M-2u_*}  \right)}\, ,
\end{equation}
which is finite for $u_*>\frac{M}{2}$. This point is a maximum for $-\Delta s$.\\
For $Q_+ =0$ we have the minimum for $-\Delta s$ which is
\begin{equation}\label{mDeltasmin}
    \Delta s = \log\left( \frac{4}{\sqrt{1+2M}} \right) + \mathrm{arccosh}\left( \sqrt{2(1+M)+(M-2u_*)^2} \right)  + \mathrm{arcsinh}\left( \frac{M-2u_*}{\sqrt{1+2M}} \right) \, .
\end{equation}
The fact that the value of $-\Delta s$ is bounded from below is already a sign of restoration of information.
For the value of $\Delta X_+$ at this point we have for $\kappa=1$
\begin{equation}
    \lim_{Q_+\rightarrow 0^{\pm}}\Delta X_+ = \pm \pi \, ,
\end{equation}
and for $\kappa=-1$
\begin{equation}
    \lim_{Q_+\rightarrow 0^{\pm}}\Delta X_+ = \mp \pi \, ,
\end{equation}
Overall taking into account the change in sign we have four different $\Delta X_+$. Whether we use $\Delta X_+^{(1)}$ or $\Delta X_+^{(2)}$ depends on the value of the charge $Q_+$. Besides that we will parametrically plot $-\Delta s$ as a function of $\Delta X_+$ for all possible signs and for all $2\pi n$ shifts. The actual contribution to $-\Delta s$ will come from whichever term is more dominant for a specific value of $\Delta X_+$.

\subsection{Specific Examples}

By setting specific values for the mass and radius of the shell we plot in fig.\ref{figGeodLengthAll} the logarithm of the two-point function $-\Delta s$ as a function of $\Delta X_+$. 
The different colours represent different branches of geodesics, see  fig.\ref{figGeodLegend} for a colour legend. The leading saddle point approximation  to the two-point function is obtained by selecting the branch with 
the highest   value of $-\Delta s$ for a given $\D X_+$.
We will see that these terms provide a suggestive picture about the shape of the geodesic and how much it penetrates the shell as we increase $\Delta X_+$ (fig.\ref{GeodPlots1}). 

\paragraph{Thermal and non-thermal behaviour:}Let us first discuss how the   thermal behaviour of the two-point function is resolved in the shell geometry.
For small values of $\Delta X_+$ the two-point function behaves as in the left-thermal geometry and is denoted with the gray line. 
At some point this thermal branch is  crossed by one of the non-thermal branches. Beyond that point we observe oscillatory behaviour instead of the exponential decay in the left-thermal solution (dotted grey line). 

Furthermore, for fixed shell mass, as we increase the radius of the shell, and bigger part of the geometry is AdS$_3$, the crossing point and the maximum of $-\Delta s$ are shifted higher as can be seen by comparing the plots fig.\ref{figGeodLengthAll}, fig.\ref{figGeodLengthAllus5} and fig.\ref{figGeodLengthAllus100}. In other words as we increase the radius of the shell, the behaviour of the geodesic length becomes more and more similar to what we would get in the case of pure AdS$_3$ spacetime.

\paragraph{Cusps:}In the plots we observe several cusps or `swallow-tail' phenomena (areas C1, C2, C3 in fig.\ref{figGeodLengthAll}). While the geodesic length changes continuously as we change $Q_+$ by moving along the coloured segments, the dominant contribution is discontinuous as we vary $\Delta X_+$. However the total two-point function should be continuous. We believe this discrepancy is due to the saddle point approximation we adopted by studying geodesics. This effect has been observed elsewhere in literature both in the study of holographic thermalization\cite{Balasubramanian:2011ur} as well as in entanglement entropy \cite{Albash:2010mv}.

We have two kinds of cusps in our plots. The first kind (C1 in fig.\ref{figGeodLengthAll}) involves also the thermal branch which crosses over the non-thermal branches. A close up view is in fig.\ref{figGeodLengthAllTC}. Although the geodesic touches the shell where the green and dashed lines meet, the contribution from geodesics entering the shell dominates
already from  an earlier value $\Delta X_+$. This point is where red and gray lines intersect. As it can be seen from the plot the red branch is more dominant than the green one. Specifically at this cusp, due to the presence of the thermal branch, the dominant contribution is continuous. The second kind (C2) is depicted in fig.\ref{figGeodLengthAllOC}. As we increase $\Delta X_+$ we move along the black line but the dominant branch jumps to the red line before the black line ends and overpassing the green line. A similar cusp, but with opposite orientation occurs when passing from the blue to the brown line, while overpassing the orange one (C3). As we will see in the next paragraph, this second kind of cusp involves transitioning from usual geodesics to geodesics with crossed external legs.

\paragraph{Behaviour of the non-thermal branches:}Let us now describe how the geodesic behaves (fig.\ref{GeodPlots1}) as we move along the dominant part of the non-thermal branches of fig.\ref{figGeodLengthAll}. At the cusps we also describe the behaviour at the non-dominant green and orange branches to present more accurately how the geodesic changes continuously and which shapes of geodesics are responsible 
for the discontinuity of the leading contribution at the cusp.

We start from the tip of the green line, which is the point the geodesic touches the shell (fig.\ref{geod1}) at $\Delta X_+=\Delta X_+^*$, $Q_+=Q_+^*$ and $\kappa=+1$. The geodesic goes deeper into the shell as we move along the green line and $Q_+$ decreases (fig.\ref{geod2}). Then we reach the end of the green line, when $Q_+=\sqrt{2M+1}$, at the point where it meets the red line. The relevant geodesic is depicted by fig.\ref{geod3}. As we move along the red line the geodesic goes deeper into the shell (fig.\ref{geod4}), until we reach the minimum of $-\Delta s$ where $\Delta X_+=\pi$ and $Q_+=0$ (fig.\ref{geod5}). This brings us to the end of the red line.

Then the geodesic starts penetrating the shell less and less (blue line and fig.\ref{geod6}). We observe that after the point where $\Delta X_+=\pi$, the legs of the geodesic for $s<0$ and $s>0$ appear to have been reversed, but that's only because we are looking at the geodesic upside down. Then we reach the end of the blue line where it meets with the orange one at $Q_+=-\sqrt{2M+1}$ (fig.\ref{geod7}). Afterwards we reach the end of the orange line where it meets with the brown one at $Q_+=-Q_+^*$ and $\Delta X_+ = 2\pi-\Delta X_+^*$. At this point the geodesic touches again the shell (fig.\ref{geod8}) as we have made a full $2\pi$ angle on the shell. 

However, there is still a $\Delta X_+^*$ that remains in order to reach $2\pi$ at the $\Delta X_+$ on the boundary. To continue beyond this point one naively might expect that the geodesic exits the shell. That's not what happens as the thermal branch at this point has decayed a lot and is far from being dominant. Instead this extra distance is covered by the branch with $\kappa=-1$ (brown line and $-Q_+^*<Q_+<-\sqrt{2M+1}$) and the external legs of the geodesic cross as seen in fig.\ref{geod9}. As we move towards the value $\Delta X_+=2\pi$ we pass to the pink line ($-\sqrt{2M+1}<Q_+<0$) and the crossing point is moved closer to the boundary as seen in fig.\ref{geod10}. When we reach $\Delta X_+ = 2\pi$ the crossing point has moved all the way to the boundary (fig.\ref{geod11}). This is the point where purple and pink lines meet in fig.\ref{figGeodLengthAll}. The purple line represents another branch with $\kappa=-1$ but with $0<Q_+<\sqrt{2M+1}$. 

As we keep increasing $\Delta X_+$, the crossing point starts moving again towards the shell (purple and black lines and figures \ref{geod12} and \ref{geod13} respectively), until we reach the end of the black line, where it meets the green one and the geodesic touches the shell once more. At this point we have $\Delta X_+ = 2\pi + \Delta X_+^*$ and $Q_+=Q_+^*$. Then the legs of the geodesic stop crossing and the cycle starts again from the beginning.

\begin{figure}
    \centering
    \includegraphics[scale=0.45]{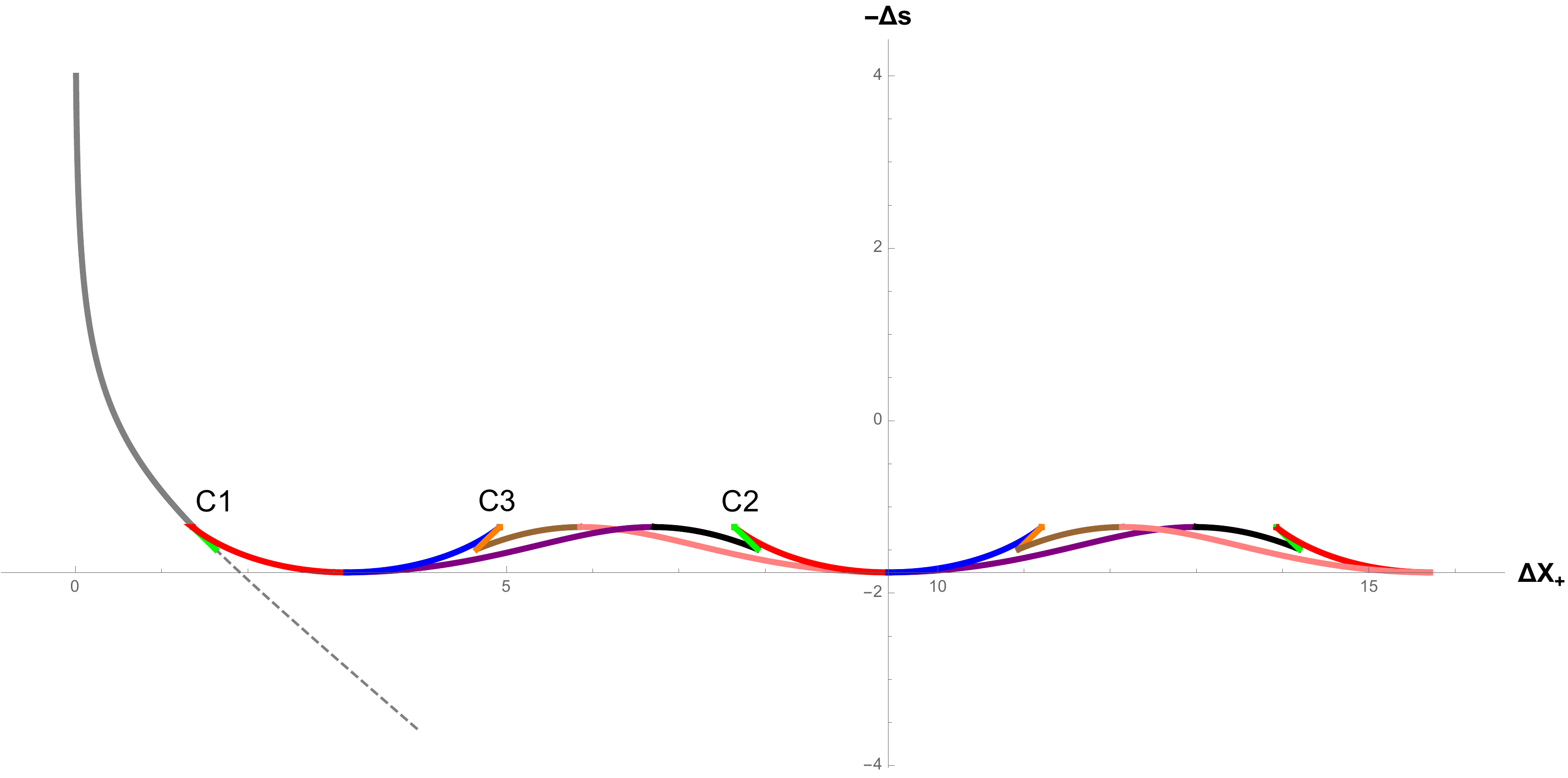}
    \caption{All branches of geodesic length as function of $\Delta X_+$ for $M=1$ and $u_*=0.95$. The initial gray line on the left is due to the thermal geometry before the geodesic crosses the shell. The dashed line represents how the thermal behaviour would continue without the presence of the shell. With $C1$, $C2$, $C3$ we denote the areas where the cusps occur. For the colour legend of the plot see fig. \ref{figGeodLegend}.}
    \label{figGeodLengthAll}
\end{figure}
\begin{figure}
    \centering
    \includegraphics[scale=0.8]{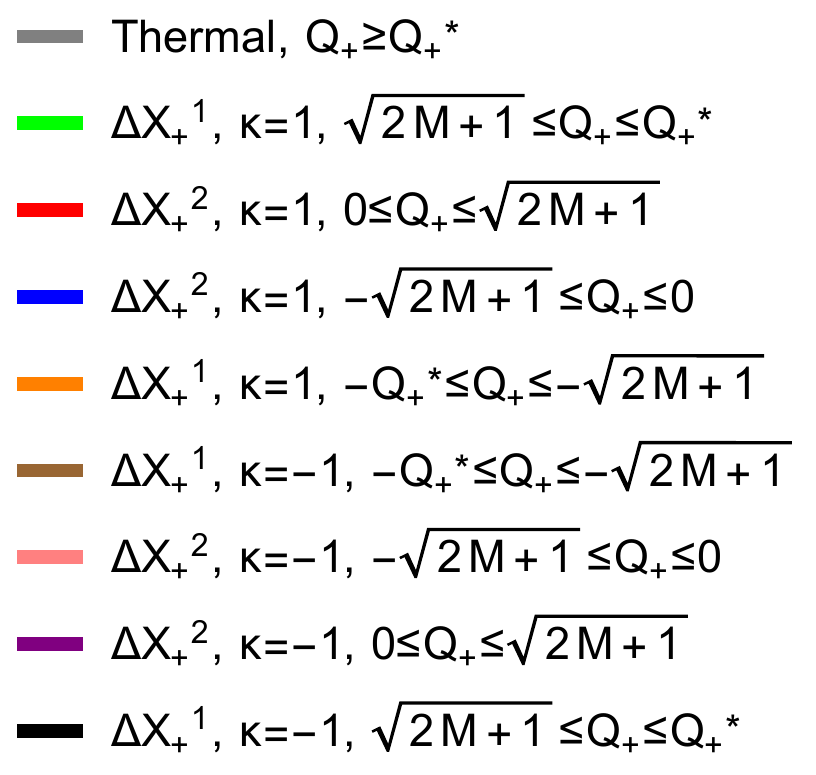}
    \caption{Colour legend for geodesic plots. We have split the plot in eight segments depending on whether we have $\Delta X_+^{(1,2)}, \, \kappa=\pm 1$ and $Q_+$ greater or smaller than zero.}
    \label{figGeodLegend}
\end{figure}
\begin{figure}
    \centering
    \includegraphics[scale=0.45]{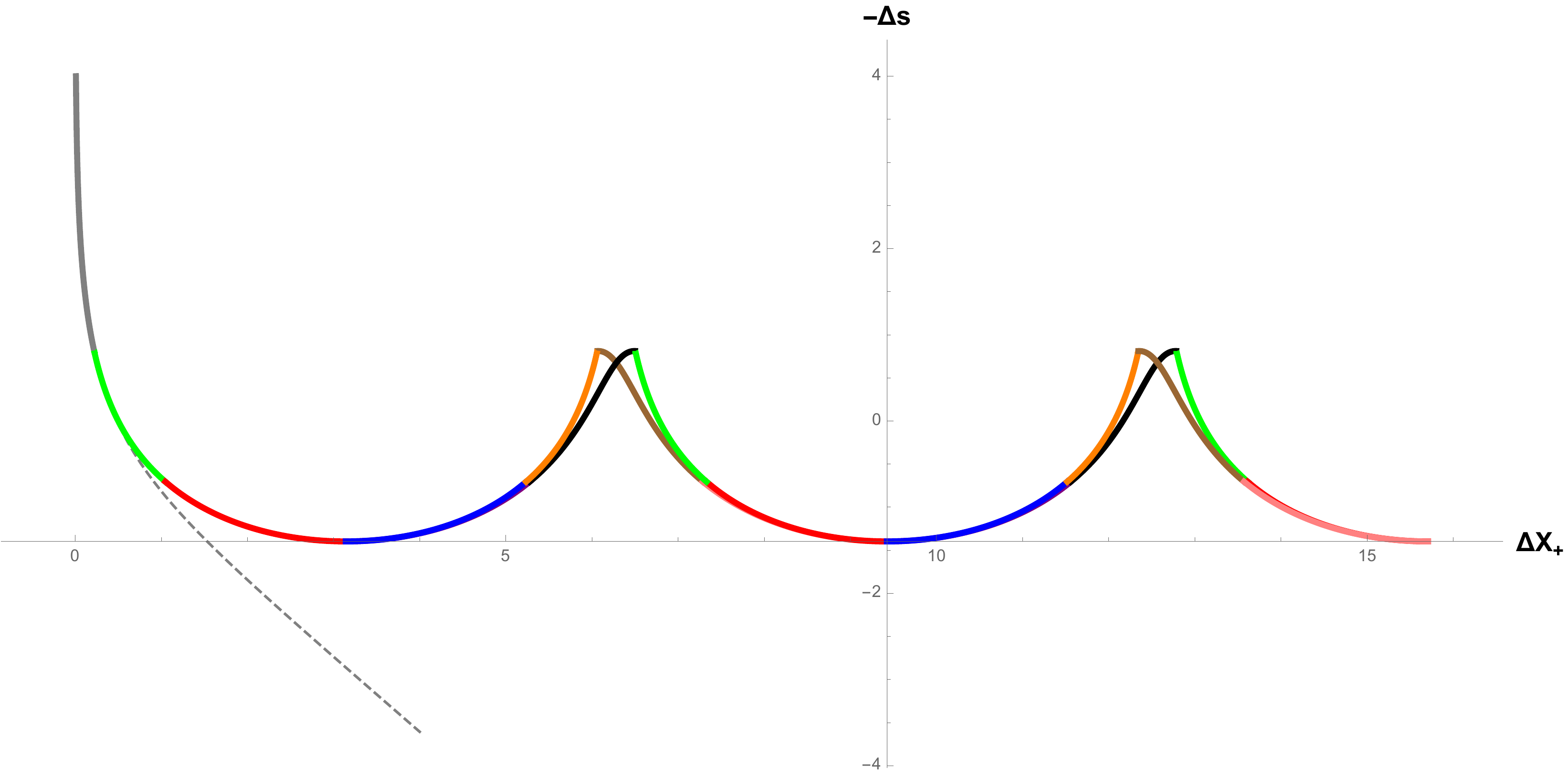}
    \caption{All branches of geodesic length as function of $\Delta X_+$ for $M=1$ and $u_*=5$. As we increase the radius the branches approach the pure $AdS_3$ case.}
    \label{figGeodLengthAllus5}
\end{figure}
\begin{figure}
    \centering
    \includegraphics[scale=0.45]{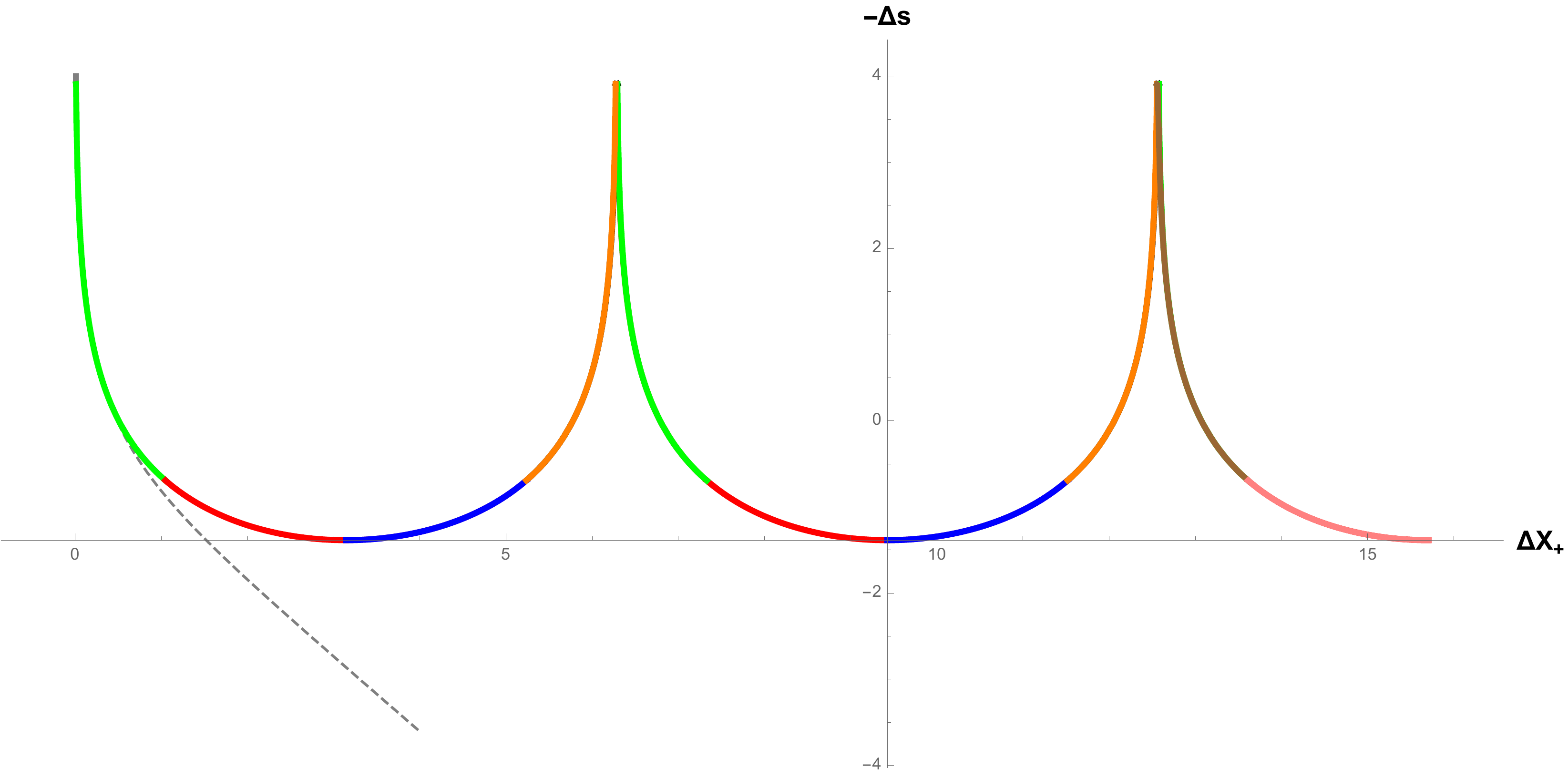}
    \caption{All branches of geodesic length as function of $\Delta X_+$ for $M=1$ and $u_*=100$. Here the geodesic length behaves almost exactly like in the pure $AdS_3$ case.}
    \label{figGeodLengthAllus100}
\end{figure}
\begin{figure}
    \centering
    \includegraphics[scale=0.48]{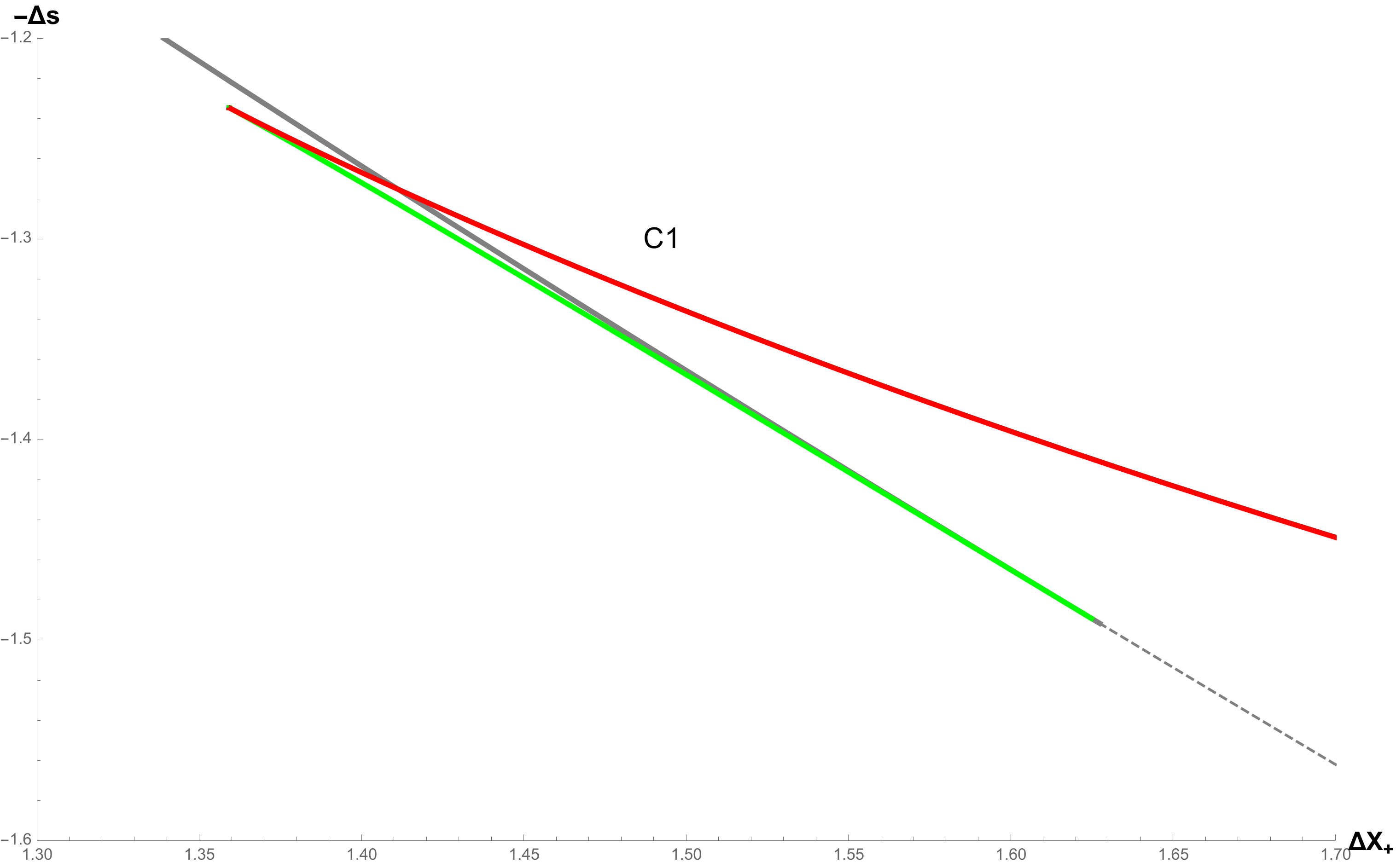}
    \caption{Crossover of geodesic lengths with the thermal branch as a function of $\Delta X_+$ for $M=1$ and $u_*=0.95$. The geodesic touches the shell where the gray and green lines meet. However it crosses the shell before that when the red line crosses the gray one.}
    \label{figGeodLengthAllTC}
\end{figure}
\begin{figure}
    \centering
    \includegraphics[scale=0.45]{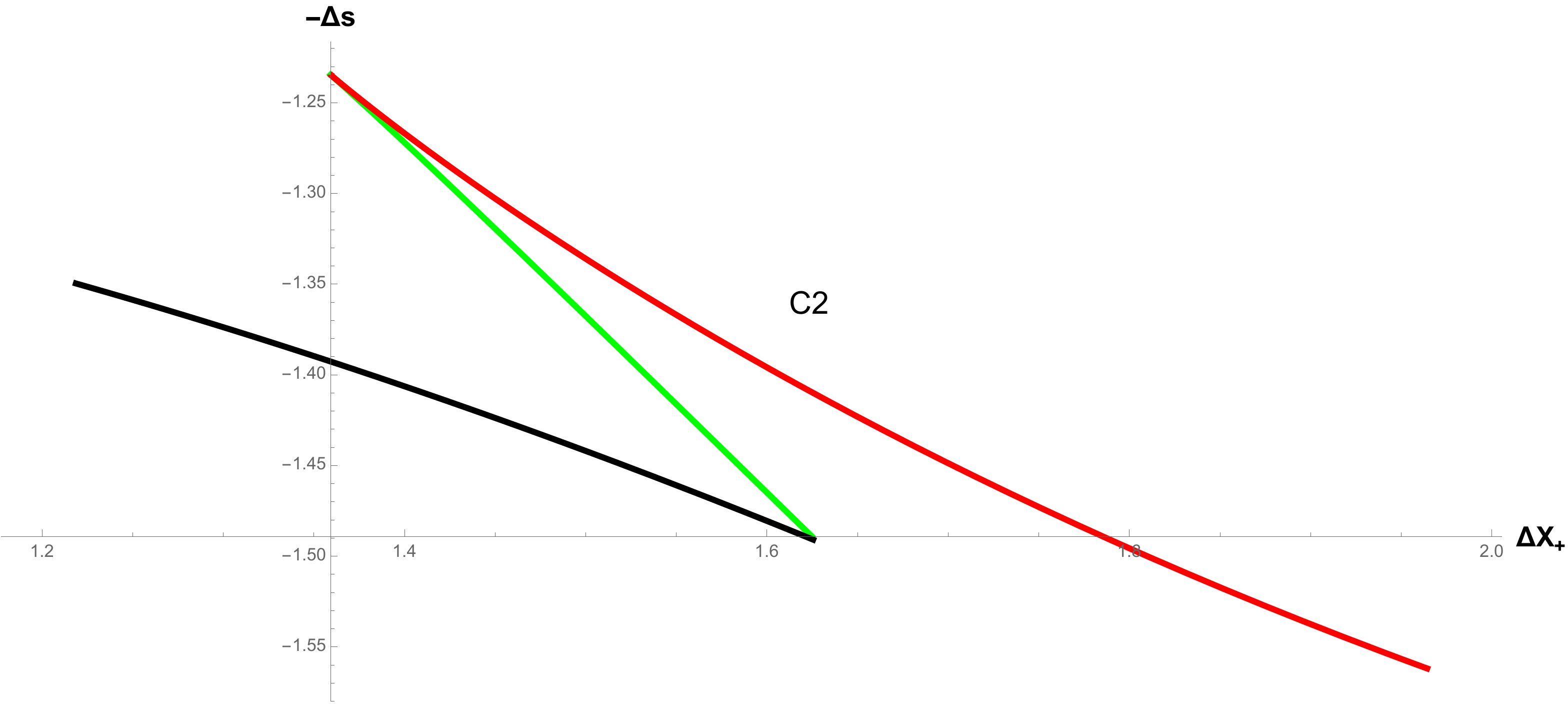}
    \caption{Cusp between minus and plus segments of geodesic lengths as a function of $\Delta X_+$ for $M=1$ and $u_*=0.95$.}
    \label{figGeodLengthAllOC}
\end{figure}
\begin{figure} 
\subfloat[]{\includegraphics[width = 2in]{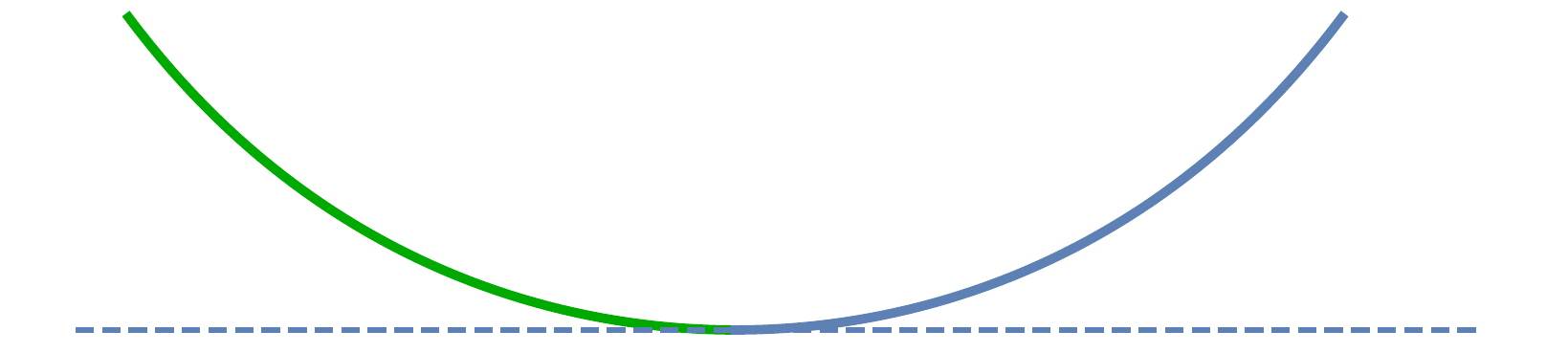}\label{geod1}}
\subfloat[]{\includegraphics[width = 2in]{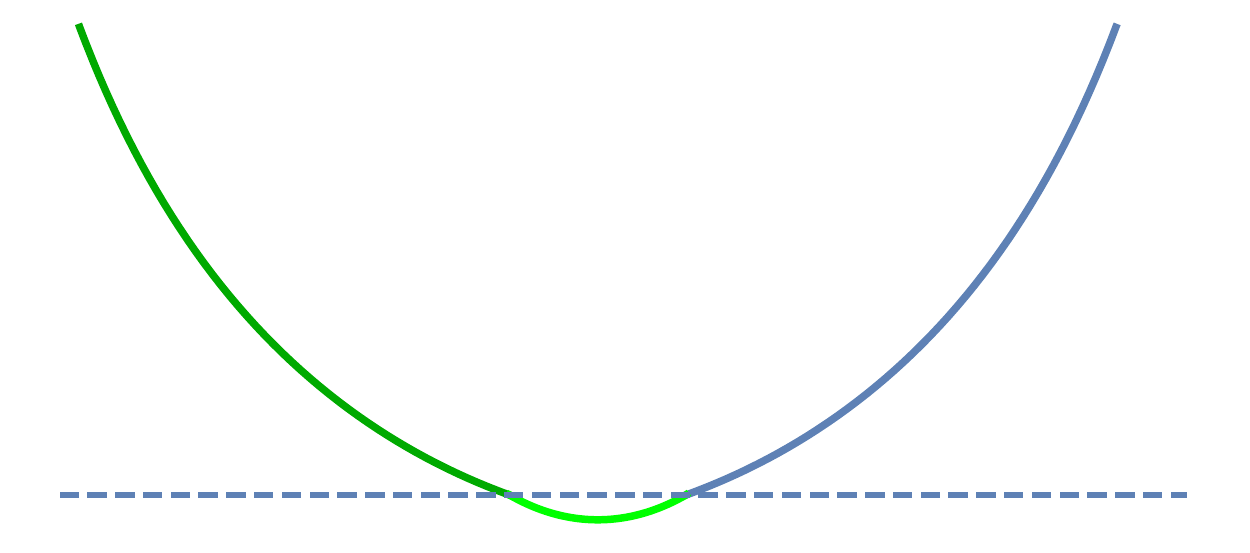}\label{geod2}} 
\subfloat[]{\includegraphics[width = 2in]{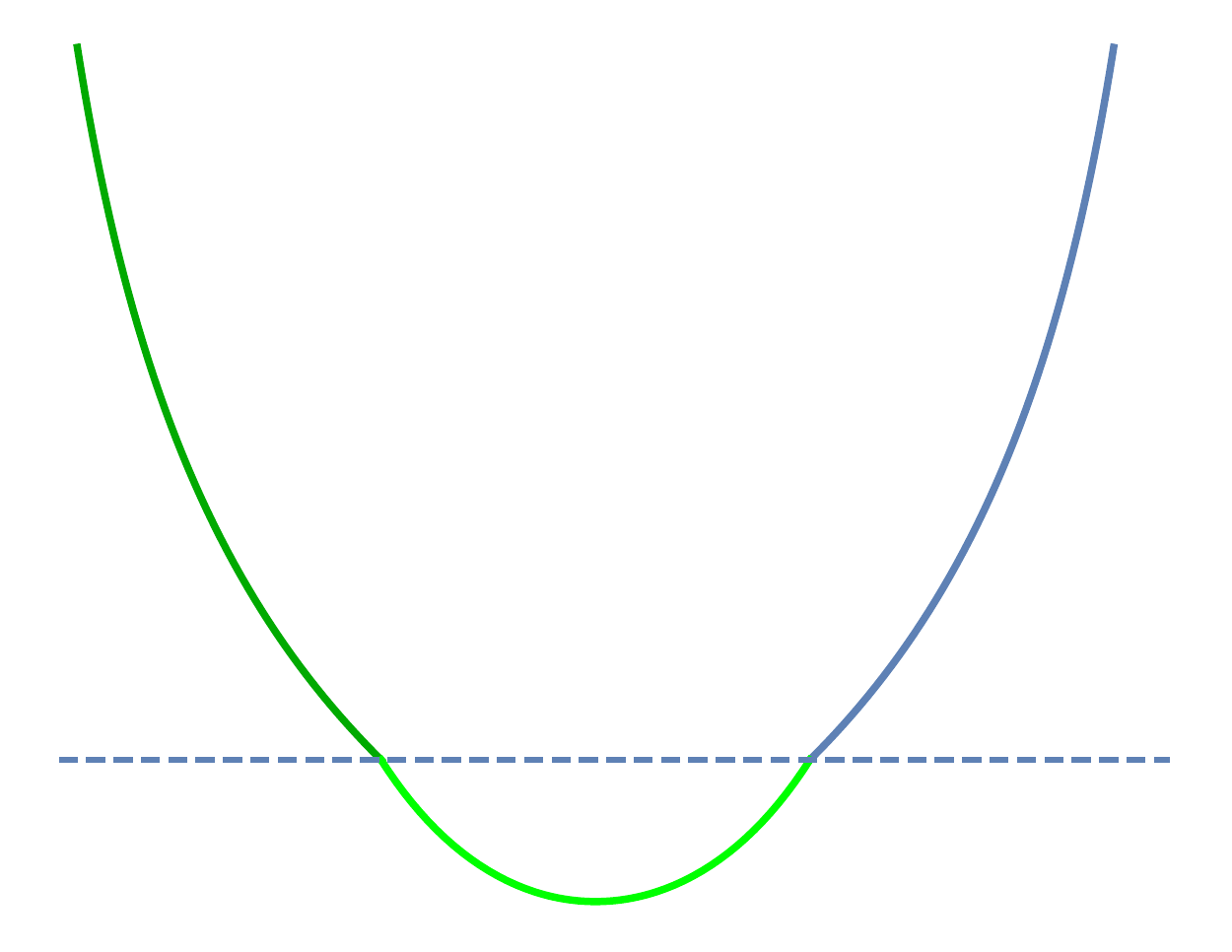}\label{geod3}}\\
\subfloat[]{\includegraphics[width = 2in]{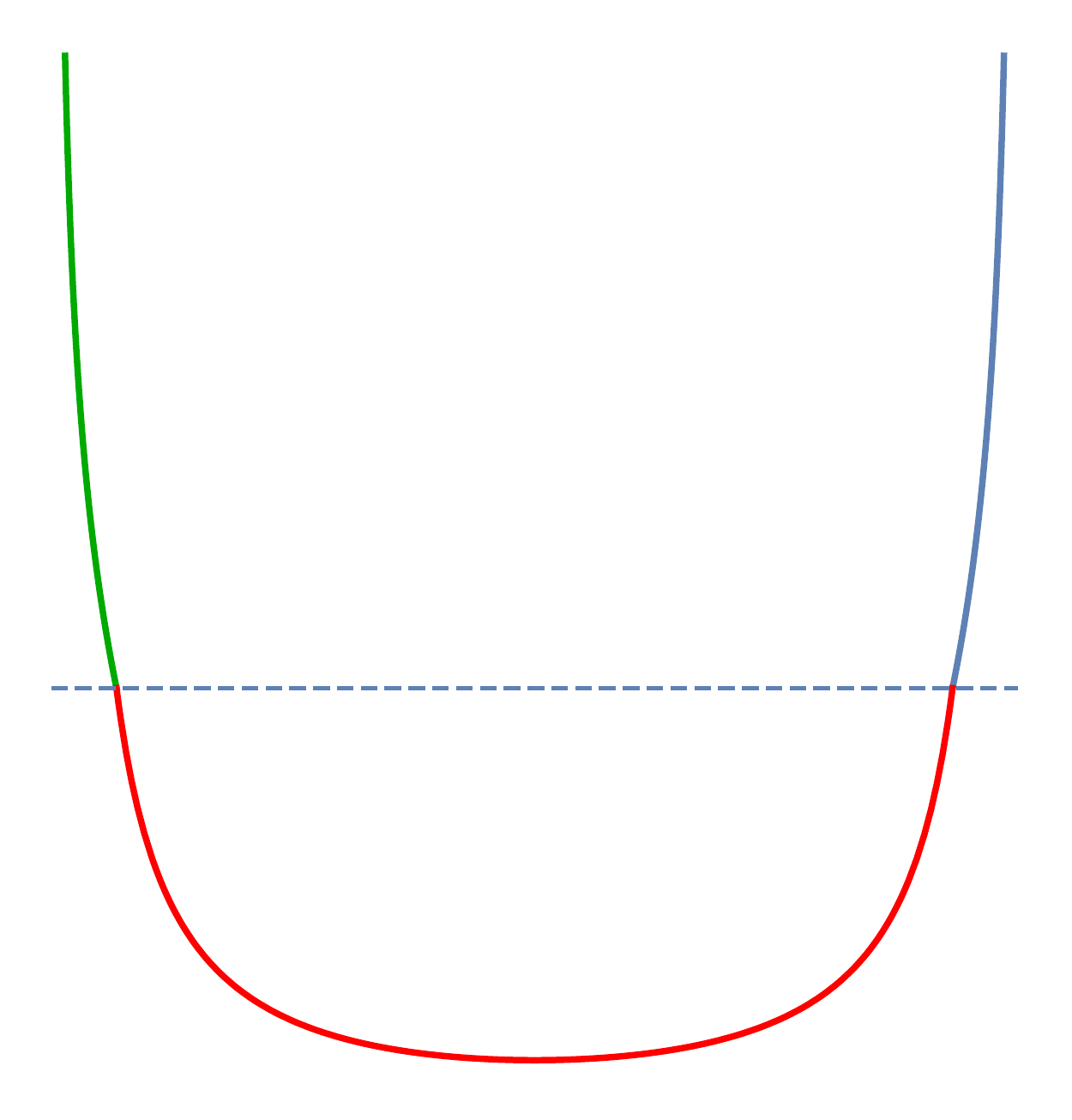}\label{geod4}} 
\subfloat[]{\includegraphics[width = 2in]{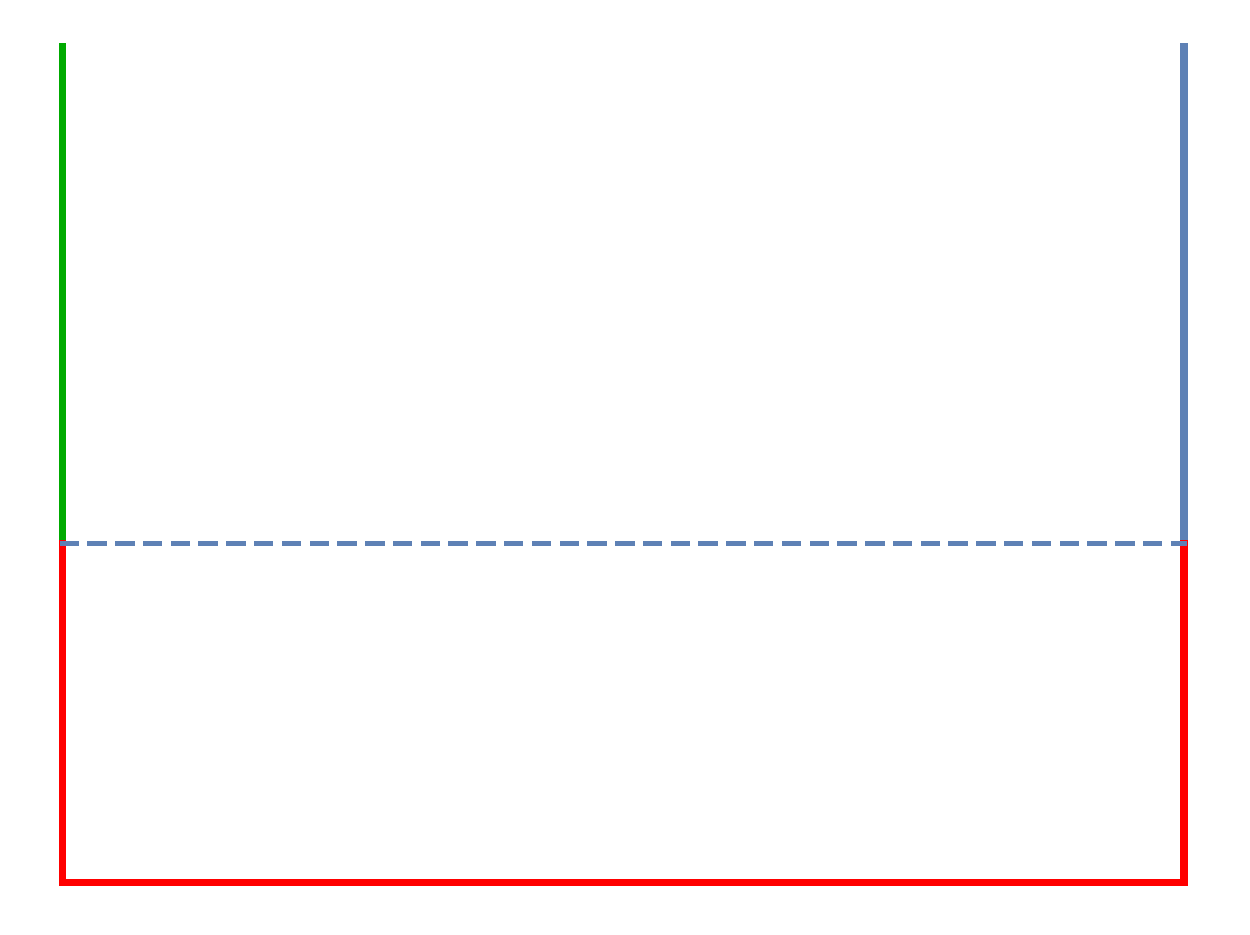}\label{geod5}}
\subfloat[]{\includegraphics[width = 2in]{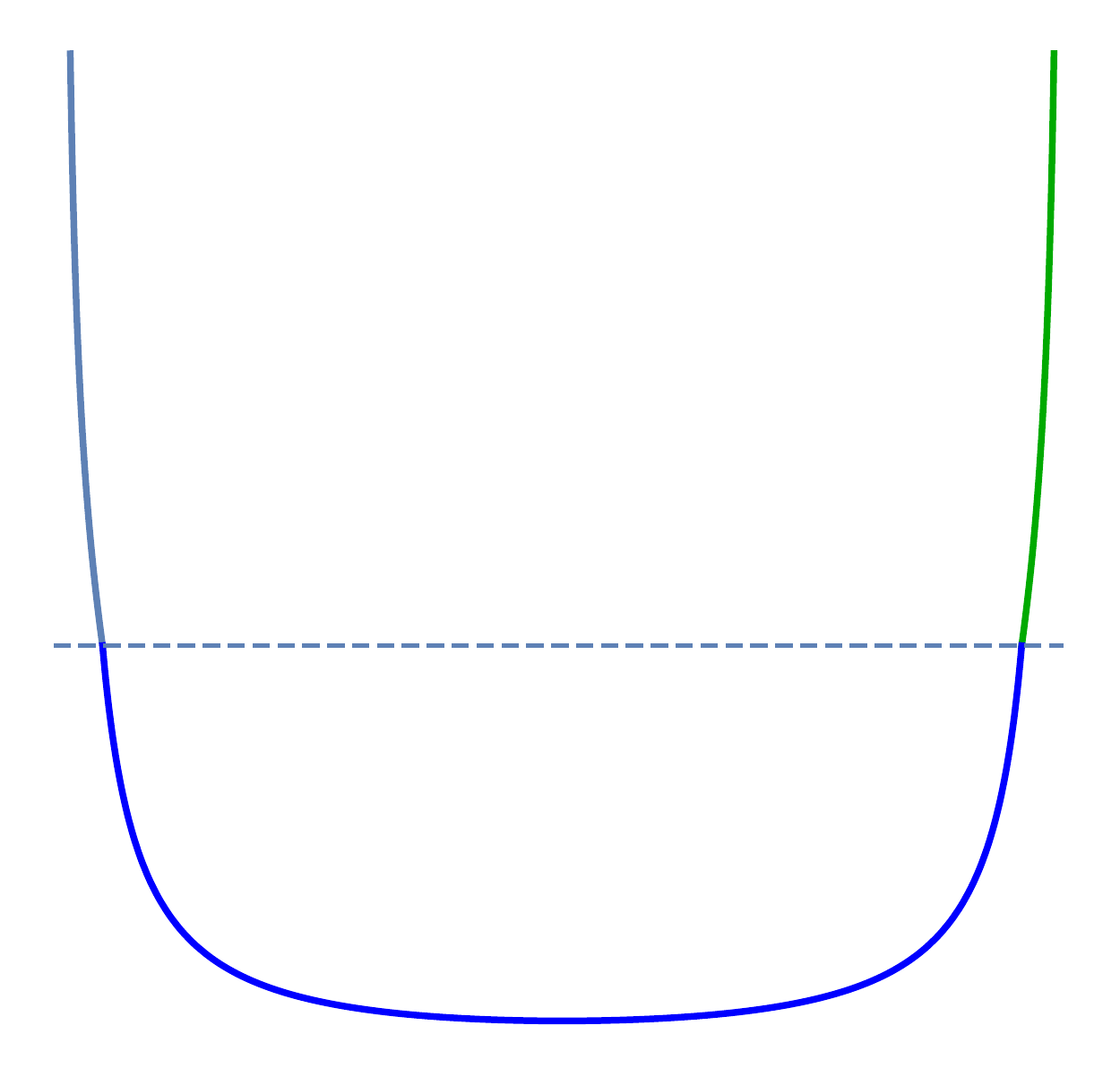}\label{geod6}} \\
\subfloat[]{\includegraphics[width = 2in]{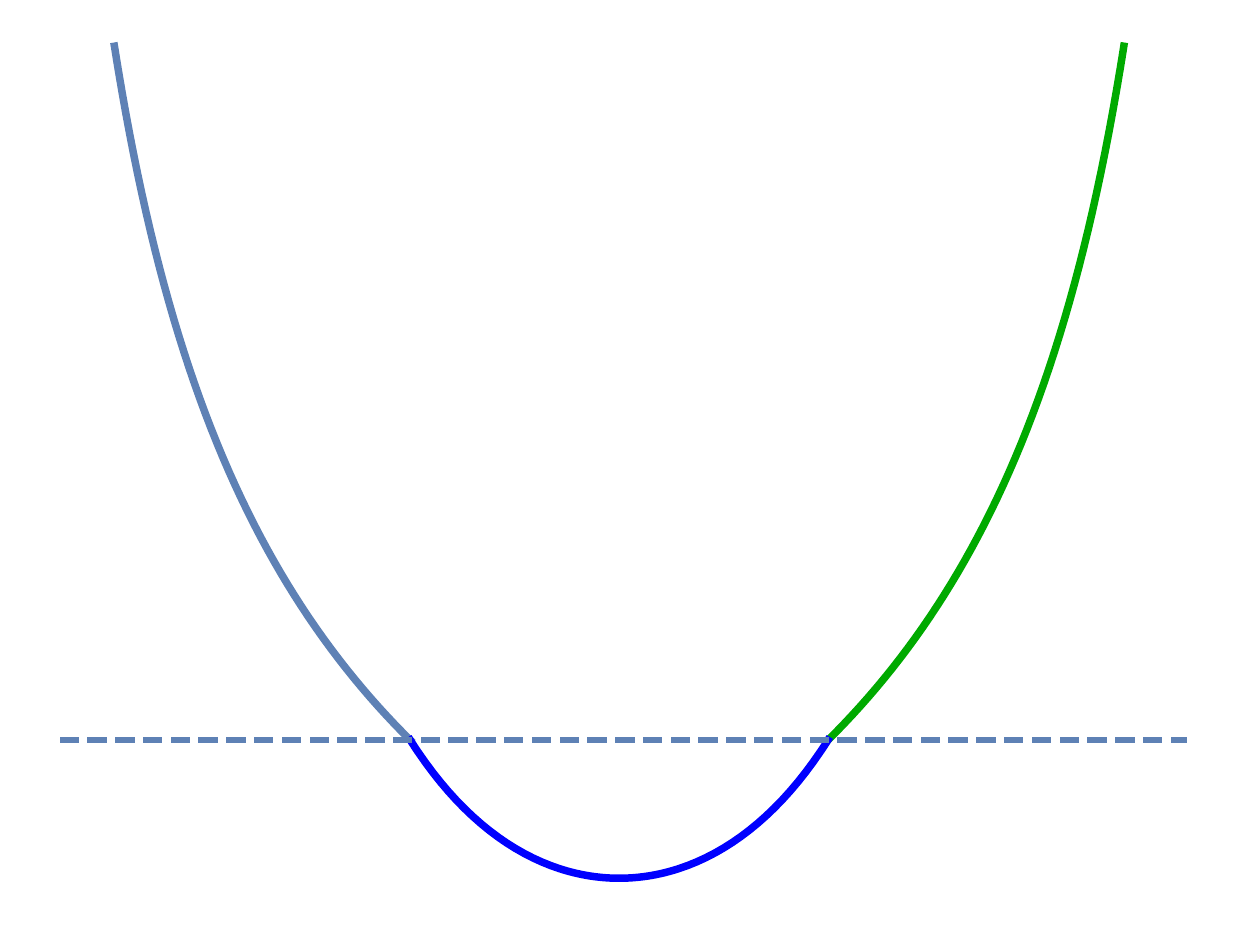}\label{geod7}} 
\subfloat[]{\includegraphics[width = 2in]{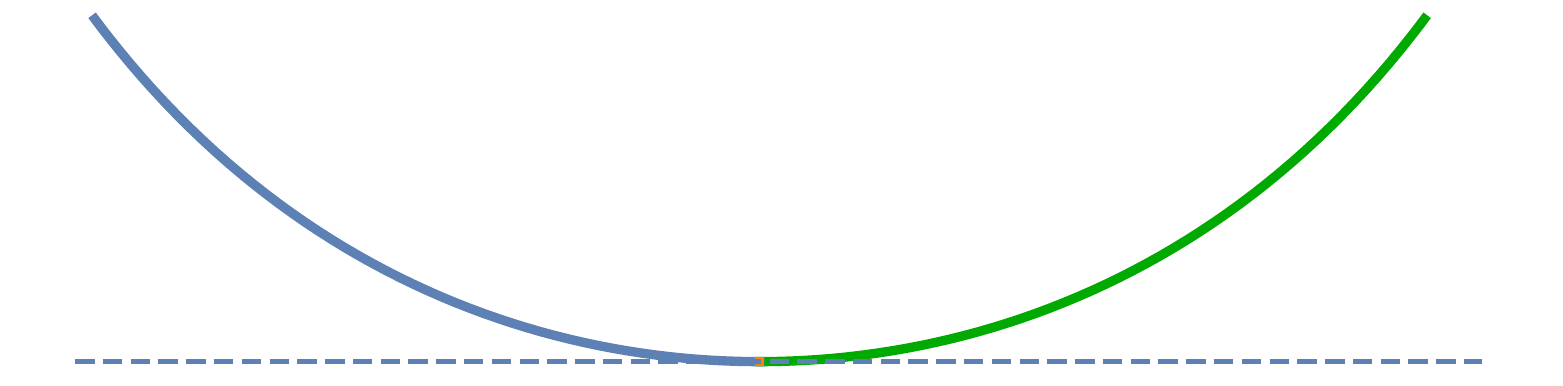}\label{geod8}} 
\subfloat[]{\includegraphics[width = 2in]{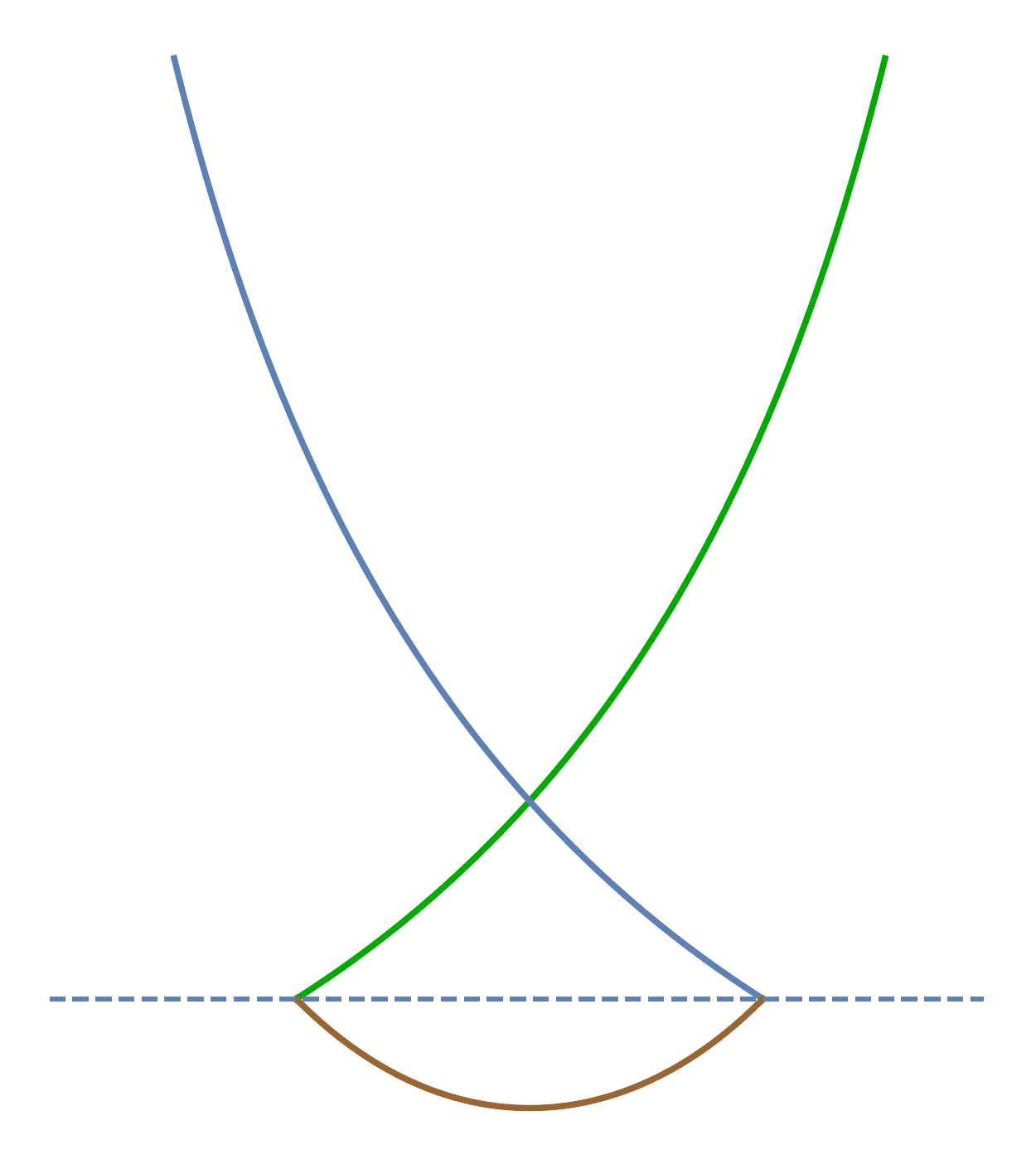}\label{geod9}}
\caption{First part of geodesic plots. The plots depict the behaviour of the dominant geodesic as we vary $\Delta X_+$. The horizontal axis is $X_+$ and the vertical axis is $U$. The dashed line is the position of the shell at $u_*$. We have set $u_*=0.95$, $M=1$. The external segment of the geodesic that comes from $s<0$ is with dark green colour (segment I in fig.\ref{Geodesic_Shell}). The segment II of the geodesic inside the shell has colour that corresponds to the colour of the branches in fig.\ref{figGeodLengthAll}.}
\label{GeodPlots1}
\end{figure}
\begin{figure}
\setcounter{figure}{11}
\setcounter{subfigure}{9}
\subfloat[]{\includegraphics[width = 2in]{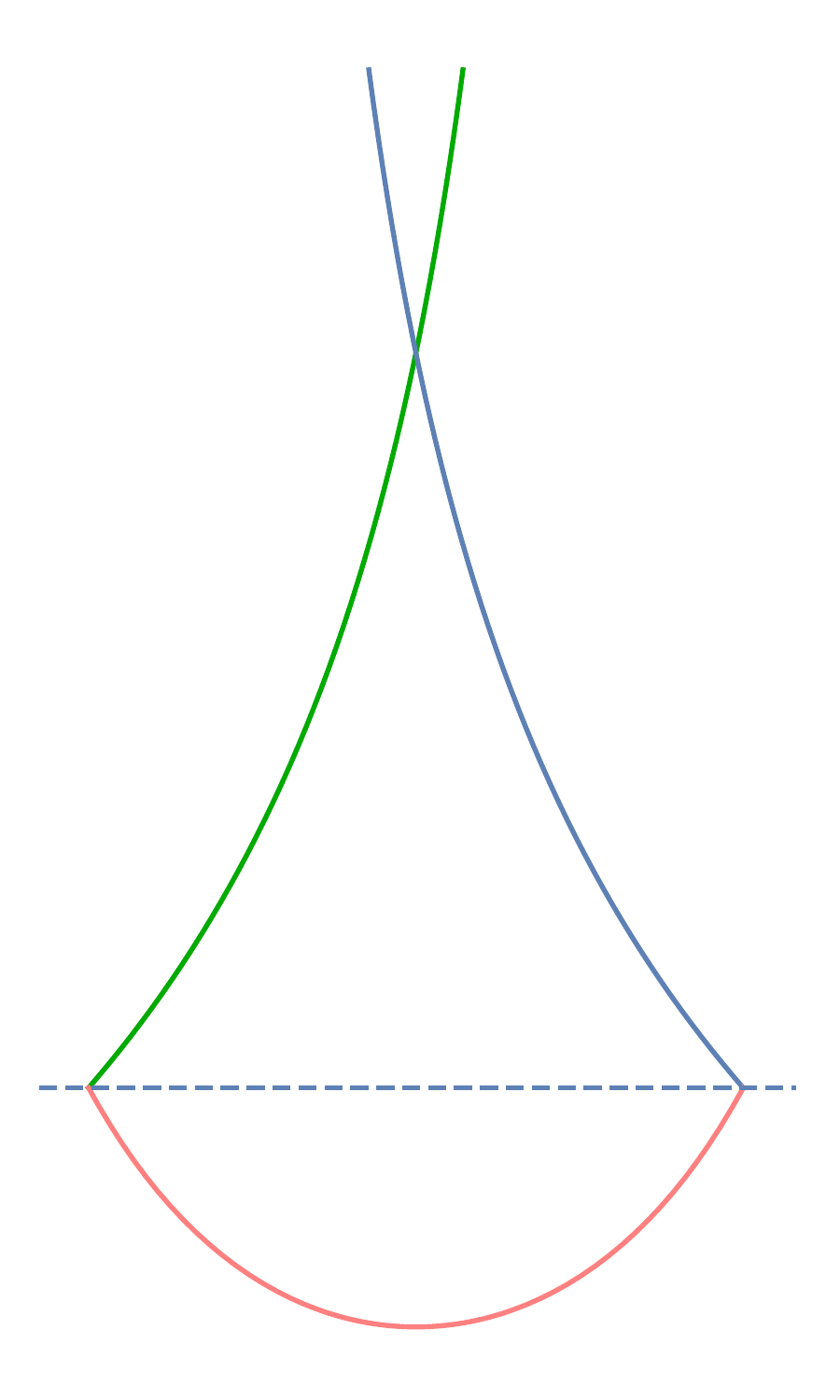}\label{geod10}} 
\subfloat[]{\includegraphics[width = 2in]{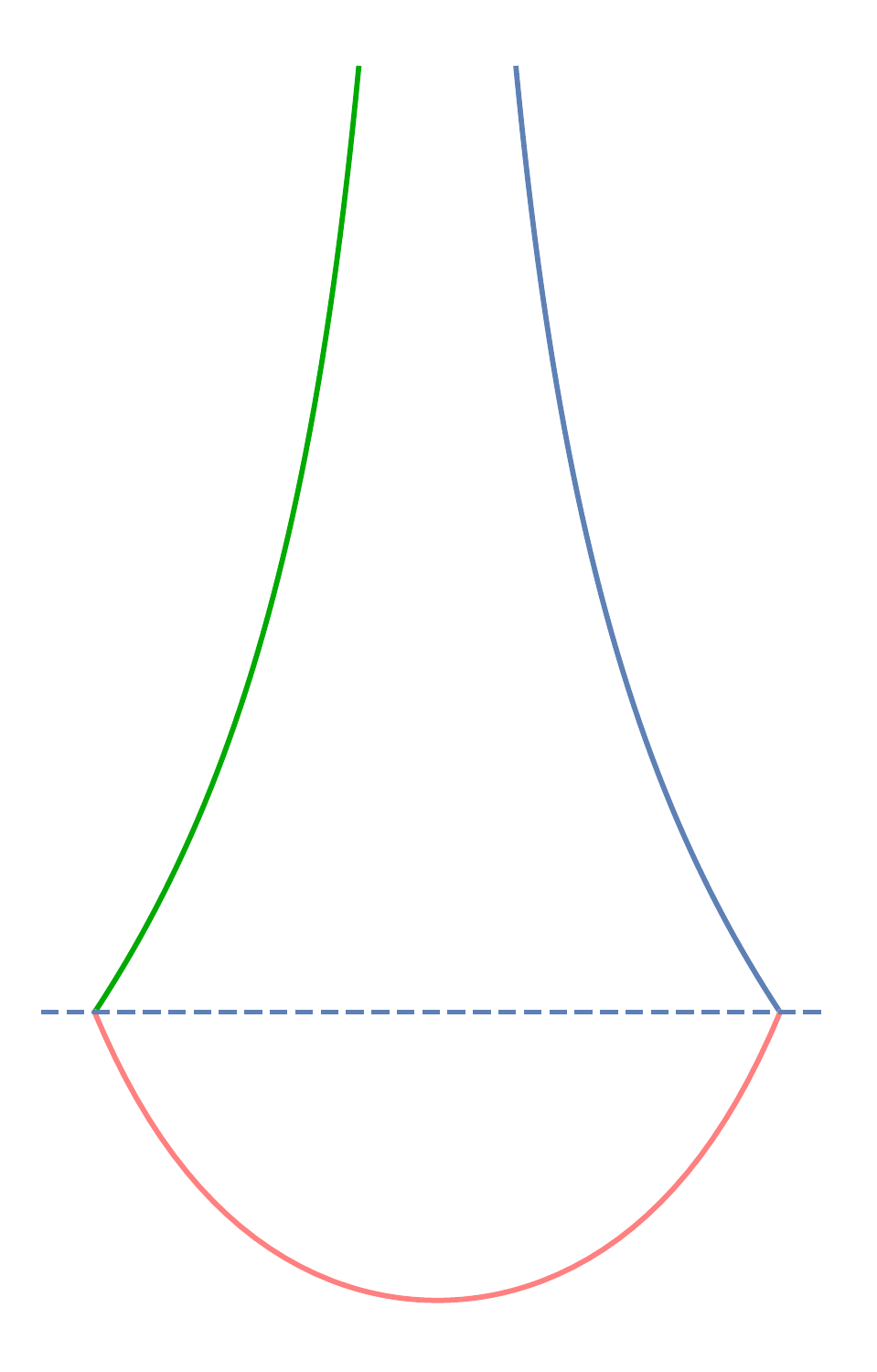}\label{geod11}} 
\subfloat[]{\includegraphics[width = 2in]{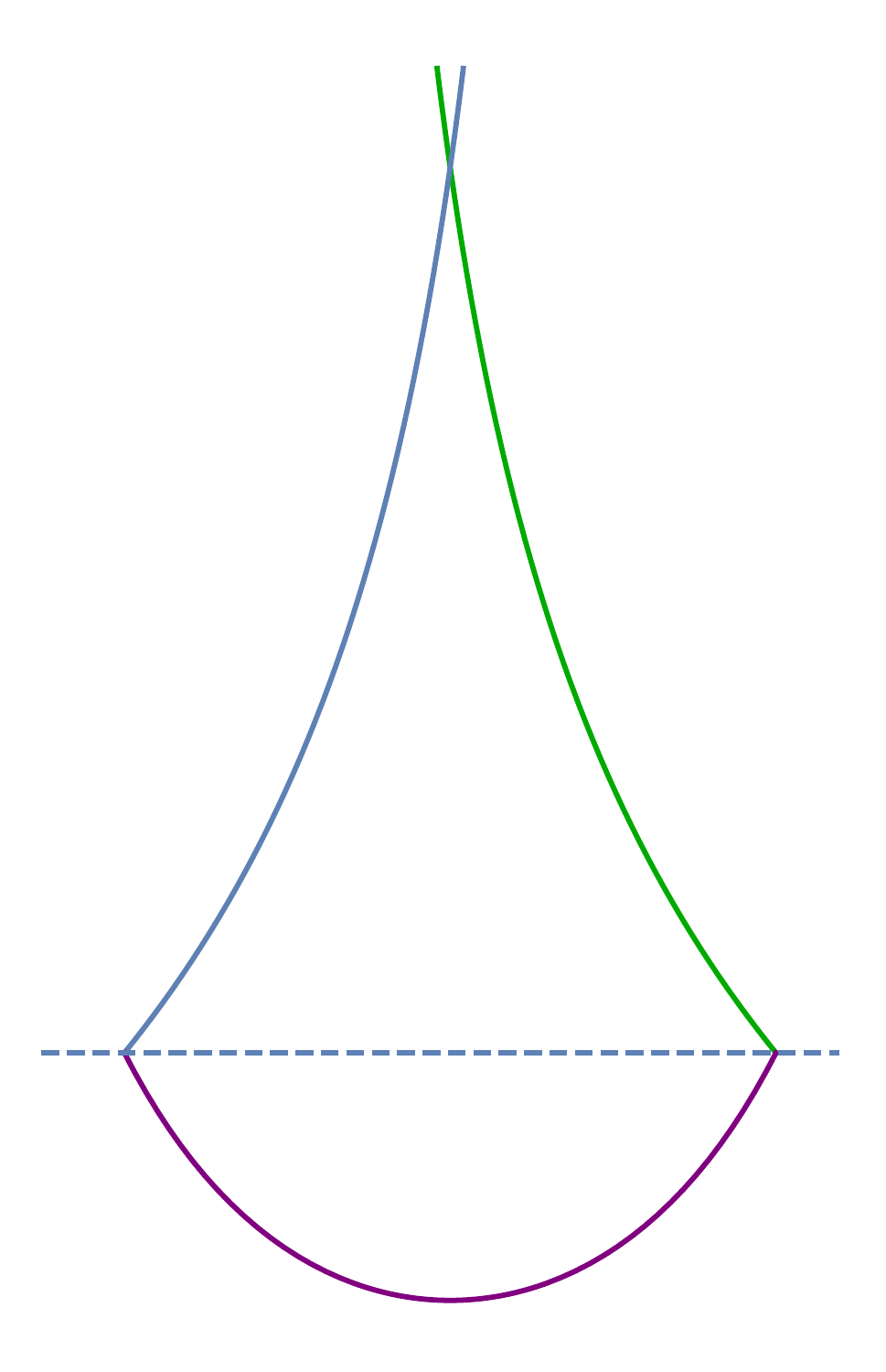}\label{geod12}} \\
\subfloat[]{\includegraphics[width = 2in]{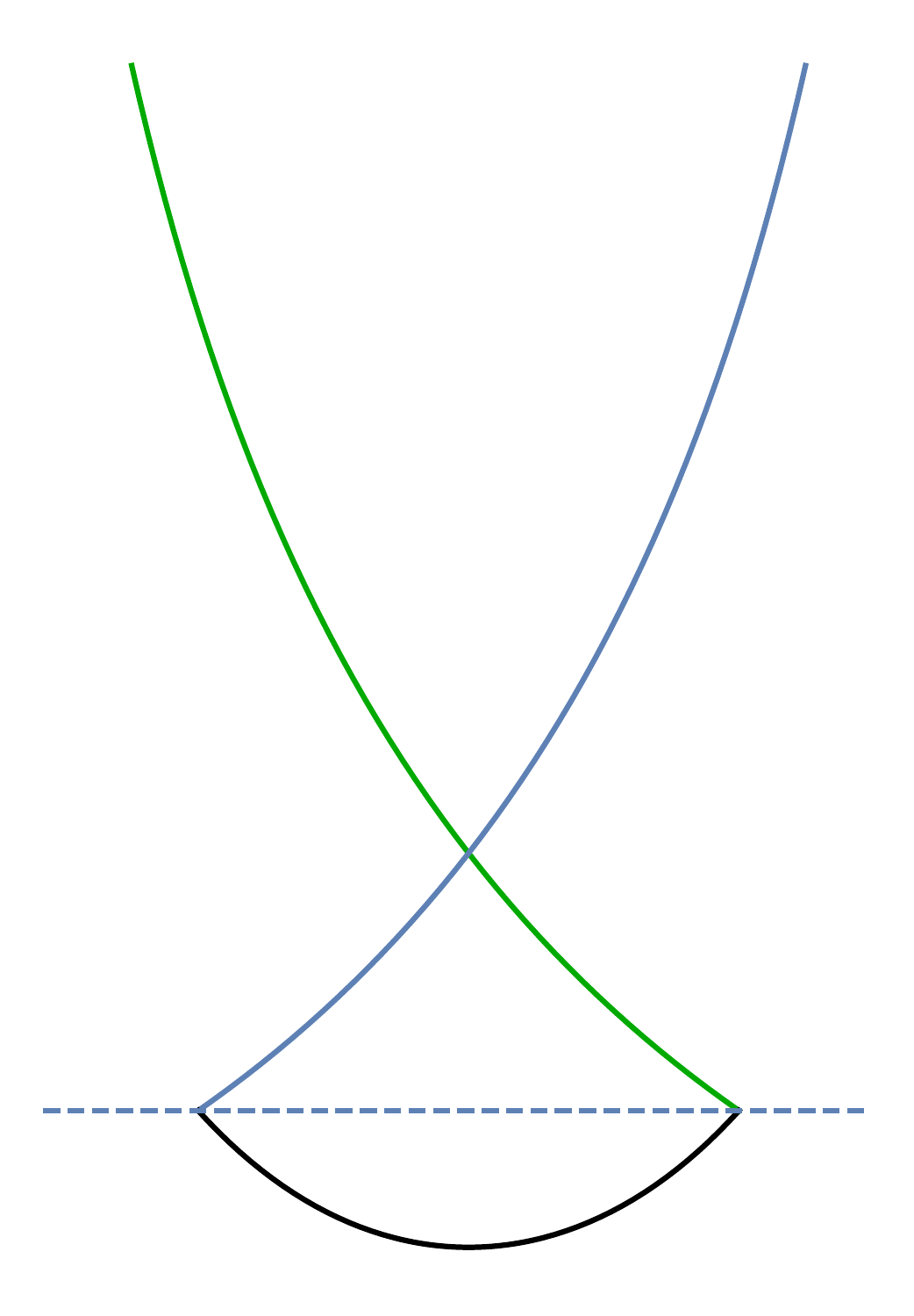}\label{geod13}} 
\subfloat[]{\includegraphics[width = 2in]{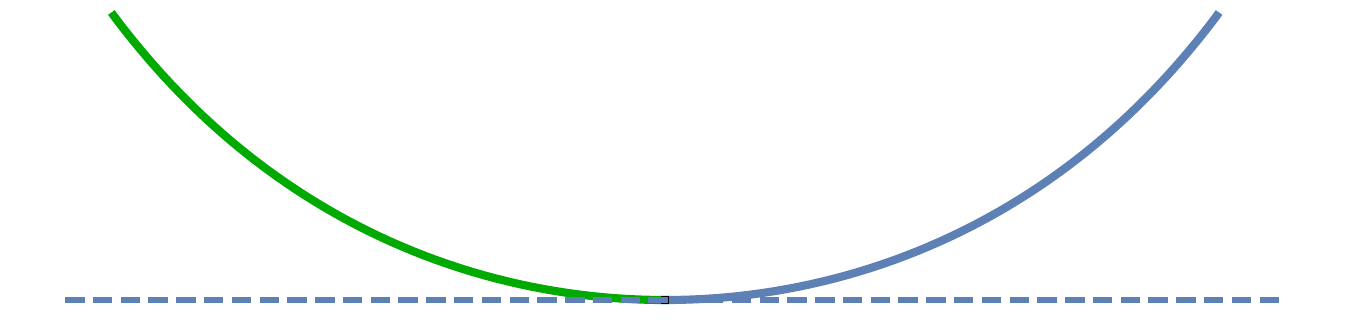}\label{geod14}} 
\caption{Second part of geodesic plots}
\label{GeodPlots2}
\end{figure}
%

\section{Discussion and outlook}\label{sec:Conclusions}

In this paper we examined the Lorentzian classical gravity geometries   associated, \`a la Banados \cite{Banados:1998gg},  to left-moving pure states in the dual CFT.
To test the validity of the classical geometric picture, we compared the behaviour of  the boundary-to-boundary two-point function evaluated in these geometries with CFT expectations.

For heavy primary states, we found that the geometry behaves as unreasonably thermal, leading to a chiral version of the information puzzle. One can argue, as in \cite{Fitzpatrick:2016ive}, that nonperturbative effects  should restore information.

We then studied a class of heavy, non-primary, pure states which are created by a dispersed collection of chiral vertex operators. The corresponding bulk geometry is a multi-centered solution with spinning particle sources, which is determined by a solution of classical Liouville theory with elliptic singularities.  We constructed the metric in the rotationally symmetric case of a circular shell of particles. It contains a throat geometry which is capped off by the shell. 

We then analyzed the probe two-point function in this background, whose late-time behaviour is governed by the capped throat geometry. When the shell is inserted  before  reaching the narrowest point of the throat (Fig. \ref{Figcap}(a)), we found that information is returned when the relevant geodesics start exploring the geometry of the cap. However, when the shell is inserted beyond the narrowest point of the throat (Fig. \ref{Figcap}(a)), the two-point function still decays. In the latter case, the classical geometry is not a reliable guide to the physics and finite $c$ corrections need to be considered to understand information recovery.   

From these considerations we conclude that the states for which the semiclassical picture gives a reliable guide to information restoration are, from the bulk point of view, loosely bound states which are spread out over a distance larger than the scale set by the narrowest size of the throat.  The states we considered in this work  do not carry the quantum numbers of an actual black hole, being above the $c/24$ treshold only on the left-moving side of the CFT. 
Though we have not shown that it is valid to do so, if we extrapolate our findings to the black hole regime, the geometries for which we found information to be returned  are star-like states spread out over super-horizon distances. As we discussed in Section \ref{sec:shellsol}, microstate geometries rather resemble the situation where   the matter shell is just outside of the narrowest point of the throat  as in Figure \ref{Figcap}(b).
In this case the geodesic grazes the shell without penetrating it and therefore we are at the limit where we can get an
understanding of information recovery from the classical picture.
This may point to  limitations of the use of microstate geometries  for understanding the return of information (somewhat similar to those raised in \cite{Raju:2018xue}), suggesting that typical microstates should rather be understood within the quantum regime. 
The importance of quantum effects within the fuzzball paradigm  has been stressed in  e.g. \cite{MathurFAQ}.

One possible caveat in the above is that some of the observed behaviour might be due the highly symmetric continuous distribution of particles operators we considered. This led to competing geodesics and swallow-tail phenomena which are especially significant when the shell is close to the narrowest point of the throat, making a complete analysis   difficult in that case.  It would  therefore be interesting to get a handle on the less symmetric situation with discrete centers.

In continuing along the lines of the present work there are various open problems which we leave for the future. Specifically we would like to address the following:
\begin{itemize}
    \item The most important issue to address is the generalisation of the constructions in the present work for spinning BTZ geometries with both sectors above the extremality bound. This would lead to the construction of classical microstates for non-extremal black holes within three-dimensional gravity.
    
    \item While the focus of this paper was on the bulk, it would be interesting to see how our results are reproduced by computations in the dual CFT at large $c$. For two-point functions in the Vaidya geometry such an analysis was performed in \cite{Anous:2016kss}.
    It would be particularly interesting to see how the CFT reproduces the various saddle points exchanging dominance;  most likely this would come from a saddle point analysis of the integral that results from summing up contributions of the vacuum block from all points of the shell, as discussed in \cite{Anous:2017tza}.
    
    \item As we mentioned above it would be interesting to explore less symmetric microstates consisting of several distinct particles in the bulk. An exact solution to such a problem is at the moment out of technical reach. However we believe one can gain some ground in this direction by using perturbative methods.
    
    \item As promoted in \cite{Castro:2018srf}, there is an alternative method for evaluating  the two-point function using Wilson lines. This method uses Chern-Simons variables which are more  suitable for treating spinning probe particles. It should be interesting to perform the analysis of the shell solution using this alternative method and confirm the results presented here.  
\end{itemize}

\begin{appendix}
\section{Left-thermal geometries as quotients of global $AdS_3$}\label{Appquotient}
To get more insight into the global properties of the overspinning BTZ metrics, it is useful to describe them as
 quotients of global AdS$_3$. This is most conveniently achieved by embedding the geometry into  a flat 4-dimensional ambient space with metric of signature $(2,2)$:
\be 
ds^2 = - (dX^0)^2+(dX^1)^2+(dX^2)^2-(dX^3)^2\label{embmetr}
\ee
Global $AdS_3$ is embedded as the submanifold
\be 
- (X^0)^2+ (X^1)^2+ (X^2)^2- (X^3)^2=-1.\label{adssubman}
\ee
The embedding coordinates can be arranged in an $SL(2, \RR)$ group element
\be 
g = \left( \begin{array} {cc} X^0 + X^1 & X^2 + X^3\\ X^2 - X^3 & X^0 - X^1 \end{array}\right)
\ee
in terms of which the metric reads
\be 
ds^2 = - \half \tr\, d g^{-1} d g.\label{embmetrmatr}
\ee
We define the left-moving temperature $T_L$ and the parameter $\t_R$ as
\be 
T_L = {\sqrt {M+J }\over 2 \p}, \qquad \t_R = {\sqrt {J-M }\over 2 \p}
\ee 
These are real and positive for the overspinning BTZ metrics of interest.
The appropriate  group element these metrics reads
\be 
g (u, x^+,x^-) = \left( \begin{array} {cc}e^{\p T_L x^+} & 0\\0 & e^{-\p T_L x^+} \end{array}\right)\left( \begin{array} {cc}\cosh \l (u) & \sinh \l (u)\\ \sinh \l (u) & \cosh \l (u) \end{array}\right) \left( \begin{array} {cc}\cos \p \t_R x^- & \sin \p \t_R x^-\\- \sin \p \t_R x^- & \cos \p \t_R x^- \end{array}\right)\label{groupel}
\ee 
where $\l(u ) $ is given by
\be
\l(u ) =\half \ln{ \sqrt{(u - \p^2 (T_L^2 - \t_R^2))^2 + 4 \p^4 T_L^2 \t_R^2 }- (u - \p^2 (T_L^2 - \t_R^2)) \over 2 \p^2 T_L \t_R}
\ee
We note that $\l (u)$ is real and finite for all $u \in \RR$; therefore the $u\leq 0$ region (with CTCs) is part of the geometry.
Substituting (\ref{groupel}) into (\ref{embmetrmatr}) we indeed find (\ref{BTZJ}), where
\be u = r^2, \qquad x^\pm = t \pm \f .\ee
Explicitly, the embedding coordinates are given by
\bea
 X^0 &=& \cosh \l (u) \cosh \p T_L x^+ \cos \p \t_R x^- +  \sinh \l (u) \sinh \p T_L x^+ \sin \p \t_R x^-\nonu
 X^1 &=& \sinh \l (u) \cosh \p T_L x^+ \sin \p \t_R x^- +  \cosh \l (u) \sinh \p T_L x^+ \cos \p \t_R x^-\nonu
 X^2 &=& -\sinh \l (u) \cosh \p T_L x^+ \cos \p \t_R x^- +  \cosh \l (u) \sinh \p T_L x^+ \sin \p \t_R x^-\nonu
 X^3 &=& \cosh \l (u) \cosh \p T_L x^+ \sin \p \t_R x^- -  \sinh \l (u) \sinh \p T_L x^+ \cos \p \t_R x^-
 \eea
 
We can now show that the overspinning geometries are regular quotients of AdS$_3$. The periodic identification $\f \sim \f + 2 \p$ implies at the level of the group element
\be g(u,x^+,x^-)\sim h_L(T_L)g(u,x^+,x^-)h_R(\t_R)
\ee where $h_L$ and $h_R$ are constant $SL(2,\RR)$ matrices of boost resp.  rotation type:
\be 
h_L = \left( \begin{array} {cc}e^{2 \p^2 T_L } & 0\\0 & e^{-2\p^2 T_L } \end{array}\right), \qquad h_R = \left( \begin{array} {cc}\cos 2 \p^2 \t_R & -\sin 2 \p^2 \t_R\\\sin 2 \p^2 \t_R & \cos 2 \p^2 \t_R \end{array}\right).
\ee
By inspecting the action on the embedding coordinates, one finds that the only fixed point is $X^0=X^1=X^2=X^3=0$, which lies outside the AdS submanifold  (\ref{adssubman}). We conclude that the overspinning BTZ geometry arises as a quotient of global AdS which acts without fixed points and is therefore completely smooth. The identification group is generated by $e^{2 \p \xi}$ where $\xi = \pa_\f$ is the infinitesimal  Killing vector. In terms of the $SO(2,2)$ Killing vectors of global AdS
\be 
J_{ab} = X_a {\pa \over \pa X^b }-  X_b {\pa \over \pa X^a },
\ee
where indices are lowered with $\h_{ab} = {\rm diag} (-1,1,1,-1)$, $\xi$ can be written as
\be \xi = \p T_L ( J_{10} + J_{23} ) + \p \t_R (J_{03}+ J_{21} )
\ee
To find out where this quotient belongs in the general classification of \cite{Banados:1992gq}, we write the generator as $ \x = \half \o_{ab} J^{ab}$ and consider the eigenvalues of $\o$. These are purely imaginary and given by
\be  
\pm i \p ( T_L + \t_R) , \pm i \p ( T_L - \t_R)
\ee 
and therefore the quotient giving rise to overspinning BTZ is of type $I_c$ in the classification of \cite{Banados:1992gq} App. A (rather than of type $I_a$ as was conjectured there).

In conclusion, in contrast to the situation higher dimensions, the overspinning BTZ geometry is a smooth manifold. The metric does contain other pathologies in the form of closed timelike curves  in the region $u<0$ where the generating Killing vector $\x$ is timelike.

\section{Coordinate transformation between AdS and BTZ}
AdS and BTZ spaces are locally isomorphic. Let us briefly recapitulate the coordinate transformation constituting this relation. In the Poincare coordinates the metric of AdS space reads
\begin{equation}
{\rm d}s^2  
=
\frac{1}{z^2} \left(  {\rm d}w_+ {\rm d}w_- + {\rm d}z^2 \right) 
\end{equation}
where we have set $\ell = 1$ and defined $w_{\pm} = x \pm t$. The desired coordinate transformation is
\begin{equation} \label{trans}
    w_{\pm} = \sqrt{\frac{r^2 - r_+^2}{r^2 - r_-^2}} \ex^{2\pi T_{L,R } (\phi \pm \tau)}
    \qquad
    z = \sqrt{\frac{r_+^2 - r_-^2}{r^2 - r_-^2}} \ex^{( r_+ \phi + r_- \tau)}
    \qquad
\end{equation}
where $T_{L,R} = \frac{r_+ \pm r_-}{2 \pi}$ are the left and right temperatures and $\phi\, , \tau$ are the boundary BTZ coordinates. Applying this coordinate transformation to the $AdS_3$ metric yields the BTZ metric.

\end{appendix}

\acknowledgments
We would like to thank Dionysios Anninos, Tarek Anous, Roberto Oliveri, Iosif Bena and Nicholas Warner for valuable discussions and correspondence.
This project was supported by the Grant Agency of the Czech Republic under the grant 17-22899S and by ESIF and MEYS (Project CoGraDS - CZ.02.1.01/0.0/0.0/15 003/0000437). O.V. would also like to thank the Erwin Schrodinger International Institute for Mathematics and Physics, for hospitality while this work was in progress.


\bibliographystyle{ytphys}
\bibliography{ref}

\end{document}